\documentclass[a4paper,11pt]{article}
\pdfoutput=1

\usepackage{jheppub} 

\usepackage[T1]{fontenc} 

\usepackage{graphicx}
\usepackage{epsfig}
\usepackage{rotating}
\usepackage{amssymb}
\usepackage{subfigure}
\usepackage{dsfont}
\usepackage{psfrag}
\usepackage{amsmath,euscript,array,mathrsfs}
\usepackage{axodraw}
\usepackage{bbold}
\usepackage{epsf}

\newcommand{\mt}[1]{\textrm{\tiny #1}}
\def\nc {N_\mt{c}}

\newcommand{\lqcd}{\Lambda_\mt{QCD}}
\newcommand{\lf}{\Lambda_\mt{flow}}

\newcommand{\sac}{\, , \qquad}
\newcommand{\qq}[1]{(\ref{#1})}

\newcommand{\SP}[1]{\begin{equation}\begin{split} #1
\end{split}\end{equation}}

\newcommand{\beq}{\begin{equation}}
\newcommand{\eeq}{\end{equation}}
\newcommand{\beqs}{\begin{eqnarray}}
\newcommand{\eeqs}{\end{eqnarray}}
\newcommand{\lsim}{\mathrel{\raisebox{-
.6ex}{$\stackrel{\textstyle<}{\sim}$}}}

\newcommand{\Tr}{{\rm Tr}}

\def\hbar{\hspace{0pt}\raisebox{1pt}{$-$} \hspace{-7pt} h}

\def\di{\mbox{d}}
\def\r{\rho}

\newcommand{\be}{\begin{equation}}
\newcommand{\ee}{\end{equation}}
\newcommand{\bea}{\begin{eqnarray}}
\newcommand{\eea}{\end{eqnarray}}

\def\lbldef#1#2{\expandafter\gdef\csname #1\endcsname {#2}}

\def\href#1#2{#2}

\newcommand{\ber}{\begin{eqnarray}}
\newcommand{\eer}{\end{eqnarray}}

\newcommand{\beqar}{\begin{eqnarray}}

\newcommand{\eeqar}{\end{eqnarray}}

\newcommand{\dsl}
  {\kern.06em\hbox{\raise.15ex\hbox{$/$}\kern-.56em\hbox{$\partial$}}}

\newcommand{\eeqarr}{\end{eqnarray}}
\newcommand{\ZZ}{{\rm \kern 0.275em Z \kern -0.92em Z}\;}

\def\CC{{\mathchoice
{\rm C\mkern-8mu\vrule height1.45ex depth-.05ex
width.05em\mkern9mu\kern-.05em}
{\rm C\mkern-8mu\vrule height1.45ex depth-.05ex
width.05em\mkern9mu\kern-.05em}
{\rm C\mkern-8mu\vrule height1ex depth-.07ex
width.035em\mkern9mu\kern-.035em}
{\rm C\mkern-8mu\vrule height.65ex depth-.1ex
width.025em\mkern8mu\kern-.025em}}}

\def\RR{{\rm I\kern-1.6pt {\rm R}}}

\def\ZZ{{\rm Z}\kern-3.8pt {\rm Z} \kern2pt}
\def\IB{\relax{\rm I\kern-.18em B}}
\def\ID{\relax{\rm I\kern-.18em D}}
\def\II{\relax{\rm I\kern-.18em I}}
\def\IP{\relax{\rm I\kern-.18em P}}

\newcommand{\bear}{\begin{eqnarray}}
\newcommand{\eear}{\end{eqnarray}}

\def\to{\rightarrow}

\def\to{\rightarrow}

\def\i{\iota}

\def\tt{\tilde{\theta}}

\def\6{\partial}

\def\bea{\begin{eqnarray}}
\def\eea{\end{eqnarray}}

\def\beqx{\begin{displaymath}}
\def\eeqx{\end{displaymath}}

\newcommand{\bmat}{\left(\begin{array}}
\newcommand{\emat}{\end{array}\right)}

\def\i{\iota}

\def\r{\rho}

\def\bo{{\raise-.3ex\hbox{\large$\Box$}}}               
\def\face{{\raise.2ex\hbox{$\displaystyle \bigodot$}\mskip-2.2mu \llap {$\ddot
        \smile$}}}                                   
\def\>{\rangle}                                      
\def\<{\langle}                                      

\def\leftrightarrowfill{$\mathsurround=0pt \mathord\leftarrow \mkern-6mu
        \cleaders\hbox{$\mkern-2mu \mathord- \mkern-2mu$}\hfill
        \mkern-6mu \mathord\rightarrow$}        
\def\dvec#1{\vbox{\ialign{##\crcr
        \leftrightarrowfill\crcr\noalign{\kern-1pt\nointerlineskip}
        $\hfil\displaystyle{#1}\hfil$\crcr}}}           
\def\Tr{{\rm Tr \,}}                                    

\def\-{\hphantom{-}}

\preprint{ICCUB-13-248, TAUP-2980/13}

\title{Multiscale confining dynamics from holographic \\ RG flows}

\author[a,b]{ Daniel Elander,}
\author[c,d]{ Anton F.~Faedo,}
\author[e]{Carlos Hoyos,}
\author[d,f]{David Mateos}
\author[c]{and Maurizio Piai}

\affiliation[a]{Department of Physics, Purdue University, 525 Northwestern Avenue, West Lafayette, IN 47907-2036, U.S.A.}
\affiliation[b]{Department of Theoretical Physics, Tata Institute of Fundamental Research,
Homi Bhabha Road, Mumbai 400 005, India}
\affiliation[c]{Department of Physics, College of Science, Swansea University,
Singleton Park, Swansea, Wales, UK}
\affiliation[d]{Departament de F\'\i sica Fonamental \&  Institut de Ci\`encies del Cosmos, Universitat de Barcelona, Mart\'{\i}  i Franqu\`es 1, E-08028 Barcelona, Spain}
\affiliation[e]{Raymond and Beverly Sackler School of
Physics and Astronomy, Tel-Aviv University, Tel-Aviv 69978, Israel
}
\affiliation[f]{Instituci\'o Catalana de Recerca i Estudis 
Avan\c{c}ats 
(ICREA), Passeig Llu\'\i s Companys 23, E-08010, Barcelona, Spain}

\date{\today}

\abstract{
We consider renormalization group flows between conformal field theories in five (six) dimensions with a string (M-theory) dual. By compactifying on a circle (torus) with appropriate boundary conditions, we obtain continuous families of confining four-dimensional theories parametrized by the ratio $\lf/\lqcd$, with $\lf$ the scale at which the flow between fixed points takes place and $\lqcd$ the confinement scale. We construct the dual geometries explicitly and compute the spectrum of scalar bound states (glueballs). We find a `universal' subset of states common to all the models. We comment on the modifications of these models, and the corresponding fine-tuning,  required for a parametrically light `dilaton' state to be present. We also comment on some aspects of these theories as probed by extended objects such as strings and branes.}

\begin{document}
\maketitle
\flushbottom

\section{Introduction}

\noindent
The study of strongly-coupled, confining gauge theories in four dimensions
is notoriously difficult. Gauge/gravity dualities~\cite{AdSCFT,reviewAdSCFT} offer an unprecedented opportunity to make progress on the analytical side. The basic idea (holography) is that instead of studying directly the strongly-coupled gauge theory, one studies a  weakly-coupled gravity theory in a curved background in five dimensions, and the field-theory observables can be computed by looking at the boundary values of fields (and extended objects) that are allowed to propagate in the bulk of the geometry. Since the two descriptions are supposed to be completely equivalent, if the gauge theory is a consistent quantum theory (an ultraviolet complete theory) then the dual description must be given by a consistent theory of quantum gravity such as string theory or M-theory. As is well known, the latter theories reduce to a tractable limit, namely classical supergravity, 
in the gauge theory limit of large number  of colors, $\nc \to \infty$, and strong coupling, $\lambda \to \infty$.

Two prominent examples of four-dimensional, confining  gauge theories
of phenomenological interest are Quantum Chromodynamics (QCD), the well-established theory of strong nuclear interactions, and Technicolor (TC)~\cite{TC,reviewsTC}, a hypothetical scenario for the dynamical origin of electro-weak symmetry breaking. Since neither of these theories is strictly speaking a large-$\nc$ theory, nor are they strongly coupled at all energy scales, they cannot be described in the  supergravity approximation of string or M-theory. It is therefore interesting to find as large a class of gauge theories as possible that share some of the fundamental properties of QCD or TC and that can be described with supergravity. Within this class of theories one then searches for generic results, the hope being that these may also be applicable to QCD or TC. Examples of this type of results are provided by the meson spectrum computed in \cite{Imoto:2010ef} for the model of \cite{Witten,SS}, which shows a remarkable  agreement with real-world QCD, or the computation of the shear viscosity to entropy density ratio for the quark-gluon plasma (see in particular~\cite{Son:2007vk} and references therein), in which leading-order calculations performed in simple holographic 
models yield results that are remarkably close to the real-QCD ones. Further directions in which this approach has been applied include the computation of the QCD and Yang-Mills spectrum \cite{Erdmenger:2007cm,Nunez:2010sf},  the behavior of QCD at finite temperature \cite{CasalderreySolana:2011us,DeWolfe:2013cua} and/or finite chemical potential \cite{Erdmenger:2007cm,Erdmenger:2008yj},  the construction of new models of electroweak symmetry breaking \cite{TCholo}, and the description of some strongly-coupled systems of interest for  the condensed matter \cite{McGreevy:2009xe} and cosmology \cite{cosmology} communities.

In this paper we revisit the glueball spectrum in four-dimensional, confining $SU(N_c)$ Yang-Mills gauge theories with large-$\nc$, intended as approximations of real QCD and/or of realistic TC models, in the limit in which the number of matter fields is small ($N_f\ll N_c$). The two main questions we want to ask are: (i) What is the spectrum of scalar glueballs,  and how does it depend on the details of the gravity model? (ii) Is it possible that one such glueball be anomalously light with respect to the rest of the spectrum, the corresponding physical state being predominantly the dilaton, and how does this depend on other dynamical features of the model?

The models we consider are generalizations of the Witten model~\cite{Witten} in that they depend on two dynamical scales. The starting point is a renormalization group flow between two conformal field theories (CFT) in five (six) dimensions. The dual supergravity solution is a domain-wall like geometry that interpolates between two AdS$_6$ (AdS$_7$) spaces. Since this geometry can be embedded into massive Type IIA string theory (M-theory),  we will refer to it as the `string model' (`M-theory model').  Each flow is characterized by the scale $\lf$ at which the `transition' between the ultraviolet (UV) and the infrared (IR) fixed points takes place. The choice of this scale is equivalent to a choice of units, and hence all the flows are physically equivalent. In order to produce a four-dimensional confining theory we then compactify the theory on a circle (torus) with appropriate  boundary conditions following \cite{Witten}. On the gravity side the solutions are constructed by placing a black hole at the bottom of the geometry and performing a double analytic continuation so that the circle (one of the circles) shrinks to zero size. The scale at which this happens is essentially the confinement scale $\lqcd$. In this way we obtain a continuous family of four-dimensional, confining theories parametrized by the dimensionless ratio $\lf/\lqcd$. 

We then focus on the first question above, namely the spectrum of 
scalar fluctuations. Interestingly, we find that this contains two types of states, one that is sensitive to the value of $\lf/\lqcd$ and one that is virtually insensitive to it. Moreover, a subset of the latter states is common to both the string and the M-theory models, suggesting a certain `universality'. 

The second  question is  related to the first one. In the context of TC, and in particular of walking TC~\cite{WTC,Yamawaki}, a  multi-scale variant of TC,
 already in  early papers~\cite{Yamawaki,Bando} it was suggested that 
  an anomalously light scalar particle might be present in the spectrum, the so-called `dilaton'.
There exists a vast
literature on the subject~\cite{HT,dilatonpheno,dilaton4,dilatonnew,dilatonandpheno,dilaton5D},
but a clear systematic understanding of what specific 
types of models give rise to a light dilaton is still missing.
The phenomenological relevance of this question arises from the fact that
such a particle might coincide with the Higgs particle discovered at the 
 LHC experiments ATLAS~\cite{ATLAS} and CMS~\cite{CMS},
because the main properties of the Higgs particle are due to 
the fact that it is itself a dilaton.

From the gauge/gravity perspective,  
there exist classes of models that resemble in many crucial aspects the dynamical properties of  walking theories~\cite{walking,NPR}. In one such class, an anomalously light 
scalar composite state has been identified~\cite{ENP}. However, the systematics behind this discovery is far from clear: it is not evident whether supersymmetry plays any role 
in the mechanism that makes the scalar light, and it is not known what other backgrounds 
would admit such a state. It is known that the lightness of the scalar is related to a large
VEV of a dimension-six operator, it is known that this is also related
to the fact that the gravity dual exhibits hyperscaling violation over some finite range of the radial direction,
and it is known that such a dimension-six operator 
is present in vast classes of gravity duals (see for instance~\cite{PZ}).

It is hence interesting to pose the same question in the class of models that we consider. We find that, generically, no parametrically light state exists regardless of the value of $\lf/\lqcd$. Presumably, the reason is that in these top-down models there is no parametric separation between the scales associated to the explicit and the spontaneous breaking of scale invariance.  
However, as we explain in the last section, it is possible to extend these models further by modifying the boundary conditions in the UV in such a way that a light state appears. As expected, though, this can only be done at the expense of introducing a considerable amount of fine-tuning.

\section{General formalism}
\label{general}

In this section we collect a set of general, technical results that are repeatedly employed in the
main body of the paper. We also set the notation and conventions used throughout the paper, and we highlight a few technical subtleties of general validity and interest.

\subsection{Five-dimensional $\sigma$-model}

The action of the five-dimensional $\sigma$-model we are interested in 
may in general involve $n$ scalars $\Phi^a$ coupled to  five-dimensional gravity.
We write the action as
\beqs
{\cal S}_5&=&\int \di^5 x \sqrt{-g_5}\left(\frac{{\cal R}_5}{4}
-\frac{1}{2}G_{ab}g^{MN}\partial_M\Phi^a\partial_N\Phi^b-V(\Phi^a)\right)\,,
\label{form}
\eeqs
where  capital roman letters refer to five-dimensional space-time
coordinates, with $a=1\,,\ldots\,,n$  the $\sigma$-model indexes,  $G_{ab}$ is the $\sigma$-model metric, $g_{MN}$
is the space-time metric, with signature $\{-\,,\,+\,,\,+\,,\,+\,,\,+\}$, $g_5=\det g_{MN}$, and ${\cal R}_5$ is the corresponding Ricci scalar.
Unless we specify otherwise, we are looking for `domain-wall-like' solutions in which the metric takes the form
\beqs
\di s_5^2 &=&e^{2A}\eta_{\mu\nu}\di x^{\mu}\di x^{\nu}+\di r^2\,,
\eeqs
with $A=A(r)$ and $\Phi^a=\Phi^a(r)$, and where the Greek indexes refer to the Minkowski directions
$\mu,\nu=0,\ldots,3$.
The classical equations of motion derived from this action are
\beqs
\Phi^{a\,\prime\prime}+4A^{\prime}\Phi^{a\,\prime}+{\cal G}^a_{\,\,\,\,bc}\Phi^{b\,\prime}\Phi^{c\,\prime}-V^a&=&0\,,\\
6A^{\prime\,2}+3A^{\prime\prime}+G_{ab}\Phi^{a\,\prime}\Phi^{b\,\prime}+2V&=&0\,,\\
6A^{\prime\,2}-G_{ab}\Phi^{a\,\prime}\Phi^{b\,\prime}+2V&=&0\,,
\eeqs
where the $^{\prime}$ denotes derivative with respect to the radial direction $r$, $V^a = G^{ab}\, \partial V/\partial \Phi^b$, and  ${\cal G}^a_{\,\,\,\,bc}$ is the $\sigma$-model connection (we follow the conventions of~\cite{EP}). 

\subsection{Spectrum of bound states}

Let us assume that the class of backgrounds described by the five-dimensional $\sigma$-model provides the dual gravity description of some four-dimensional, confining gauge theory. The confinement scale is set by
the end-of-space of the geometry. We want to compute the mass spectrum 
of scalar bound states of such a theory by fluctuating the bulk equations and boundary conditions.
The fluctuations of the  scalars $\Phi^a$ and of the metric couple in a quite non-trivial way.
Furthermore, because of the diffeomorphism invariance of the gravity theory, many combinations of the
original scalar and gravity fluctuations are spurious (pure gauge), and must be removed.
Also, given that the backgrounds are often known only numerically (since the equations
 for the fluctuations must be solved numerically), and because some of the background 
functions diverge in the UV and in the IR, it is necessary to introduce two regulators $0<r_I \ll r_U$,
 solve for fixed $r_I$ and $r_U$, and then take the limit $r_U\rightarrow +\infty$ and
$r_I\rightarrow 0$.

In order to proceed, it is convenient to use the gauge-invariant formalism  developed in~\cite{BHM}, and extended in~\cite{E} to the general case in which the $\sigma$-model does not admit a superpotential. We apply at $r_I$ and $r_U$ the boundary conditions discussed in~\cite{EP}, taking the special limit in which divergent boundary masses are added for all the $\sigma$-model scalars. After the regulators are removed, this effectively enforces the correct boundary conditions (regularity in the IR and normalizability in the UV) dictated by holography. Roughly speaking, the reason is that a large boundary mass penalizes the non-normalizable mode in the UV and the singular one in the IR. 
We refer the reader to~\cite{EP} for the derivation of the boundary conditions from the boundary actions of the $\sigma$-model itself.

Given a $\sigma$-model with $n$ scalars and 
action of the form \qq{form}, the scalar physical fluctuations are given by gauge-invariant functions denoted by $\mathfrak{a}^a$ and satisfying the bulk equations~\cite{E}
\beqs
\label{Eq:diffeq}
	0&=&\Big[ {\cal D}_r^2 + 4 A' {\cal D}_r + e^{-2A} \Box \Big] \mathfrak{a}^a \\ \nonumber
	&& - \Big[ V^a_{\ |c} - \mathcal{R}^a_{\ bcd} \bar \Phi'^b \bar\Phi'^d + \frac{4 (\bar \Phi'^a V_c + V^a \bar \Phi'_c )}{3 A'} + \frac{16 V \bar \Phi'^a \bar \Phi'_c}{9 A'^2} \Big] \mathfrak{a}^c ,
\eeqs
while the boundary conditions are given by~\cite{EP}
\beqs
\label{Eq:BCa}
	&\left[ \delta^a_{\ b} + e^{2A} \Box^{-1} \left( V^a - 4 A' \Phi'^a - \lambda^a_{\ |c} \bar \Phi'^c \right) \frac{2 \bar \Phi'_b}{3 A'} \right] {\cal D}_r \mathfrak a^b \Big|_{r_i} = \\
\nonumber	&  \left[ \lambda^a_{\ |b} + \frac{2 \bar \Phi'^a \bar  \Phi'_b}{3 A'} + e^{2A} \Box^{-1} \frac{2}{3 A'} \left( V^a - 4 A' \bar \Phi'^a - \lambda^a_{\ |c} \bar\Phi'^c \right) \left( \frac{4 V \bar \Phi'_b}{3 A'} + V_b \right) \right] \mathfrak a^b \Big|_{r_i}.
\eeqs
The meaning of all the notation is explained in detail in~\cite{EP}.\footnote{In particular, note that $\mathcal{R}^a_{\ bcd}$ is the Riemann tensor of the $\sigma$-model metric $G_{ab}$, not of the space-time metric $g_{MN}$.}

From the point of view of the gravity calculation, the two matrices $\lambda^a_{\ |c}$
(one defined in the UV and the other in the IR) are completely arbitrary, and come from 
the fact that one can add a localized mass term for any of the $\sigma$-model scalars,
without this affecting the boundary conditions for the background functions.
However, as explained above, by taking $\lambda\rightarrow \pm\infty$, and then taking the two regulators to  their physical values, one recovers the standard requirements of normalizability and regularity, and hence we do so.
The boundary conditions take in this case the simpler form:
\beqs
\label{Eq:BCb}
	&\left[  e^{2A} \Box^{-1}  \, \frac{2  \bar \Phi'^c }{3 A'} \right] 
	\left( \bar \Phi'_b{\cal D}_r -\frac{4 V \bar \Phi'_b}{3 A'} - V_b \right)\mathfrak a^b \Big|_{r_i} =  
	-\mathfrak a^c \Big|_{r_i}.
\eeqs

All the calculations of the spectra are performed in the following way. We choose  a third value $\bar{r}$ of the radial coordinate, with 
 $r_I \ll \bar{r} \ll r_U$, and replace $\Box \to M^2$ in the equations above for a fix trial value $M^2 > 0$, where $M^2$ is the four-dimensional mass squared of the fluctuations. 
 We select $n$ independent $n$-vectors $u_j^a$ (in the internal space of the $\sigma$-model), 
so that $\det u_j^a \neq 0$.
 We impose the IR boundary conditions on the $n$ fluctuations $\mathfrak{a}_{Ij}(r_I)$ and their derivatives,
 using the $n$-vector $u_j^a$, together with Eq.~(\ref{Eq:BCb}),
  and use the bulk equations
to evolve the resulting $n$ independent solutions up to $\bar{r}$, obtaining $\mathfrak{a}_{Ij}(\bar{r})$. 
 The UV ($r=r_U$) is treated similarly, and evolving down we obtain $\mathfrak{a}_{Uj}(\bar{r})$. 
With this data, we build the $2n\times 2n$ real matrix
\beqs
\mu (M^2)&=&\left(\begin{array}{ccc}
\mathfrak{a}_{Ij}(\bar{r})  &|& \mathfrak{a}_{Uk}(\bar{r})\cr
&|& \cr\hline &|&  \cr
\partial_{r}\mathfrak{a}_{Ij}(\bar{r}) &|& \partial_{r}\mathfrak{a}_{Uk}(\bar{r})\cr
\end{array}\right)\,,
\eeqs
and compute $\det \mu$. We then iterate by varying  $M^2$, 
and look to find  the zeros of $\det \mu(M^2)$.
At this point we have the spectrum, for the given choices of $r_I$ and $r_U$.
We can then repeat the whole procedure for different values of the regulators, and study the
extrapolation to the physical limits.

\subsubsection{The probe approximation}

We briefly digress in this subsection, to explain an important property of the system of equations and boundary conditions for the fluctuations.
The crucial observation we start with is that the gauge-invariant variables $\mathfrak{a}^a$ 
are defined by~\cite{BHM}:
\beqs
\mathfrak{a}^a&=&\varphi^a\,-\,\frac{\bar \Phi^{\prime\,^a}}{6A^{\prime}} h\,,
\eeqs
where $\varphi^a$ is the fluctuation of the scalars with respect to their background values, i.e.~ 
\be
\Phi^a(x^{\mu},r)=\bar \Phi^a(r)+\varphi^a(x^{\mu},r) \,,
\ee
while $h(x^{\mu},r)$ is related to the (four-dimensional) trace of the fluctuation of the metric. In particular, $h$ couples to the trace of the stress-energy tensor at the boundary. 
In this sense, if the lightest state in the theory
 is predominantly $h$, up to small mixing terms with the sigma-model scalars,
it is the pseudo-dilaton.\footnote{In models with  
spontaneous breaking of exact scale invariance, one has to treat the  
massless dilaton separately. This is not a realistic situation, and  
hence we do not discuss it in this paper.}$^{,\,}$\footnote{We emphasize that, since only the combinations $\mathfrak{a}^a$ are gauge-invariant, both $\varphi^a$ and $h$ can be separately changed by a gauge transformation. However, fluctuations of scalar fields whose background is constant do not mix with the metric and are gauge-invariant (and ``non-dilatonic'') by themselves. For this reason, 
as long as the pseudo-dilaton is parametrically lighter than the other bound states, its properties are well approximated by those of the fluctuation $h$. This is reminiscent of what happens in a spontaneously-broken gauge theory, where processes mediated by the massive gauge bosons are for all practical purposes (and as long as the mass is small) determined by the physics of the corresponding would-be Goldstone bosons. This holds true irrespectively of the fact that the Goldstone mode can be, for instance, gauge-fixed to zero.} 

All the complications in the bulk equations and in the boundary conditions come from the mixing
of the fluctuations of the scalars with those of the metric, specifically with $h$.
Let us hence do the following exercise: what do the bulk equations and boundary conditions look like 
in a case where the probe approximation is legitimate?
A way to define what is meant by probe approximation is that we can think of situations in which
\beqs
\nonumber
\frac{\bar \Phi^{\prime\,a}}{A^{\prime}}&\ll&1\,
\eeqs
for all $a=1\,,\,\ldots\,,\,n$ and for every $r>0$. 
We also make the simplifying assumption that the $\sigma$-model metric be 
flat, in the sense that ${\cal G}^{a}_{\,\,\,\,bc}=0$.
We then use these two assumptions to rewrite the bulk equations and boundary conditions 
in the drastic approximation in which $\bar \Phi^{\prime\,a}/A^{\prime}=0$.
This is equivalent to stating that the scalars do not back-react on the geometry.

In this limit, we see that the bulk equations become
\beqs
\left[\partial_{r}^2+4A^{\prime}\partial_{r}+M^2 e^{-2A}\right] \mathfrak{a}^a\,-\,G^{ab}\frac{\partial^2 V}{\partial \Phi^b\partial\Phi^c}\,\mathfrak{a}^c&=&0\,,
\eeqs
and the boundary conditions reduce simply to
\beqs
\left.\frac{}{}\mathfrak{a}^a\right|_{r_i}&=&0\,.
\eeqs
These are the familiar equations one is used to (from the AdS/CFT literature) for a system of probes,
the dynamics of which is influenced by the background, but not the other way around.
Notice in particular that the boundary conditions reduce simply to Dirichlet,
and that only a mass term (second  field-derivative of the potential) enters the bulk equations, besides the 
three terms that come from the Laplacian in curved space.

To some extent, this digression is intended to illustrate the fact that in more familiar cases these seemingly complicated  equations reduce to the usual results.
But we also want to make a far more important point.
The probe approximation is defined by the very assumption that the dilaton decouples from the  fluctuations of the scalars. 
Indeed, in the limit we are considering in this digression, there is no remnant of $h$ to be found anywhere.
In particular, if a light pseudo-dilaton is present, it cannot be found in the probe approximation.
It is hence necessary to use the complete boundary conditions in order 
to even ask one of the questions we are interested in, and there is no such a 
thing as a limit in which one can neglect the back-reaction. 

Notice another very important fact, visible from the complete form of the boundary conditions and bulk equations.
When dealing with a set of $n$ scalars, there exist two distinct sources of mixing between their fluctuations:
besides the obvious one coming from the scalar potential (and from the $\sigma$-model metric, if curved),
there is a less obvious one that arises due to the fact that for any of the fields having a non-trivial ($r$-dependent)
bulk profile, their fluctuations mix with $h$, and hence all of the fields that have a 
role in determining the bulk geometry ultimately mix with one another.
This also means that in order to compute the spectrum of glueballs
of a Yang-Mills theory (or closely related) one must use the bulk equations and boundary conditions
we reported here, since it is understood the the bulk dynamics is supposed to be 
determined by the dynamics of the glue.

\subsection{Wilson loops}
\label{Sec:Wilson}

Given a supergravity background that can be obtained as the low-energy theory of 
a superstring theory, one can compute the expectation value of the rectangular Wilson loop 
extended over a time interval $T$ and a space interval $L_{QQ}$
by using the standard prescription~\cite{Rey:1998ik}. 
As we will see, it is more natural to work in a new radial coordinate $\r$, which
will be defined for all the models we are interested in later on.
We embed an open string with 
endpoints separated along a space-like direction on a regulating brane at some large value of the radial coordinate $\r_U$, and allow the string to fall  into the radial direction.

We start from the string-frame metric in ten dimensions and define the functions (for convenience, we follow the notation of~\cite{NPR}):
\beqs
F^2&\equiv&g_{tt}g_{xx}\,,\\
G^2&\equiv&g_{tt}g_{\r\r}\,,\\
V_{\rm eff}^2(\r,\hat{\r}_o)&\equiv&\frac{F^2(\r)}{F^2(\hat{\r}_o)G^2(\r)}\left(F^2(\r)-F^2(\hat{\r}_o)\right)\,,\\
{\cal Z}&\equiv& \partial_{\r}\left(\frac{G(\r)}{\partial_{\r}F(\r)}\right)\label{DefZ}\,,\\
L_{QQ}(\hat{\r}_o)&=&2\int_{\hat{\r}_o}^{\r_U}\di \tilde{\r}\frac{1}{V_{\rm eff}(\tilde{\r},\hat{\r}_o))}\,,\\
E_{QQ}(\hat{\r}_o)&=&2\int_{\hat{\r}_o}^{\r_U}\di \tilde{\r}\sqrt{\frac{F^2(\tilde{\r})G^2(\tilde{\r})}{F^2(\tilde{\r})-F^2(\hat{\r}_o)}}\,.
\eeqs
The origin of $F$ and $G$ is the Nambu-Goto action:
\beqs
{\cal S}&=&\frac{1}{2\pi\alpha^{\prime}}\int_{[0,T]}\di \tau \int_{[0,2\pi]} \di \sigma \sqrt{- g}\,.
\eeqs
Making the choice of parameterization $t=\tau$, $x=x(\sigma)$ and $\r=\r(\sigma)$,
which means that the string is stretched along the space-like directions, yields
\beqs
{\cal S}&=&\frac{T}{2\pi\alpha^{\prime}} \int \di \sigma \sqrt{F^2x^{\prime\,2}+G^2 \rho^{\prime\,2}}\,,
\eeqs
where the prime refers to derivatives with respect to the world-sheet coordinate $\sigma$.
From the classical equations of motion derived from this action one finds
that the profile of the string is determined by the minimum value $\hat{\r}_o$ 
reached by the string profile, and the Euclidean distance $L_{QQ}$ and energy $E_{QQ}$
are given by the expressions we wrote above, 
with ${2\pi\alpha^{\prime}}=1$.

Let us remind ourselves of some important general results.
In order for the calculation of the Wilson loop to be doable, and to yield a 
confining linear potential at long distances, one must find that (see also \cite{Kinar:1998vq})
\begin{itemize}
\item $F^2$ is monotonically increasing,
\item $\lim_{\r\rightarrow +\infty}V_{eff}=+\infty$,
\item $V_{eff} \propto \r$ for $\r\rightarrow 0$.
\end{itemize}
If all of this is verified, then one also finds that the string tension is
\beqs\label{stringtension}
\sigma&\equiv&\left.\frac{\di E_{QQ}}{\di L_{QQ}}\right|_{L_{QQ}\rightarrow \infty}\,=\,F(0)\,.
\eeqs

Furthermore, if ${\cal Z}<0$ for any $\r>0$, 
then the function $L_{QQ}(\hat{\r}_o)$ is monotonically decreasing. In particular, this means that $L_{QQ}$ is a single-valued function of $\hat{\r}_o$ and hence that there are no phase transitions as $L_{QQ}$ varies from zero to infinity \cite{Faedo:2013ota}.

Let us explain this point in some more detail.
A dramatic signature of the presence of an intermediate region where the physics changes drastically
(such as is the case for walking theories) would be the emergence of a first-order quantum phase transition
in the (weakly-coupled) system living on the probe.
This has been observed to take place in the walking solutions 
within the conifold~\cite{NPR}.
Namely, for  $\hat{\r}_0$ in the walking region, one finds that the string configurations are unstable.
In terms of the separation this happens because the same value of $L_{QQ}$
 corresponds to three distinct configurations, with different energy $E_{QQ}$, and thus inevitably $L_{QQ}(\hat{\r}_0)$ is not monotonic.
But confinement necessarily means that $\di L_{QQ}/\di \hat{\r}_{0} < 0 $ near the end of space, when $\hat{\r}_0\rightarrow 0$.
Also, UV-completeness requires that   
$\di L_{QQ}/\di \hat{\r}_{0} < 0 $ in the far-UV, when $\hat{\r}_0\rightarrow \r_U$.
Hence, there  can be only two possibilities: either $\di L_{QQ}/\di \hat{\r}_{0} < 0 $  for every possible $\hat{\r}_0$,
in which case the behavior of the quark-antiquark potential resembles qualitatively that of a QCD-like theory,
or there may exist some intermediate range of $\hat{\r}_0$ for which $\di L_{QQ}/\di \hat{\r}_{0} > 0 $.
It turns out that this is comparatively easy to check, and does not require to actually integrate $L_{QQ}$.
From the careful analysis of the expression of $L_{QQ}$ one finds that the 
sign of $\di L_{QQ}/\di \hat{\r}_{0}$ is controlled by the sign of ${\cal Z}$ defined in Eq.~(\ref{DefZ}). Hence, by simply looking at the sign of ${\cal Z}$ we can exclude the presence of such non-trivial discontinuous behaviors. We anticipate here that in all the models we are going to discuss in the paper, ${\cal Z}<0$, so that there is no phase-transition of this special type in any of them.
\newline

For later reference we close this section by reproducing the famous calculation of~\cite{Rey:1998ik},
namely the calculation of the rectangular Wilson loop in ${\rm AdS}_5\times S^5$, with $T\gg L_{QQ}$.
The ten-dimensional metric of Type IIB is given by 
(there is no distinction between string-frame and Einstein-frame, since the dilaton
is constant and we fix it to $\Phi=0$):
\beqs
\di s^2_{10}&=&\frac{r^2}{R^2}\di x^2_{1,3}\,+\,\frac{R^2}{r^2}\left(\di r^2 + r^2\di \Omega^2_5\right)\,\\[2mm]
&=&e^{2\frac{\r}{R}}\di x_{1,3}^2\,+\,\di \r^2\,+\,R^2\di \Omega_5^2\,,
\eeqs
where $R$ is the radius and $\di \Omega_5^2$ is the metric of $S^5$. The background functions $F$ and $G$ are given by
\beqs
F^2&=&\frac{r^4}{R^4}\,=\,e^{4\frac{\r}{R}}\,,\\
G^2&=&\frac{r^2}{R^2}\,=\,e^{2\frac{\r}{R}}\,.
\eeqs
The separation $L_{QQ}$ is finite in the limit $\r_U\rightarrow +\infty$, and depends only on $\hat{\r}_0$:
\beqs
L_{QQ}(\hat{\r}_0)&=&\frac{2 \sqrt{\pi } R \Gamma \left(\frac{3}{4}\right)
   e^{-\frac{\hat{\r}_0}{R}}}{\Gamma \left(\frac{1}{4}\right)}\,.
\eeqs
As for $E_{QQ}$, there is a divergent term, that we need to subtract:
\beqs
E_{QQ}(\hat{\r}_0)&=&
R \left({2}{}e^{\frac{\r_U}{R}}-\frac{2 \sqrt{\pi } \Gamma \left(\frac{7}{4}\right)
   e^{\frac{\hat{\r}_0}{R}}}{3 \Gamma \left(\frac{5}{4}\right)}\right)
\eeqs
After subtracting the divergent part, the finite energy $\bar{E}_{QQ}$ is
\beqs
\bar{E}_{QQ}&=&-\frac{\pi  R^2 \Gamma \left(\frac{3}{4}\right)^2}{\Gamma \left(\frac{1}{4}\right)
   \Gamma \left(\frac{5}{4}\right)}\,\frac{1}{L_{QQ}}\,,
\eeqs
in perfect agreement with~\cite{Rey:1998ik}, up to the fact that 
we took $\alpha^{\prime}=\frac{1}{2\pi}$ instead of $\alpha^{\prime}=1$ (or, equivalently, up to the fact that our 
definition  of $R$ differs by factors of $\alpha^{\prime}$).

\subsection{Running gauge couplings and glueballs}
\label{running}

In this section we remind the reader of a few aspects of the gauge theory dual that will be important in subsequent sections. 

As is well known, in the simple case of ${\cal N}=4$ SYM, which is dual to Type IIB string theory on ${\rm AdS}_5 \times S^5$, the gauge theory coupling is given by
\beqs
\frac{1}{g_{YM}^2}&=&\frac{ T_3(2\pi\alpha^{\prime})^2e^{-\Phi}}{2}\,,
\label{analogous}
\eeqs
where $T_3$ is the tension of a D3-brane.
This result can be derived, for example, by placing a D3-brane probe in the ${\rm AdS}_5 \times S^5$ background, expanding the Dirac-Born-Infeld action to quadratic order in the gauge fields, and reading off the coefficient in front of this term. 

With the replacement $T_p=(2\pi)^{-p}(\alpha^{\prime})^{\frac{-p-1}{2}}$, modified with respect to~\cite{Bertolini:2003iv} 
 by the convention from~\cite{Itzhaki:1998dd}
that $g_s=e^{\Phi_{\infty}}$, one has\footnote{
In~\cite{Bertolini:2003iv} one finds that $\frac{T_p}{k}=\frac{(2\pi)^{-p}(\alpha^{\prime})^{\frac{-p-1}{2}}}{g_s}$,
and the dilaton $\Phi$ is chosen so that $\Phi_{\infty}=0$. Which is, of course, equivalent to the present expressions.}
\beqs
\frac{1}{g_{YM}^2}&=&\frac{ e^{-\Phi}}{4\pi}\,=\,\frac{ e^{\Phi_{\infty}-\Phi}}{4\pi g_s}\,.
\eeqs
This shows that the running gauge coupling is given by the exponential of the ten-dimensional dilaton. In the specific case of ${\cal N}=4$ SYM the field $\Phi$ is constant, as expected for a CFT, and hence this translates into the relation  $4\pi g_s=g_{YM}^2$, the 't Hooft coupling being $\lambda\equiv g_{YM}^2 N_c=4\pi g_s N_c$. This is the famous result implying that the 't Hooft large-$N_c$ limit is the classical limit in which $g_s\rightarrow 0$. 

The gauge theory operator dual to $\Phi$ is actually the ${\cal N}=4$ Lagrangian, 
\be
{\cal L}= \Tr F_{\mu\nu}F^{\mu\nu} + \cdots \,.
\ee
Normalizable states of $\Phi$ on the gravity side thus correspond to poles in the two-point function of $\cal L$. Normalizable modes of other supergravity fields correspond to poles of two-point functions of other operators. In particular, normalizable modes of the scalars in the  five-dimensional effective actions (the $\sigma$-models) that we will consider correspond to poles of scalar operators. Since all these states correspond to bound states of gluons and adjoint matter in the gauge theory, we will collectively refer to them as `glueballs'. 

Consider now a higher-dimensional case that will be of interest in this paper, namely the five-dimensional gauge theory living on the world volume of $\nc$ D4-branes. The dual effective five-dimensional string-frame metric can be written as  
\beqs
\di \tilde{s}^2_5 &=&\tilde{g}_{xx}\di x^2_{1,3}\,+\,\tilde{g}_{\eta\eta} \di \eta^2\,,
\eeqs
where $\eta$ is the fifth coordinate. The five-dimensional gauge coupling is given by an expression analogous to (\ref{analogous}), namely by 
\beqs
\frac{1}{g_5^2}&=&\frac{T_4(2\pi\alpha^{\prime})^2}{2}e^{-\Phi}\,=\,\frac{e^{-\Phi}}{16\pi^3 \sqrt{\alpha^{\prime}}}\,=\,\frac{e^{-\Phi}}{16\pi^3 \ell_s}\,,
\label{5Dcoupling}
\eeqs
where $\ell_s$ is the string length. Note that this is dimensionful, as expected for a gauge theory in five dimensions. If $\eta$ is periodically identified with period $2\pi R_5$, then the effective four-dimensional physics at energies below $\lqcd = 1/R_5$ is controlled by the four-dimensional coupling 
\beqs
\frac{1}{g_4^2 }&=&\frac{T_5(2\pi \alpha^{\prime})^2}{2}e^{-\Phi}\int\di \eta \sqrt{\tilde{g}_{\eta\eta}}\,=\,\frac{2\pi R_5}{g_5^2}\sqrt{\tilde{g}_{\eta\eta}}\,=\,\frac{e^{-\Phi} R_5}{8\pi^2}\,\sqrt{\frac{\tilde{g}_{\eta\eta}}{\alpha^{\prime}}}\,.
\label{4Dcoupling}
\eeqs
This is dimensionless, as expected. We will see later that, in models in which confinement is generated dynamically, $\lqcd$ sets the confinement scale, hence our choice of nomenclature. 

We conclude with three important observations. First of all, in the case $D>4$, for example in the D4-brane theory,  the effective four-dimensional gauge coupling entails other functions besides the dilaton. It is hence necessary to include the $\sigma$-model scalars they involve in the calculations yielding the properties of the glueballs. Secondly, the ten-dimensional dilaton $\Phi$ is in  general not one of the scalars of the five-dimensional $\sigma$-model, but some complicated
 combination of them whose determination requires knowing the lift to ten dimensions. And finally, the dilaton and the functions appearing in the metric are not constant if the background is dual to a confining theory 
 (for which there is a running of the gauge coupling), and hence one cannot use the probe approximation.
 A particularly striking illustration of all of these aspects is provided by the Klebanov-Strassler system~\cite{KS},
 for which the dilaton is constant, and yet the dual gauge couplings run.

\subsection{Embedding probe D$8$-branes}

All the Type IIA backgrounds we will consider have an internal space that 
consists of a circle parameterised by an angle $\eta$ and an internal four-dimensional manifold
parameterised by the angles $\theta$, $\varphi$, $\psi$ and $\xi$ (the model-dependent details will be explained 
in due time).
It is hence possible, in principle, to study the embedding of a probe D8-brane
that extends in the four Minkowski directions and wraps the internal four-dimensional manifold.
The embedding is then characterised by the functions $\r(\sigma)$ and $\eta(\sigma)$.
The idea of~\cite{SS} is to find embeddings of this type which have a U-shape in the $(\r,\eta)$-plane,
study the physics of fluctuations of such embeddings and interpret the results in terms of 
the mesons of the dual (chiral symmetry-breaking) strongly-coupled  large-$N_c$ theory with a small number of fundamental quarks.   

We will show that this program can be repeated in several relevant examples later on.
We report here the general setup of the derivation of the embedding functions.
The starting point is the DBI action, in terms of the (string-frame) induced metric $\tilde{g}_{MN}$:
\beqs
{\cal S}_{\rm D8}&=&-T_{8}\int \di^4 x \di\theta\di \varphi\di \psi \di \xi \di \sigma \,e^{-\Phi}\sqrt{-{\rm det}\left(\tilde{g}+B_2 +2\pi\alpha^{\prime} F_2\right)}\,,
\eeqs
We assume that $F_2=B_2=0$ for simplicity, and use the embedding ansatz $\rho=\rho(\sigma)$ and 
$\eta=\eta(\sigma)$.
After integrating over the other eight dimensions, and fixing the constants so that the 
normalisations cancel, the action takes the form
\beqs
{\cal S}_{\rm D8}&=&- \int \di \sigma \, \sqrt{\tilde{F}^2\eta^{\prime\,2}\,+\,\tilde{G}^2\r^{\prime\,2}}\,,
\label{d8}
\eeqs
which, aside from the replacements $F,G\rightarrow \tilde{F},\tilde{G}$ has the same form 
as the action we looked at for the rectangular Wilson loop.
Hence, all the same considerations apply, once this replacement is done.
In particular, this will allow us to straightforwardly check whether the U-shape embeddings
exist for any values of the parameter $\hat{\r}_0$, which (as in the case of the string probe) is the turning point of the configuration in the radial direction $\r$.

\section{A string theory model}

In this section we study a multi-scale, confining four-dimensional theory with a string theory dual. On the gauge theory side, the 4D theory is obtained by compactifying a 5D CFT on a circle with appropriate boundary conditions.  On the gravity side, the starting point is massive Type IIA supergravity \cite{RomansIIA}. Compactification of this theory on $S^4$ as in \cite{IIAEmbedd} yields the six-dimensional  $F(4)$ gauged supergravity constructed by Romans~\cite{Romans:1985tw}. This six-dimensional theory admits two AdS$_6$ backgrounds (one of which is supersymmetric), and can be further truncated to a 6D $\sigma$-model containing only one scalar $\phi$  coupled to gravity.  A class of solutions driven by $\phi$ that interpolates between the two AdS$_6$ geometries was constructed in~\cite{5DRGFlow}. Reducing the 6D $\sigma$-model on a circle finally produces the 5D $\sigma$-model that we will use in our calculations, and that contains a second scalar $\chi$ coming from the size of the circle. As already explained, fluctuations of these scalars will correspond to a subsector of the $0^{++}$ spectrum of glueballs. The quantum numbers can be fixed by considering how these modes couple to a D4-brane, as was done for the Witten model for instance in~\cite{gravityspectrum2}. In the case of $\chi$, the argument for assigning positive parity and charge is parallel to that in~\cite{gravityspectrum2}, while the new scalar $\phi$ is directly related to the dilaton in 10D, the couplings of which were detailed in sec.~\ref{running}.

We note that, thanks to recent progress on non-abelian T-duality~\cite{Tduality}, the Type IIA truncation can be dualized to a Type IIB truncation, which provides an alternative lift of the 5D $\sigma$-model to ten dimensions on which we will comment  following~\cite{Jeong:2013jfc}. We display in Fig.~\ref{Fig:Ob} the relations between all the theories mentioned  here. 

\begin{figure}[t]
\begin{center}
\includegraphics[height=5.5cm]{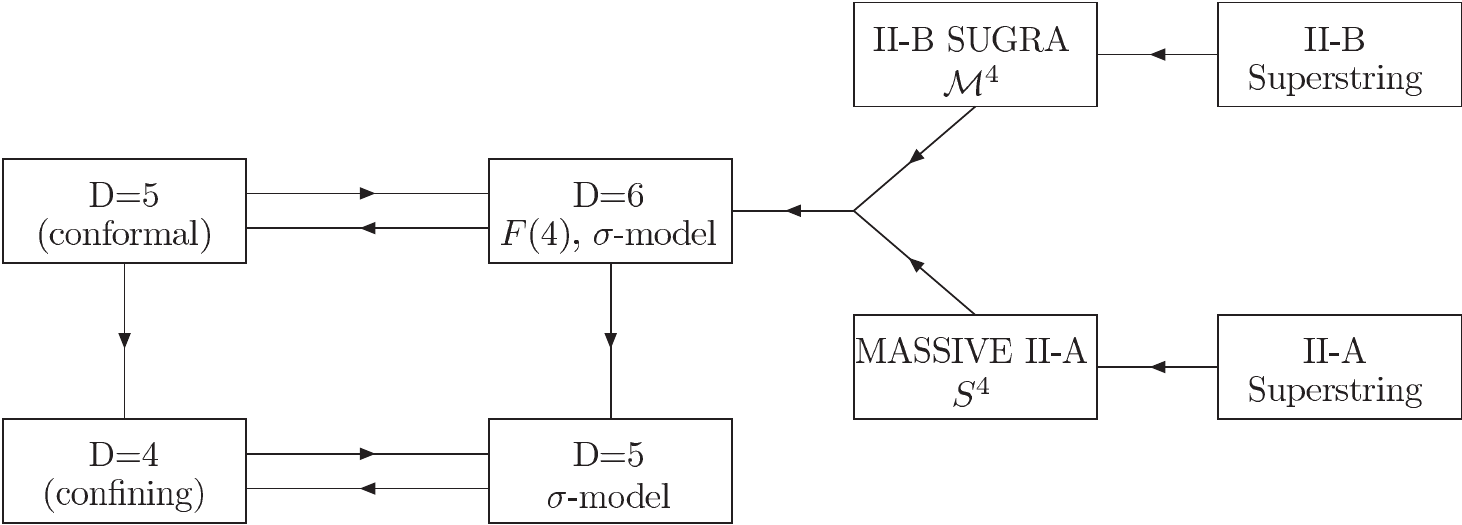}
\end{center}
\caption{Diagramatic representation of the relations between various theories 
in reference to the compactification of $F(4)$, $D=6$ supergravity.
}
\label{Fig:Ob}
\end{figure}

\subsection{The model}

\subsubsection{The $\sigma$-model in five dimensions}

We start from the five-dimensional formulation of the model, useful for the computation of the spectrum, and then present the successive steps needed for its embedding in string theory. The action is given by \qq{form} supplemented by appropriate boundary terms, for which we follow the conventions in~\cite{EP}. In the specific case we are interested in, there are $n=2$ scalars $\Phi^a=(\phi,\chi)$,
and the $\sigma$-model metric is
\beqs
G_{ab}&=&{\rm diag}\,\left\{2,6\right\}\,,
\eeqs
so that ${\cal G}^a_{\,\,\,\,bc}=0$. The potential is
\beqs
V&=&e^{-2\chi}\,{\cal V}(\phi)\,,
\eeqs
with 
\beqs
{\cal V}(\phi)&=&\frac{1}{9}\left(\frac{}{}e^{-6\phi}-9e^{2\phi}-12e^{-2\phi}\right)\,.
\eeqs

\subsubsection{The $\sigma$-model in six dimensions}

The same system can be rewritten also in six dimensions.  
This is actually its natural definition,
since the model is a consistent truncation of the $F(4)$
gauged supergravity in $D=6$ constructed by Romans, that 
has been later discovered to be the $S^4$-reduction of massive Type IIA. Let us briefly summarize here the 6D gravity theory.
It is understood that the equations we will get must agree with those of the 5D case.
The $\sigma$-model consists of only one scalar $\phi$, for which we write the action
as
\beqs
{\cal S}_6&=&\int \di^6 x \sqrt{-g}\left[\frac{{\cal R}_6}{4}\,-\,g^{MN}\partial_M \phi \partial_N \phi \,-\, {\cal V}(\phi)\right]\,,
\eeqs
where the potential ${\cal V}(\phi)$ has been defined earlier.
The ansatz for the six-dimensional metric that takes the 6D $\sigma$-model to the 5D $\sigma$-model is
\beqs
\di s_6^2=e^{-2\chi}\di s_5^2+e^{6\chi} \di \eta^2\,,
\label{subs}
\eeqs
where $0\leq \eta < 2\pi$  is the coordinate of the circle, whose size is parameterized by the second scalar $\chi$ appearing in the 5D model. This metric can be rewritten in a number of ways that will be useful below: 
\beqs
\di s_6^2
&=&e^{2A-2\chi} \di x_{1,3}^2\,+\,e^{-2\chi}\di r^2\,+\,e^{6\chi} \di \eta^2\,\\[1.5mm]
&=&e^{2A-2\chi} \left(\di x_{1,3}^2\,+\,e^{-2A+8\chi}\di \eta^2\right)\,+\,\di \r^2\,\\[1.5mm]
&=&e^{2\hat{A}} \left(\di x_{1,3}^2\,+\,e^{-2\hat{A}+6\chi}\di \eta^2 \right)\,+\,\di \r^2\label{DWansatz}\,.
\eeqs
Note that we made the change of radial variable $e^{-\chi} \di r=\di \r$,
 that $\hat{A}=A-\chi$ is the six-dimensional warp-factor, and that the particular value $\hat{A}=3\chi\propto \r$, or equivalently $A=4\chi\propto \r$, 
yields (locally) the AdS$_6$  geometry.

\subsubsection{Lift to massive Type IIA}

The 10D background within massive Type IIA contains the metric, the dilaton $\Phi$, the four-form $F_4$ and the mass parameter $m$, as detailed in \cite{IIAEmbedd,massiveIIA}.
The Einstein-frame metric reads
\beqs
\di s_{10}^2 &=&(\sin\xi)^{\frac{1}{12}}X^{\frac{1}{8}}\Delta^{\frac{3}{8}}\left[\frac{}{}
\di s_6^2\,+\,\frac{1}{g^2}\di \tilde{\Omega}_4^2\right]\,,
\eeqs
where we fix the radius of the sphere to $g=1$ in order to be consistent with the conventions we adopted for the 6D and 5D  $\sigma$-model equations. The various functions appearing in the metric are
\beqs
X&=&e^{\phi}\,,\\
\Delta&=&X^{-3}\sin^2\xi\,+\,X\cos^2\xi\,,\\
\di \tilde{\Omega}_4^2 &=&X^2 \di \xi^2 \,+\, X^{-1}\Delta^{-1}\cos^2\xi \frac{1}{4}\left[\frac{}{}\di \theta^2+\sin^2\theta\di \varphi^2
+(\di \psi+\cos\theta \di \varphi)^2\right]\,,
\eeqs
where the angles describing the sphere are chosen to have the ranges \be
0\leq \theta \leq \pi \sac  
0\leq \varphi <2\pi \sac
0\leq \psi < 4\pi \sac 
-\frac{\pi}{2} \leq \xi \leq \frac{\pi}{2} \,.
\ee
However, notice that the $(\sin \xi)^{\frac{1}{12}}$ factor in the metric yields a singularity at the equator in $\xi$. One has to restrict to $0<\xi \leq \frac{\pi}{2}$ and thus consider just one of the hemispheres. The boundary at $\xi=0$ corresponds merely to the presence of an O8 plane \cite{massiveIIA}. The scalar in $X$ parameterizes an inhomogeneous deformation of the $S^4$ in which a three-sphere is preserved.

The Ramond-Ramond (RR) four-form is given by 
\beqs
F_4&=&\frac{2}{3g^3} \Delta^{-2}(\sin \xi)^{\frac{1}{3}}
\left(\frac{}{}6 X^{-2}\,-\,X^{-6}\sin^2\xi \,+\,3X^2\cos^2\xi\,-4X^{-2}\cos^2\xi\right)\epsilon_{(4)}\nonumber\\
&+&\frac{1}{2g^3}\Delta^{-2}X^{-3}(\sin \xi)^{\frac{4}{3}}\cos^4\xi\, \epsilon_{(3)}\wedge \di X\,,
\eeqs
where $\epsilon_{(4)}$ and $\epsilon_{(3)}$ 
are the volume forms of the four- and three-sphere, respectively:
\beqs
\epsilon_{(4)}&=&\frac{1}{8}\cos^3\xi \sin \theta\, \di \xi\wedge \di \theta\wedge \di \varphi \wedge\di \psi\,=\,\frac{1}{8}\cos^3\xi \,\di \xi\wedge\epsilon_{(3)}\,,\\
\epsilon_{(3)}&=&\sin\theta \,\di \theta\wedge \di \varphi \wedge\di \psi\,.
\eeqs
For later convenience, we define 
\beqs\label{G1}
G_1&=&\frac{1}{12g^3} \Delta^{-2}\cos^3\xi (\sin \xi)^{\frac{1}{3}}
\left(\frac{}{}6 X^{-2}\,-\,X^{-6}\sin^2\xi \,+\,3X^2\cos^2\xi\,-4X^{-2}\cos^2\xi\right)\,\di \xi\nonumber\\
&-&\frac{1}{2g^3}\Delta^{-2}X^{-3}(\sin\xi)^{\frac{4}{3}} \cos^4\xi\,\di X\,,
\eeqs
so that $F_4=G_1\wedge\epsilon_{(3)}$. In addition, the background is supported by a non-vanishing Romans mass
\beqs
m&=&\frac23\,g\,.
\eeqs
The parameters $g$ and $m$ are related respectively to the number of D4 and D8 branes. In principle they can be taken to be independent as in \cite{massiveIIA}, but we chose to identify them for simplicity and to make contact with \cite{IIAEmbedd}. 

Finally, the ten-dimensional dilaton is given by
\beqs
e^{\Phi}&=&(\sin \xi)^{-\frac{5}{6}}\Delta^{\frac{1}{4}} X^{-\frac{5}{4}}\,,
\eeqs
which allows us to rewrite the metric in string frame:
\beqs
\di s^2_s&=&e^{\Phi/2} \di s^2_{10}\,=\,
(\sin\xi)^{-\frac{1}{3}}X^{-\frac{1}{2}}\Delta^{\frac{1}{2}}\left[\frac{}{}
\di s_6^2\,+\,\frac{1}{g^2}\di \tilde{\Omega}_4^2\right]\,.
\eeqs
We conclude by noticing that if we set $\phi=0$ then $X=1=\Delta$ and $\di X=0$. In this case the internal space is the round four-sphere and $F_4$ is proportional to its volume form.

\subsubsection{Lift to Type IIB}

The same $\sigma$-model can be uplifted to Type IIB as shown in~\cite{Jeong:2013jfc}. This is due to the possibility of performing a non-abelian T-duality along an $SU(2)\subset SO(4)$ isometry of the internal $S^3\subset S^4$. The ten-dimensional metric takes the same form as in Type IIA except for the fact that the metric on the internal space is a different deformation of $S^4$ that reads
\beqs
\di \tilde{\Omega}_4^2 &=&X^2\di \xi^2 \,+\,X^{-1}\Delta^{-1} \cos^2\xi \,\frac{1}{4}\left[e^{-4\Theta}\di \sigma^2 +\frac{\sigma^2}{\sigma^2+e^{4\Theta}}\left(\di \theta^2 +\sin^2\theta \di \varphi^2\right)\right]\,,
\eeqs
where
\beqs
e^{\Theta}&=&\frac{1}{2g}X^{-\frac{3}{4}}\Delta^{-\frac{1}{4}}(\sin \xi)^{-\frac{1}{6}}\cos \xi\,,
\eeqs
and the angles $\theta$, $\varphi$ and $\xi$ have the same meaning as in the massive IIA case. The new 
coordinate $\sigma$  no longer has a clear geometric interpretation and its range has not been determined. We only observe that for fixed $\Theta$ (or equivalently for fixed $\xi$) 
asymptotically the three-dimensional manifold interpolates between $\mathbb{R}^3$ and $\mathbb{R}^1\times S^2$. In addition to the singularity at $\xi=0$ present in the type IIA  geometry, T-duality generates a further one at $\xi=\frac{\pi}{2}$, so one needs to remove also the would-be pole. It would be interesting to understand if these singularities are also due to the presence of orientifold planes. The first steps towards the identification of the dual to the supersymmetric fixed point were given in \cite{Lozano}.

The dilaton $\Phi$ and NS $B$-field are given by
\beqs
e^{-2\Phi}&=&\frac{1}{4g^2}\Delta^{-1}X (\sin \xi)^{\frac{4}{3}}\cos^2\xi (\sigma^2+e^{4\Theta})\,,\\
B&=&-\frac{\sigma^3}{\sigma^2+e^{4\Theta}}\,\sin\theta \di \theta \wedge \di \varphi\,,
\eeqs 
while the RR sector reads
\beqs
F_1&=&m\,  \sigma \, \di \sigma \,-\,G_1\,,\\
F_3&=&\frac{\sigma^2}{\sigma^2 +e^{4\Theta}}\left[\frac{}{}\sigma G_1 + m e^{4\Theta}\di \sigma\right] \wedge \sin\theta \di \theta \wedge \di \varphi\,,\\
F_5&=&0\,.
\eeqs
with $G_1$ defined in (\ref{G1}). Notice that in this case the non-trivial fluxes call for an interpretation in terms of a number of D5- and D7-branes, in the appropriate limit, as opposed to the D4- and D8-branes of the massive Type IIA case \cite{massiveIIA}.

\subsection{Finding solutions}

Once we have defined the model in all relevant dimensions (five, six and ten), we need to solve the corresponding equations and hence fix the backgrounds of interest.

\subsubsection{Fixed points}
First of all, we notice that the potential ${\cal V}$ of the six-dimensional theory has two critical points at
\beqs\label{UVpoint}
\phi_{UV}\,=\,0 \,\,\,\,\,\,\,\,\,\, 
&\rightarrow& \,\,\,\,\,\,\,\,\,\, 
{\cal V}(\phi_{UV})\,=\,-\frac{20}{9}\,,
\label{UV}
\eeqs
and at
\beqs\label{IRpoint}
\phi_{IR}\,=\,-\frac{\log 3}{4} \,\,\,\,\,\,\,\,\,\, 
&\rightarrow & \,\,\,\,\,\,\,\,\,\, 
{\cal V}(\phi_{IR})\,=\,-\frac{4}{\sqrt{3}}\,.
\label{IR}
\eeqs
These define the two different AdS$_6$ solutions of $F(4)$ supergravity~\cite{Romans:1985tw}, the first one being supersymmetric.
We call them UV and IR, respectively, because the second one has a lower value of 
\be
v\equiv {\cal V}(\phi) \,,
\label{littleV}
\ee
which allows for the existence of a flow joining the two, as first constructed in \cite{5DRGFlow}.

With our conventions, the curvature of the AdS$_6$ spaces is related to the value of the potential at the extrema as $R^2=-5/v$. This gives the radii  
\begin{equation}
R_{UV}^2\,=\,\frac94 \sac R_{IR}^2\,=\,\frac{5\cdot3^{1/2}}{4} \,.
\end{equation}
From the expansion of the potential at quadratic order around the extrema one reads off the mass of the scalar
\begin{equation}
m_{UV}^2R_{UV}^2\,=\,-6 \sac m_{IR}^2R_{IR}^2\,=\,10\,.
\end{equation}
Correspondingly, the operator dual to the scalar $\phi$ has dimensions
\begin{equation}
\Delta_{UV}\,=\,3 \sac \Delta_{IR}\,=\,\frac12\left(5+\sqrt{65}\right)\,.
\end{equation}
Note that whereas $\Delta_{UV}$, $5-\Delta_{UV}$ and $\Delta_{IR}$ are positive, $5-\Delta_{IR}$ is not. This signals that, for the flow to exist, one has to tune the coefficient of $\Delta_{IR}$ (corresponding to the VEV in the IR) to vanish in such a way that the fixed point is reached along the direction of an irrelevant operator.

The five-dimensional $\sigma$-model is described by the equations
\beqs
&&-\frac{1}{18} e^{-2 \chi (r)} \left(-6 e^{-6 \phi (r)}+24 e^{-2
   \phi (r)}-18 e^{2 \phi (r)}\right)+4 A'(r) \phi '(r)+\phi''(r)\,=\,0\,,
   \nonumber\\[1.5mm]
&&   \frac{1}{27} e^{-2 \chi (r)} \left(e^{-6 \phi (r)}-12 e^{-2 \phi
   (r)}-9 e^{2 \phi (r)}\right)+4 A'(r) \chi '(r)+\chi''(r)\,=\,0\,,
   \nonumber\\[1.5mm]
&&      6 A'(r)^2+2 \phi '(r)^2+6 \chi '(r)^2+\frac{2}{9} e^{-2 \chi (r)}
   \left(e^{-6 \phi (r)}-12 e^{-2 \phi (r)}-9 e^{2 \phi (r)}\right)+3
   A''(r)\,=\,0\,,\nonumber \\[1.5mm]
&&   3 A'(r)^2- \phi '(r)^2-3 \chi '(r)^2+\frac{1}{9} e^{-2 \chi (r)}
   \left(e^{-6 \phi (r)}-12 e^{-2 \phi (r)}-9 e^{2 \phi (r)}\right)\,=\,0\,,
\eeqs
where the $'$ denotes derivatives  with respect to $r$. Changing radial coordinate according to
\beqs
\di r &=& e^{\chi} \di \r\,,
\eeqs
we find that the equations become:
\beqs
\label{eq11}
\partial_{\r}^2\phi+\left(4\partial_{\r} A -\partial_{\r}\chi\frac{}{}\right)\partial_{\r}\phi&=&
\frac{1}{2}\frac{\partial {\cal V}}{\partial \phi}\,,\\
 \label{eq12}
\partial_{\r}^2\chi+\left(4\partial_{\r} A -\partial_{\r}\chi\frac{}{}\right)\partial_{\r}\chi&=&
-\frac{1}{3} {\cal V}\,,\\[1.5mm]
 \label{eq2}
3\partial_{\r}^2 A \,+\,6(\partial_{\r}A)^2\,+\,2(\partial_{\r}\phi)^2\,+\,6(\partial_{\r}\chi)^2
-3\partial_{\r}A\,\partial_{\r}\chi&=&-2{\cal V}\,,\\[1.5mm]
 \label{eq3}
3(\partial_{\r}A)^2\,-(\partial_{\r}\phi)^2\,-\,3(\partial_{\r}\chi)^2&=&-{\cal V}\,.
\eeqs
Note that the combination 
\beqs
0&=&-12(\ref{eq12})+(\ref{eq2})+2(\ref{eq3})\,\nonumber
\\ &=&3  \left[\left(4 \partial_{\r}A - \partial_{\r}\chi
   \right)\left(\partial_{\r}A-4 \partial_{\r}\chi
   \right)+\partial_{\r}^2A-4 \partial_{\r}^2\chi\right]\,,
\eeqs
is solved by 
\beqs
A&=&4\chi\,.
\eeqs
This gives rise to the two AdS$_6$ solutions. Indeed, in this case  we are left with the following three equations for $\phi$ and $\chi$:
\beqs
\partial_{\r}^2\phi  \,+\,15 \partial_{\r}\phi  \partial_{\r}\chi &=&
 \frac{1}{2}\frac{\partial {\cal V}}{\partial \phi}\,,\\
 \partial_{\r}^2\chi  \,+\,15 (\partial_{\r}\chi)^2  &=&
 -\frac{1}{3}\, {\cal V}\,,\\[1.5mm]
- (\partial_{\r}\phi  )^2+45 (\partial_{\r}\chi  )^2&=&
   -{\cal V}\,.
\eeqs
The last two equations can also be combined into
\beqs
\partial_{\r}^2 \chi&=&-\frac{1}{3}(\partial_{\r} \phi)^2\,,
\eeqs
which shows that $\chi$ and $A=4\chi$ are both linear in $\rho$ at the critical points where $\partial_{\r}\phi=0$, as anticipated above for the AdS$_6$ solutions. In summary, the AdS$_6$ solutions take the form
\beqs
\phi&=&\phi_0\,=\,\left\{\frac{}{}0\,,\,-\frac{\log 3}{4}\right\}\,,
\nonumber \\[1.3mm]
\chi&=&\chi_0\,+\,\chi_1 \r\,,
\nonumber \\[1.9mm]
\chi_1&=&\left\{\frac{2}{9}\,,\,\frac{2}{3 \sqrt[4]{3} \sqrt{5}}
\right\}\,,
\label{nono}
\eeqs
where the integration constant $\chi_0$ is completely free for the time being. 

Substituting (\ref{nono}) into (\ref{subs}) we see that the the five-dimensional metric takes the form
\beqs
\di s_5^2&=&e^{8\chi_0} e^{8\chi_1 \r} \di x_{1,3}^2\,+\,e^{2\chi_0} e^{2\chi_1 \r} \di \r^2\,\\[1.5mm]
&=&\left(\chi_1r\right)^8\di x_{1,3}^2\,+\,\di r^2\,\\[1.5mm]
&=&{z^{-\frac{8}{3}}}\left[\left(3\chi_1\right)^8\di x_{1,3}^2\,+\,\di z^2\right]
\,\\[1.5mm]
&=& z^{-2+\frac{2}{3}\theta}\left(\di x_{1,3}^2\,+\,\di z^2\right)
\,,
\eeqs
where we have changed coordinates according to 
\be
\chi_1 r =e^{\chi_0+\chi_1 \r} \sac r=3 z^{-1/3} \,,
\ee
and we have performed an obvious rescaling of the Minkowski coordinates. The boundary (UV) lies at $r\rightarrow +\infty$ or $z\rightarrow 0$ respectively. As expected from the compactification of an AdS space on a circle, the 5D metric exhibits hyperscaling violation with hyperscaling coefficient $\theta=-1$ satisfying the null energy condition~\cite{Dong:2012se}.

\subsubsection{Flows between the two fixed points}

The first type of backgrounds we are interested in realize an RG flow between the two fixed points described above. The geometry dual to this flow interpolates between the AdS$_6$ solutions while preserving Poincar\'e invariance in the $x_{1,3}$ and $\eta$ directions. This means that the metric takes the form (\ref{DWansatz}) with $\hat{A}=3\chi$. The equations for the remaining scalar in this geometry are:
\beqs
\partial_{\r}^2\phi+5\partial_{\r}\hat{A}\partial_{\r}\phi&=&\frac12\frac{\partial {\cal V}}{\partial {\phi}}\,,\\
\partial^2_{\r} \hat{A}+(\partial_{\r}\phi)^2 &=&0 \\[1.5mm]
5(\partial_{\r}\hat{A})^2-(\partial_{\r}\phi)^2&=&-{\cal V}\label{constraint}\,.
\eeqs
Only two equations in this system are independent. Notice that (\ref{constraint}) takes the form of a constraint, that together with one of the first two equations implies the other one. We resort to numerics for solving the system, tuning the IR boundary conditions in such a way that asymptotically
\beqs
\phi(\r)&\simeq&\phi_{IR}+\tilde{\phi} \, 
e^{-(5-\Delta_{IR})\frac{\r}{R_{IR}}}\,,\\ [1.5mm]
\hat{A}(\r)&\simeq&\frac{\r}{R_{IR}}-\frac{1}{4}\, \tilde{\phi}^2\, 
e^{-2(5-\Delta_{IR})\frac{\r}{R_{IR}}}\,.
\eeqs
From the set of three integration constants that characterize the solutions we have fixed two to vanish. The first one is a harmless additive constant in $\hat{A}$ and the second one corresponds to the VEV of the operator dual to $\phi$ in the IR, that has to be set to zero for the flow to exist, as explained above.  

\begin{figure}[h]
\begin{center}
\begin{picture}(250,150)
\put(20,0){\includegraphics[height=4.6cm]{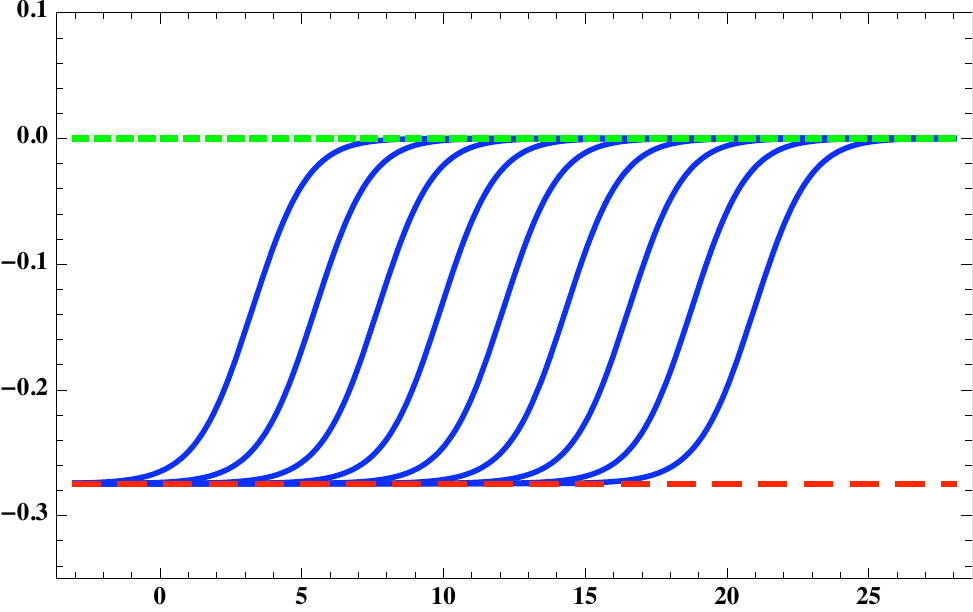}}
\put(11,120){$\phi$}
\put(225,0){$\r$}
\end{picture} 
\caption{The continuous, blue curves are members of the 
one-parameter family of solutions of the $\sigma$-model that 
interpolate between the critical values $\phi_{UV} $ and $\phi_{IR}$. The long-dashed, red line corresponds to the IR fixed point.
The short-dashed, green line corresponds to the UV fixed point.}
\label{Fig:scalar5Dflow}
\end{center}
\end{figure}

The remaining constant $\tilde{\phi}>0$ labels the family of solutions interpolating between the two critical points (\ref{UVpoint}) and (\ref{IRpoint}) as shown in Fig.~\ref{Fig:scalar5Dflow}. By varying $\tilde\phi$ we change the scale at which the transition between the CFTs occurs. However, since this is the only scale in the theory, all solutions in this one-parameter family are physically equivalent, the only difference between them being a choice of units.

Another important point is that, generically, in the UV both the $\Delta_{UV}=3$ relevant operator and its VEV will be turned on and given in terms of $\tilde\phi$. Then, in principle it is not possible to interpret the flow as being purely due to either explicit or spontaneous breaking of conformal invariance.

\subsubsection{Simple confining solutions}
\label{Sec:Simpconf}
Let us focus now on solutions that have constant $\phi=\phi_0$ corresponding to either of the fixed points \qq{UVpoint}-\qq{IRpoint} of the six-dimensional potential. In each of these solutions we can place a horizon at the bottom of the AdS$_6$ geometry and perform a  double analytic continuation to obtain a confining solution \cite{Witten}, the so called AdS-soliton. These solutions take the form:
\beqs
	\phi &=& \phi_0, \\[1.5mm]
	\chi &=& \chi_0 +\frac{\log 2}{15}-\frac{1}{5} \log \left[
	\cosh \left(\frac{\sqrt{-5v}}{2} \rho \right)\right]+\frac{1}{3} \log
   \left[ \sinh \left(\frac{\sqrt{-5 v}}{2} \rho \right)\right]\,, \\[1.5mm]
  A &=& A_0+\frac{4\log 2}{15}+\frac{4}{15} \log \Big[ 
  \sinh \left(\sqrt{-5v} \rho \right) \Big]
  + \frac{1}{15} \log \left[ \tanh
   \left(\frac{\sqrt{-5v}}{2} \rho \right) \right],
   \eeqs
where we have fixed an integration constant so that the space ends at $\rho=0$ and we recall that $v\equiv {\cal V}(\phi_0)$ is the value of the potential at either fixed point. Since the integration constant $A_0$ can be absorbed in a rescaling of the Minkowski coordinates, in the following we will conventionally set $A_0=\chi_0$. The integration constant $\chi_0$ is fixed by the requirement that the 6D metric be regular at $\rho=0$. Near this point the metric takes the form 
\beqs
\di s^2_2&\propto&e^{6\chi}\di \eta^2\,+\,\di \r^2 \nonumber\\[1.5mm]
&=& e^{6 \chi_0} \, 2^{2/5} \left( - \frac{5v}{4} \right) 
\r^2\di \eta^2\,+\,\di \r^2\,+\,\cdots \,.
\eeqs
Since $\eta$ is periodically identified with periodicity $2\pi$, in order to avoid a conical singularity we must fix
\beqs
\chi_0&=&\frac{1}{6} \log \left(- \frac{4}{2^{2/5} \, 5v}
\right)\,.
\label{got}
\eeqs
Specifying to the values of $v$ \qq{UV}-\qq{IR} at each of the fixed points this expression yields
\bea
\chi_0 &=& \frac{1}{6} \log \frac{9}{25 \times 2^{2/5}} \,\,\, \mbox{(UV)}\,,
\label{uvuv}
\\[1.5mm]
\chi_0 &=& \frac{1}{6} \log \frac{\sqrt{3}}{5 \times 2^{2/5}} \,\,\, \mbox{(IR)}
\,.\label{irir}
\eea
\noindent
Notice that for $\r\gg 1$ we have  $A\simeq 4 \chi$, namely in the UV we recover the AdS$_6$ geometry.  The general solution in which $\phi$ interpolates between the two fixed points can be found only numerically,
and it is the main subject of the study of the spectrum we are going to carry out.

\subsubsection{Multi-scale confining solutions}
Above we obtained confining solutions by placing a black hole at the bottom  of each of the two AdS$_6$ solutions of our model. Consider now the one-parameter family of solutions which interpolate between these two fixed points. 
On the gauge theory side these are RG flows between two CFTs. A given flow is parametrized by the scale at which the `transition' between the UV CFT and the IR CFT takes place. In each flow we can again place a black hole at the bottom of the geometry  and perform a double analytic continuation to obtain a confining theory. This is now characterized by two scales, the confinement scale and the scale at which the flow takes place, but the physics will only depend on the ratio of these two scales. 

In order to find the solutions, we consider the IR expansion around the point where the space smoothly closes off. As above, without loss of generality we choose coordinates  so that the space ends at $\r=0$. This means that  
the scalar $\phi$ is regular at this point, while the scalar $\chi$ 
and the metric function $A$ have a logarithmic divergence. Under these circumstances the solution near $\r=0$ takes the form
\SP{
	\phi &= \left(\tilde{\phi }-\frac{\log (3)}{4}\right)-\frac{e^{-6 \tilde{\phi }} \left(3-4 e^{4 \tilde{\phi }}+e^{8
   \tilde{\phi }}\right) \rho ^2}{4 \sqrt{3}}
   \\ & +\frac{1}{36} e^{-12 \tilde{\phi }} \left(-12+28 e^{4
   \tilde{\phi }}-17 e^{8 \tilde{\phi }}+e^{16 \tilde{\phi }}\right) \rho ^4+ \mathcal  O\left(\rho ^6\right), \\
   \chi &=\chi_0+ \frac{1}{60} \left(20 \log (\rho )+4\log (2)+5 \log \left(\frac{25}{3}\right)\right)-\frac{e^{-2 \tilde{\phi
   }} \left(\sinh(4 \tilde{\phi })+2\right) \rho ^2}{9 \sqrt{3}}
   \\ &+\frac{5}{162} e^{-4 \tilde{\phi
   }} \left(\sinh(4 \tilde{\phi })+2\right)^2 \rho ^4+ \mathcal O\left(\rho ^6\right), \\
   A &= \chi_0+\frac{1}{60} \left(20 \log (\rho )+5 \log \left(\frac{25}{3}\right)+32 \log (2)\right)+\frac{7 e^{-2 \tilde{\phi }}
   \left(\sinh(4 \tilde{\phi })+2\right) \rho ^2}{18 \sqrt{3}} \\
   &+\frac{e^{-4 \tilde{\phi }}}{324} 
   \left(108 \cosh(4 \tilde{\phi })-2 \left(20 \cosh(8 \tilde{\phi })+52 \sinh(4
   \tilde{\phi })+59\right)+27 \sinh(8 \tilde{\phi })\right) \rho ^4+ \mathcal O\left(\rho ^6\right) \,.
   \label{Eq:IRF4}
}
The constants in the expression for $\chi$ have been chosen in such a way that $\chi_0$ is given by \qq{irir}. The parameter $\tilde \phi$ determines the scale at which the flow takes place. The actual solutions cannot be found explicitly in closed form, but only numerically.
Interestingly, by noticing that ${\cal V}$ is smaller at the non-supersymmetric fixed point,
one can see that 
the profile of $\phi$ looks like a kink (cf.~Fig.~\ref{Fig:scalar5Dflow}) in which in the UV the scalar $\phi$ is close to the 
supersymmetric fixed point ($\phi=0$), while in the IR it is close to the non-supersymmetric fixed point ($\phi=-\frac{1}{4}\log 3$). In particular, this means that we must focus in the range $0\leq \tilde{\phi} \leq \frac{1}{4}\log 3$, so it is convenient to parametrize $\tilde \phi$ as 
\SP{
	\tilde \phi = \frac{\log 3}{8} \left[ 1 - \tanh \left( \frac{s_*}{2} \right)\right]\,.
	\label{analogous}
}
The limit $s_{\ast}\rightarrow +\infty$ corresponds to $\tilde \phi \to 0$. In this limit we see from \qq{Eq:IRF4} that the IR value of $\phi$ is $\phi(\rho=0)\to -\frac{1}{4}\log 3$. In other words, the scalar reaches the value of the IR fixed point before the theory confines. In the gauge theory language this means that the confinement scale is much smaller than the scale at which the transition between fixed points takes place. As a consequence there is some energy range in the IR in which the dynamics is that dictated by the IR CFT, and below this range the theory eventually confines. In the gravitational language this limit corresponds to the case in which the size of the black hole at the bottom of the flow is much smaller than the radial position of the kink in the geometry. In the opposite limit, $s_{\ast}\rightarrow -\infty$, we see that 
$\tilde \phi \to \frac{1}{4}\log 3$ and therefore $\phi(\rho=0) \to 0$. In this limit the confinement scale is much larger than the scale at which the flow would have taken place, the scalar $\phi$ remains approximately constant and equal to its UV value, and confinement takes place before the theory can probe any physics associated to the IR fixed point. 

\begin{figure}[h]
\begin{center}
\begin{picture}(290,450)
\put(26,300){\includegraphics[height=5cm]{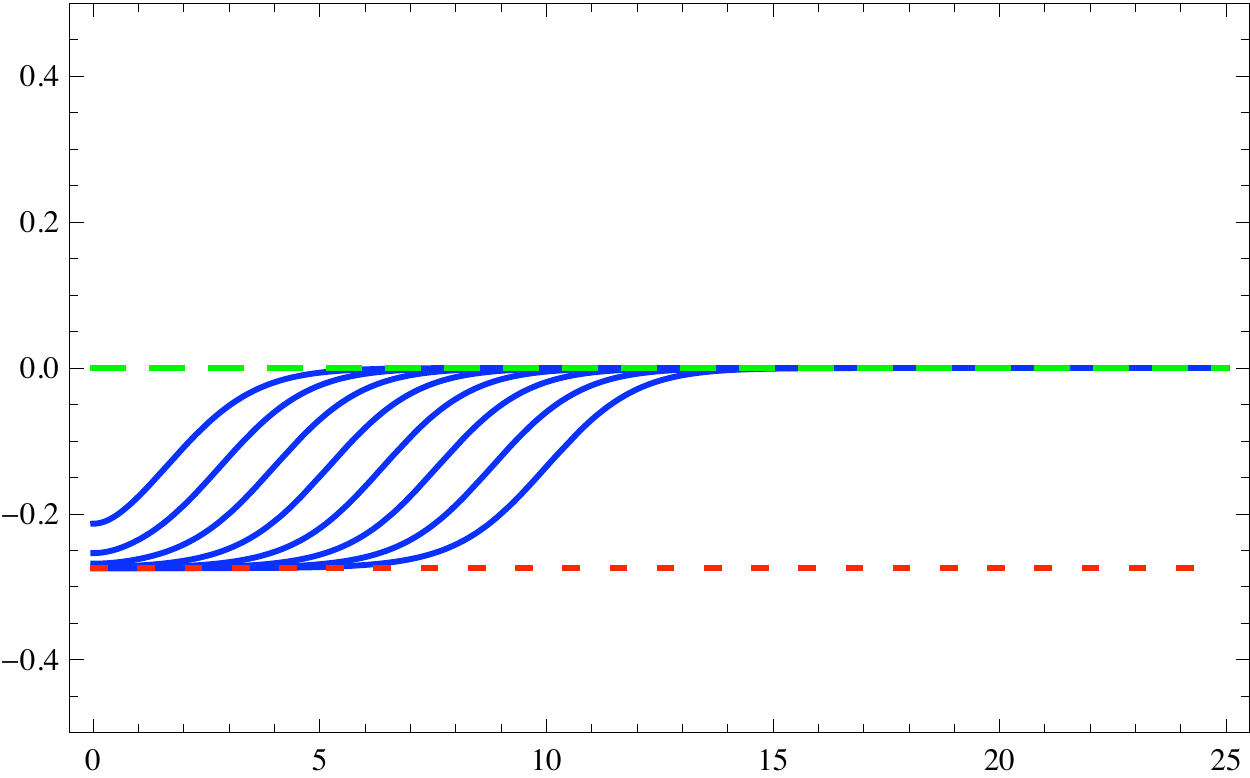}}
\put(25,150){\includegraphics[height=5cm]{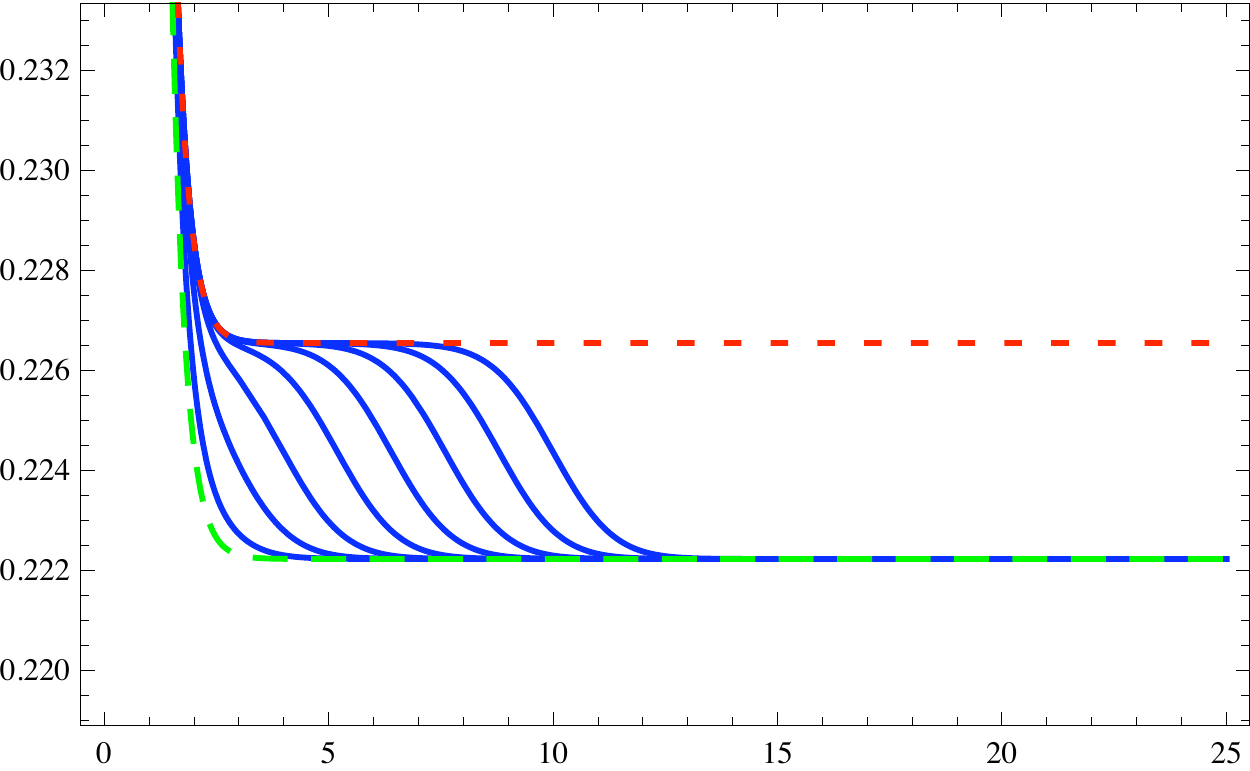}}
\put(27,0){\includegraphics[height=5cm]{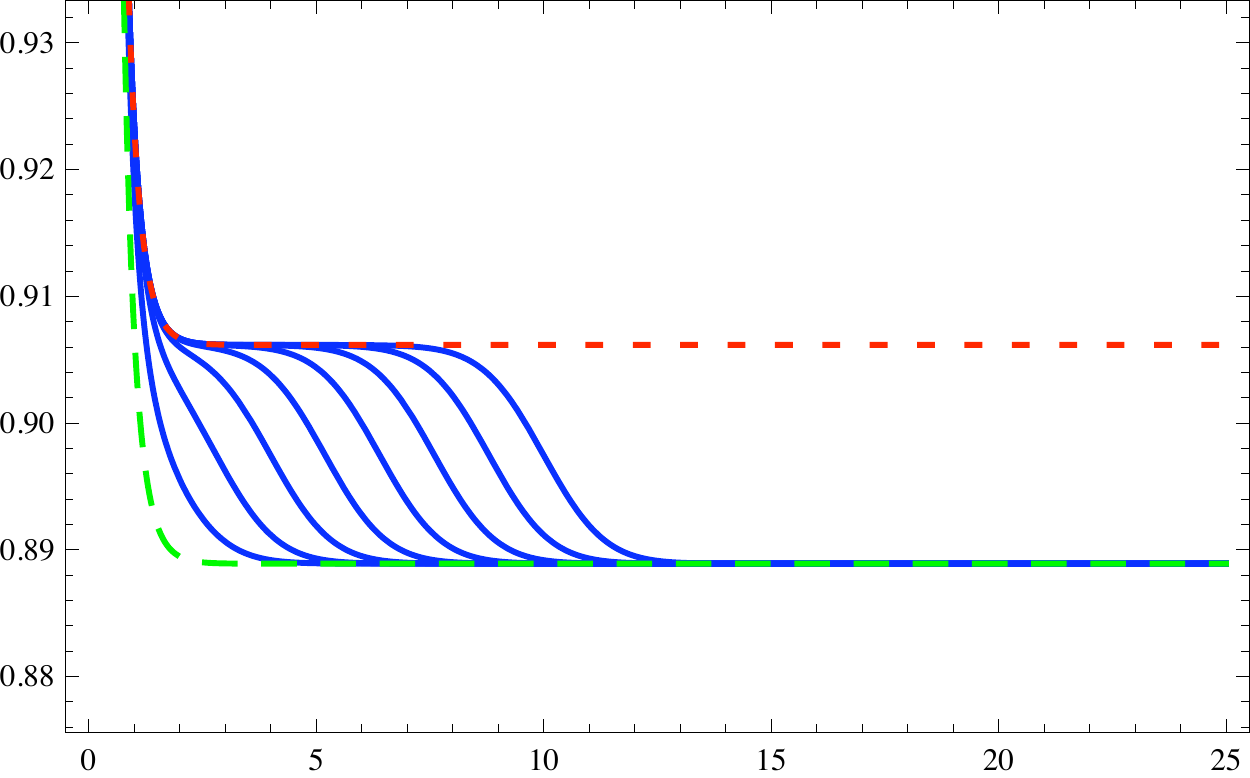}}
\put(15,430){$\phi$}
\put(0,280){$\partial_{\r}\chi$}
\put(0,130){$\partial_{\r}A$}
\put(260,6){$\r$}
\put(260,156){$\r$}
\put(260,306){$\r$}
\end{picture} 
\caption{ The solutions for the functions $\phi$, $\partial_{\r}\chi$ and $\partial_{\r}A$
to the system of equations in the five-dimensional $\sigma$-model that descends from massive Type IIA string theory,
as a function of the radial direction $\r$.
The  continuous blue curves show various solutions  that differ by the choice of $s_{\ast}$,  the long-dashed green curve is the solution with $\phi=0$, and the short-dashed red curve is the solution with $\phi=-\frac{1}{4}\log 3$.}
\label{Fig:plotF4solutions}
\end{center}
\end{figure}
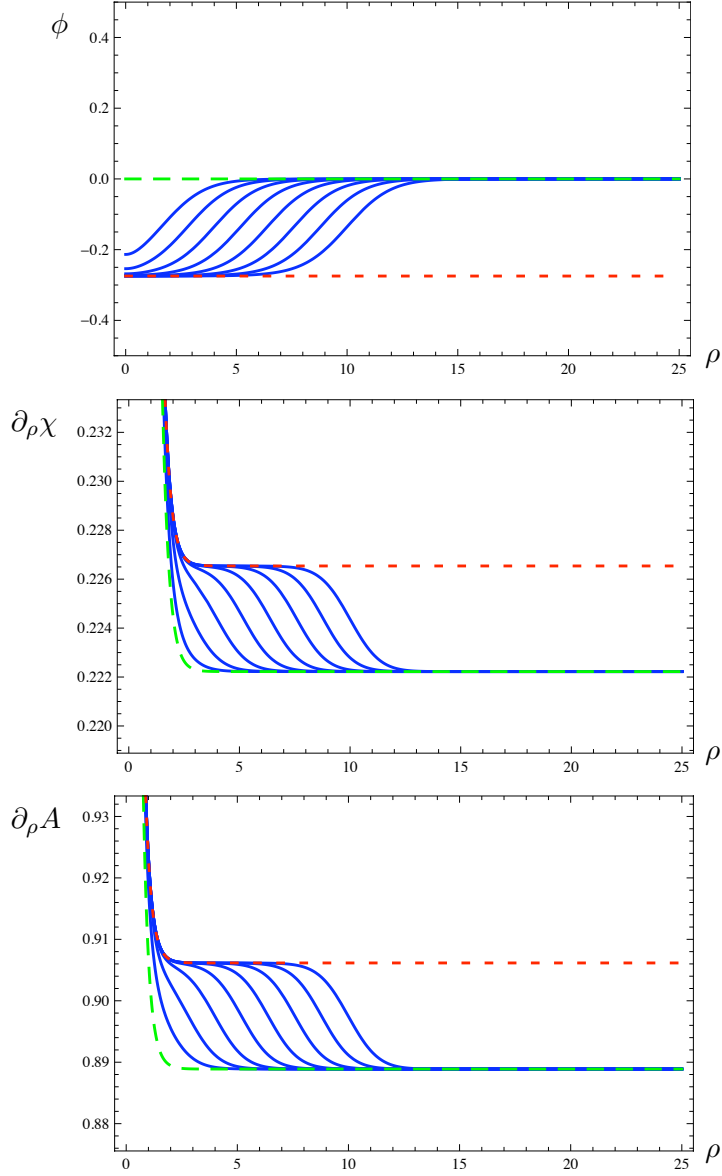

The numerical solutions of the equations are shown in Fig.~\ref{Fig:plotF4solutions}.
In the profile for $\phi$ we can clearly see the kink form anticipated, with the position of the 
kink approximately given by $s_{\ast}$. We always set up the numerics in such a way that all 
the solutions end at $\r=0$.
The scale $s_{\ast}$ is visible also in $\partial_{\r}\chi$ and $\partial_{\r}A$.
However, notice the vertical scale of the latter two plots: because the values of ${\cal V}$ at the two fixed points
in the six-dimensional gravity theory are so close to one another, the functions $A$ and $\chi$ 
change only very little along the one-parameter family of solutions.

\subsection{Spectrum of four-dimensional scalar bound states}
\label{sect1}

Since we have the model formulated in five-dimensional language, we can proceed to compute the spectrum following Sec.~\ref{general}. Let us rewrite the system in a form which is more 
convenient for our purposes, starting from the bulk equations.
First of all, the $\sigma$-model metric is $G_{ab}={\rm diag}\left\{{2},6\right\}$, 
which means that the covariant derivative 
with respect to the $\sigma$-model is just the partial derivative, but also that ${\cal R}^a_{\,\,\,\,bcd}=0$.
We have to replace $\Box=M^2>0$, the four-dimensional mass, and change radial variable according to 
$\frac{\partial}{\partial_r}=e^{-\chi}\frac{\partial}{\partial_{\r}}$.
We then have
\beqs
\label{eq:eomsflucs1}
0&=&\Big[ e^{-2\chi}\partial_{\r}^2 + e^{-2\chi}\left(4  \partial_{\r}A -\partial_{\r}\chi\right)\partial_{\r}
+ e^{-2A} M^2 \Big] \mathfrak{a}^a\,-\,(m^2)^{a}_{\,\,\,\,c} \,\mathfrak{a}^c\,,
\eeqs
where we defined
\be
\label{eq:eomsflucs2}
(m^2)^{a}_{\,\,\,\,c}\equiv  G^{ab}\frac{\partial^2V}{\partial \Phi^c\partial \Phi^b}  
+ \frac{4}{3 \partial_{\r}A}  \left( \partial_{\r}\Phi^a \frac{\partial V}{\partial \Phi^c} 
+ G^{ab}\frac{\partial V}{\partial \Phi^b} \partial_{\r}\Phi^d G_{dc} \right)
+ \frac{16 V}{9 (\partial_{\r}A)^2}  \partial_{\r}\Phi^a \partial_{\r}\Phi^b G_{bc} \,.
\ee
The boundary conditions read:
\be
\label{Eq:bcinf}
	 e^{-2\chi} \partial_{\r}\Phi^c    \partial_{\r}\Phi^d G_{db} \,\partial_{\r} \mathfrak a^b \Big|_{r_i} 
	 =   \left[ -\frac{3\partial_{\r} A}{2}e^{-2A}M^2\delta^c_{\,\,\,\,b} +   \partial_{\r} \Phi^c  
	 \left( \frac{4 V \partial_{\r}  \Phi^dG_{db}}{3\partial_{\r}  A} + V_b \right) \right] \mathfrak a^b \Big|_{r_i}.
\ee
Notice an important subtlety: the correct form of the boundary conditions has been multiplied by $M^2$ in
this expression, which is fine as long as $M^2\neq 0$. In particular, a superficial reading of these equations and boundary conditions might yield the incorrect conclusion that there is an exactly massless mode.
This is not the case, and one should focus only on $M^2\neq 0$. Another technical remark is that, due to the specific form of the potential $V=e^{-2\chi}{\cal V}(\phi)$,
by looking at the equations and boundary conditions we notice that $A$ and $\chi$ appear only
in the combination $e^{2A-2\chi}$, which means that the equations, and the spectrum,
for the confining solutions obtained from the expansion in Eq.~(\ref{Eq:IRF4})
do not depend on the integration constant $\chi_0$.

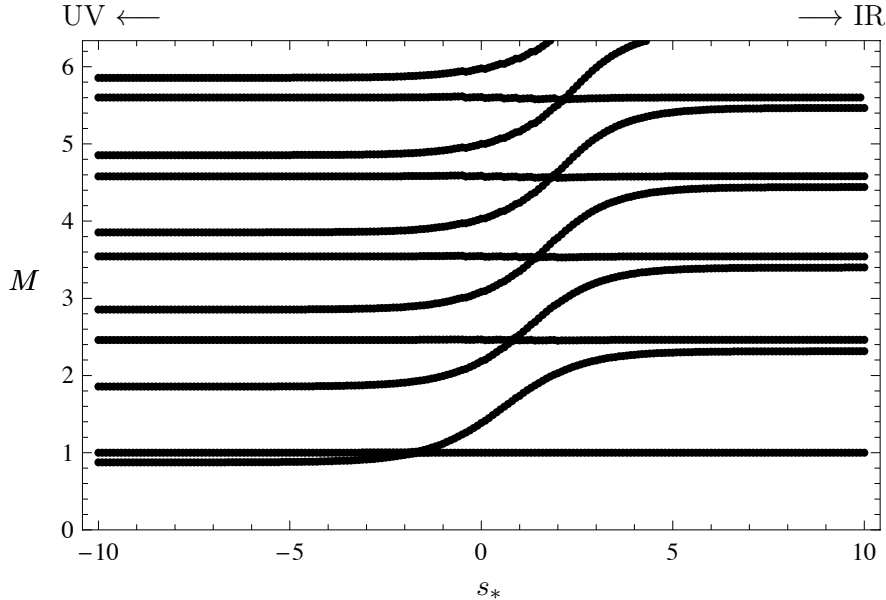
\begin{figure}[t!]
\begin{center}
\begin{picture}(320,210)
\put(5,6){\includegraphics[height=7cm]{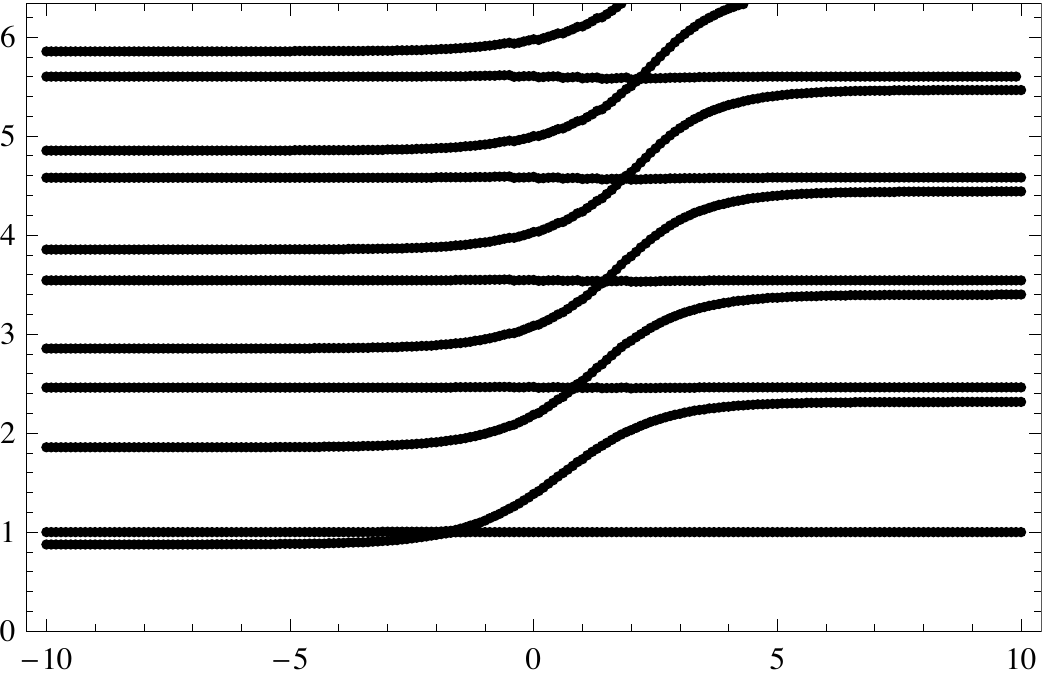}}
\put(-15,110){$M$}
\put(280,210){$\longrightarrow$ IR}
\put(5,210){UV $\longleftarrow$}
\put(-15,110){$M$}
\put(160,-5){$s_*$}
\end{picture} 
\caption{Spectrum for the confining solutions in the string theory model 
as $s_*$ is varied, with $\r_I = 10^{-3}$ and $\r_U = 8$. The spectrum at the two fixed points is recovered in the limits $s_\ast \to -\infty$ (UV fixed point) and $s_\ast \to + \infty$ (IR fixed point), as indicated in the figure. The overall normalization is chosen so that the first state among those that are insensitive to $s_*$ has unit mass.}
\label{Fig:F4SpectrumMvsr0}
\end{center}
\end{figure}

We  perform a numerical study of the complete spectrum (fluctuating both $\phi$ and $\chi$) for different values of $s_{\ast}$. For $s_{\ast}\rightarrow \pm \infty$ this calculation enables us to recover the result 
for the cases of constant $\phi$. The outcome is shown in Fig.~\ref{Fig:F4SpectrumMvsr0}.
Recall that we perform the calculation with regulators $\r_I$ and $\r_U$ 
kept finite. Hence, the spectrum does depend on the choices of these two scales.
However, we performed an extensive numerical study of its dependence (see the appendix),
in order to choose values such that the spectrum we show is in agreement (within numerical precision)
with the results obtained by extrapolating towards the physical limits $\r_I\rightarrow 0$
and $\r_U\rightarrow +\infty$.

As expected, the spectrum smoothly interpolates between the case in which $\phi=0$ (for $s_{\ast}\ll 0$) and $\phi=-\frac{\log 3}{4}$ (for $s_{\ast} \gg 0$). Notice that there is a major difference between fluctuations associated with $\phi$ and with $\chi$.
At the two fixed points, the fluctuations of $\phi$ effectively decouple from the rest of the system,
since $\phi^{\prime}=0$ on the background, but this is not the case for generic values of $\phi$. 
Nevertheless, one of the scalar fluctuations has a mass that depends very strongly on  $s_{\ast}$,
and is mostly due to the fluctuations of $\phi$, while the other (which results mostly from the mixing of 
$h$ with $\chi$) is virtually insensitive to $s_{\ast}$. 
As a result, the spectrum contains two towers of states, one that strongly depends on $s_*$ and one that is virtually insensitive to it. Motivated by this fact, and given that only ratios of masses are physically meaningful, we have chosen the overall normalization so that the first state among those that are insensitive to $s_*$ has unit mass. The numerical values of the masses at each of the fixed points are given in Table \ref{masses1}. As a consistency check, we note that, within numerical accuracy, the ratios at the UV fixed point agree with those of the scalar gravitational modes in the AdS$_6$ soliton shown in Table 1 of \cite{Cardoso:2013vpa}.  
\begin{table}[tbp]
   \centering
\begin{tabular}{ | c | c| }
  \hline
  IR ($s_\ast \to + \infty$) & UV ($s_\ast \to -\infty$) \\ \hline
  & 0.86 \\ 
 1 & 1 \\ 
  & 1.82 \\
  2.29 &  \\
  2.46 & 2.46 \\
   &  2.81 \\
  3.37 &   \\
  3.54 & 3.54 \\
   & 3.81 \\
  4.41 &  \\
  4.58 & 4.58\\
    & 4.80\\
  5.43 &  \\
  5.60 & 5.60 \\
          & 5.80 \\
  \hline
\end{tabular}
 \caption{Numerical values of the masses of scalar bound states at each of the two fixed points in the string model.}
   \label{masses1}
\end{table}

\subsection{Other observable quantities}

Here we discuss some results for other physical observables, all of which 
are related to the probe-approximation treatment of the system 
of extended objects allowed to propagate from the boundary into the bulk of the ten-dimensional geometry. For concreteness, we restrict to the Type IIA description.

\subsubsection{Wilson loop}

First we focus on strings with endpoints localized at $\rho\to\infty$, dual to a quark-antiquark pair. We start by rewriting explicitly the ten-dimensional metric in string frame:
\beqs
\nonumber
\di s^2_{s}&=&(\sin \xi )^{-\frac{1}{3}} X^{-\frac{1}{2}} \Delta^{\frac{1}{2}}\left[\frac{}{}e^{2A-2\chi}\di x_{1,3}^2\,
+\,\di \r^2 \,+\,e^{6\chi}\di \eta^2 \,+\,\frac{1}{g^2}\di \tilde{\Omega}_4^2\right]\,.
\label{ineq}
\eeqs
The background quantities defined in the general discussion of Sec.~\ref{Sec:Wilson} read:
\beqs
F^2&\equiv&g_{tt}g_{xx}\,=\,(\sin \xi)^{-\frac{2}{3}}X^{-1} \Delta \,e^{4A-4\chi}\,,\\
G^2&\equiv&g_{tt}g_{\r\r}\,=\,(\sin \xi)^{-\frac{2}{3}}X^{-1} \Delta \,e^{2A-2\chi}\,.
\eeqs
These are independent of the internal angles with the exception of $\xi$. Dependence on this coordinate makes the computation of the Wilson loop cumbersome because generically the string will not be located at constant $\xi$. This is most easily seen in the limit in which the quark-antiquark separation $L_{QQ}$ is very large. In this case the dominant contribution to the string energy comes from the very long part of the string that sits at the bottom of the geometry and runs parallel to the Minkowski directions. In other words, it comes from $F$ evaluated at the end-of-space. Therefore, regardless of the value of $\xi$ at which the string end-points sit at the boundary, the string will move towards the angular value that minimizes $F(0)$, extend there for a long distance, and then run up again to the boundary. Now, the minimum of $F$ as a function of $\xi$ depends on the value of $\phi$. Since we have that $X=e^\phi$ takes values between $X=1$ and $X\simeq 0.76$, the position of the minimum of $F(\xi)$ varies correspondingly from $\xi=\pi/2$ to $\xi=\pi/6$. This shows that, for interpolating geometries, the minimum-energy configuration of the string does not lie at constant $\xi$. For this reason, we will simply give the asymptotic behavior of the potential for the multi-scale confining solutions.

Consider first a string that remains in the vicinity of the UV fixed point, where the background is close to the AdS$_6$ given by the first case of Eq.~(\ref{nono}). The energy is minimized at an angular value $\xi=\pi/2$ and therefore the relevant functions are 
\beqs
F^2&=&e^{4A_0-4\chi_0}\,e^{\frac83 \r}\,,\\
G^2&=&e^{2A_0-2\chi_0}\,e^{\frac43 \r}\,.
\eeqs
We can hence reutilize the general AdS$_5$ result reviewed in Sec.~\ref{Sec:Wilson} provided we make the replacements
\beqs
\nonumber
R&\rightarrow &\frac{3}{2}\,,\\
\nonumber
L_{QQ}(R)&\rightarrow&e^{-A_0+\chi_0}L_{QQ}\left(\frac{3}{2}\right)\,,\\
\nonumber
E_{QQ}(R)&\rightarrow&e^{A_0-\chi_0-2\phi_0}E_{QQ}\left(\frac{3}{2}\right)\,.
\eeqs
The energy as a function of the quark-antiquark separation is then:
\beqs
E_{QQ}&=&3\,e^{A_0-\chi_0-2\phi_0}\,e^{\frac23 \r_U}\,-\,\frac{9\,\pi\Gamma\left[\frac{3}{4}\right]^2}
{4\,\Gamma\left[\frac{1}{4}\right]\Gamma\left[\frac{5}{4}\right]}\,
\frac{1}{ L_{QQ}}\,.
\label{Eq:hyperscalingWilson}
\eeqs
We see that in the far UV, for small separation, we recover the behavior of a conformal theory, and as a result one gets the same Coulombic potential as for the AdS case.

In the opposite limit one has long strings that are probing the geometry near the end-of-space. We know that the theory is confining, so the computation of the Wilson loop yields a potential linear in the separation, as can be seen using the simple confining soultions in Sec.~\ref{Sec:Simpconf}. Multi-scale confining backgrounds will also have a linear potential, with a string tension interpolating between the values corresponding to the fixed points. Indeed, the string tension can be exactly computed using Eq.~(\ref{stringtension}). As we argued, the configuration will tend to minimize $F(0)$ as a function of the angle $\xi$. The minimum is a function of the scale $\tilde{\phi}$ (or alternatively $s_{\ast}$) and has a transition from $\xi=\pi/2$ for $\frac{\log2}{4}\le\tilde{\phi}<\frac{\log3}{4}$ to $\cos\xi=\left(\frac23+\frac{2}{6-3e^{4\tilde{\phi}}}\right)^{-\frac12}$ for $0<\tilde{\phi}<\frac{\log2}{4}$. The resulting tension is thus:
\begin{equation}
\sigma\,=\,\frac{\di \bar{E}_{QQ}}{\di L_{QQ}}\,=\,F(0)\,=\, \left\{ 
  \begin{array}{l l}
    2^{\frac35}\sqrt{3}\,\left(\frac{3}{3-e^{4\tilde{\phi}}}-1\right)^{-\frac16} & \quad \quad \quad0<\tilde{\phi}<\frac{\log2}{4}\\[8mm]
    2^{\frac{14}{15}}\sqrt{3}\,e^{-2\tilde{\phi}} & \quad \quad \frac{\log2}{4}\le\tilde{\phi}<\frac{\log3}{4}
  \end{array} \right.
\end{equation}
Given the maximal and  minimal values of $\phi$, we find that $1.9\lsim \sigma\lsim 2.9$.
Notice that these numerical results are affected by the specific choice we made of the 
additive integration constant in the warp factor $A$.

By comparing with the spectrum, we learn another interesting thing.
We have found that it consists of two towers of states, 
the masses in one of which are almost completely independent of the scale
$s_{\ast}$. However, the string-tension does depend on $s_{\ast}$.
In particular this means that in units of the string-tension the spectrum of models
that sits close to the the non-supersymmetric critical point ($\tilde{\phi}=0$) are actually lighter than in the 
case in which the model sits always close to the supersymmetric fixed point.

\subsubsection{Gauge coupling}

Let us focus again on the equations for the lift to ten-dimensional massive Type IIA. We know that the dilaton reads
\beqs
e^{\Phi}&=&(\sin\xi)^{-\frac{5}{6}}\Delta^{\frac{1}{4}}X^{-\frac{5}{4}}
\,,
\eeqs
from which we find that the five-dimensional gauge coupling is (see eqn.~\qq{5Dcoupling})
\beqs
\frac{1}{g_5^2}&=&\frac{e^{-\Phi}}{16\pi^3\ell_s}\,=\,
\frac{(\sin\xi)^{\frac{5}{6}}\Delta^{-\frac{1}{4}}X^{\frac{5}{4}}}{16\pi^3\ell_s}\,=\,
\frac{e^{2\phi}}{16\pi^3 \ell_s}\,,
\eeqs
where in the last step we have set $\xi=\pi/2$ for concreteness. On the other hand, the component of the string-frame induced metric along the circle takes the form
\beqs
\tilde{g}_{\eta\eta}&=&(\sin\xi)^{-\frac{1}{3}}X^{-\frac{1}{2}}\Delta^{\frac{1}{2}}\,e^{6\chi}\,.
\eeqs
Using \qq{4Dcoupling} we also find the four-dimensional coupling
\beqs
\frac{1}{g_4^2}&=&\frac{e^{-\Phi} \sqrt{\tilde{g}_{\eta\eta}} \, R_5}{8\pi^2\ell_s}\,=\,
\frac{(\sin\xi)^{\frac{2}{3}}Xe^{3\chi} \,R_5}{8\pi^2\ell_s}\,=\,\frac{e^{\phi+3\chi}\, R_5}{8\pi^2 \ell_s}\,,
\eeqs
where again we have set $\xi=\pi/2$ in the last step. Note that the formulas for both couplings were motivated in Sec.~\ref{running} by placing a D4-brane in the ten-dimensional geometry. We emphasize that we use the action of a D4-brane placed at a constant value of $\xi$ as a tool to read off the gauge couplings, but this does not mean that the D4-brane will stay there at rest: since the metric depends on $\xi$, generically there will be a force on the D4-brane. 

\begin{figure}[h]
\begin{center}
\begin{picture}(290,300)
\put(11,150){\includegraphics[height=5cm]{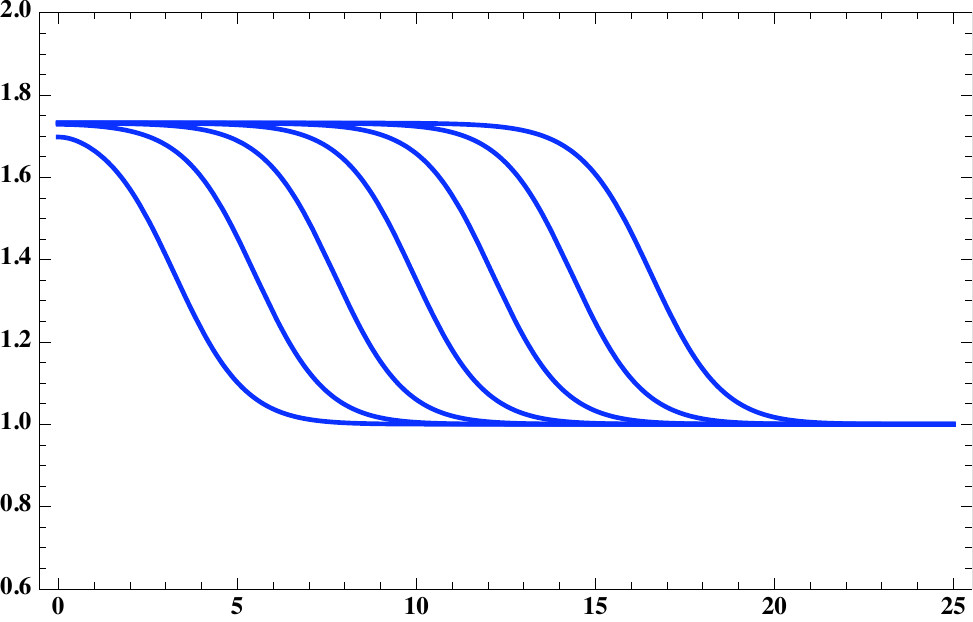}}
\put(14,0){\includegraphics[height=5cm]{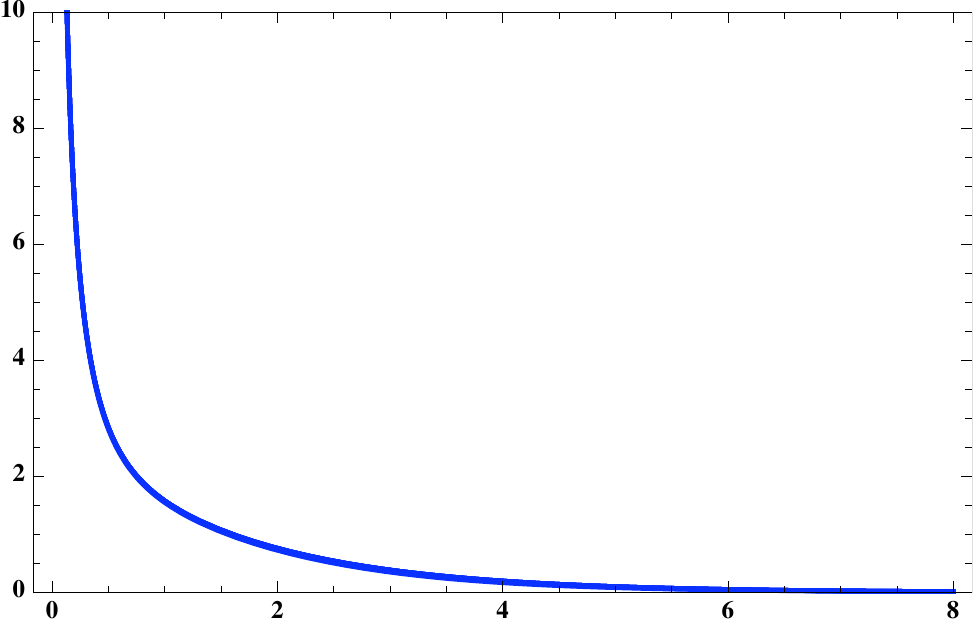}}
\put(-20,280){$\frac{g_5^2}{16\pi^3\ell_s}$}
\put(-16,130){$\frac{g_4^2\,R_5}{8\pi^2\ell_s}$}
\put(240,156){$\r$}
\put(240,6){$\r$}
\end{picture} 
\caption{The five-dimensional gauge coupling $g_5^2$ and the four-dimensional gauge coupling $g_4^2$ in the string model
as a function of $\r$, for a sample of confining solutions that differ by the choice of scale $s_{\ast}$.}
\label{Fig:gaugeF4}
\end{center}
\end{figure}
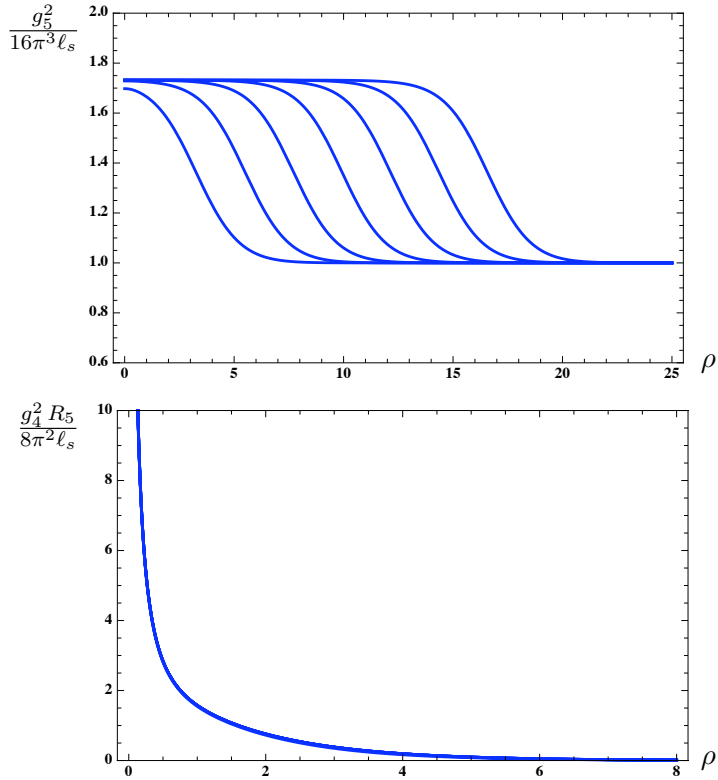

We show the running couplings in Fig~\ref{Fig:gaugeF4}.
First of all, notice that the five-dimensional coupling does behave as one would expect: since the
six-dimensional theory is the dual of a flow between two fixed points, the five-dimensional coupling 
shows the expected behavior.
The four-dimensional coupling exhibits three important features.
Unsurprisingly, it diverges at the end of the space in the IR.
Surprisingly though, the effect of the scale $s_{\ast}$ is virtually invisible: the various curves, obtained 
with different choices of $s_{\ast}$ are almost identical, with small discrepancies
that can barely be resolved.

The third feature deserves some explanation. Given that the four-dimensional gauge coupling goes to zero 
in the UV, one might think that this is the dual of an asymptotically free theory.
However, this not the case because the four-dimensional coupling is the gauge coupling of the lightest gauge bosons obtained by compactifying on a circle the five-dimensional dual theory. 
Smallness of the coupling in the far-UV is just due to the blowing up of the circle, the volume of which diverges
when $\r\rightarrow +\infty$. On the one hand, this is welcomed, since it means that the theory in the far-UV 
is indeed five-dimensional. On the other hand, it also means that the whole KK tower of
gauge bosons is becoming light, and hence while the individual coupling may be small, the collective effect results in the interaction being strong, as is clear from the behavior of the five-dimensional coupling. In other words, the four-dimensional coupling looses  meaning at high energies. 

\subsubsection{The U-shaped embedding of  probe D$8$}

Let us now consider the embedding function for probe D8-branes in the massive Type IIA lift.
We start by rewriting the ten-dimensional metric in string frame, specifying explicitly the internal metric and the dilaton $\Phi$:
\beqs
\nonumber
\di s^2_{s}&=&(\sin \xi )^{-\frac{1}{3}} X^{-\frac{1}{2}} \Delta^{\frac{1}{2}}\left[\frac{}{}e^{2A-2\chi}\di x_{1,3}^2\,
+\,\di \r^2 \,+\,e^{6\chi}\di \eta^2 \,+\,\frac{1}{g^2}\di \tilde{\Omega}_4^2\right]\,,\nonumber\\
\di \tilde{\Omega}_4^2&=&X^2\di \xi^2 +X^{-1}\Delta^{-1}\cos^2\xi\,\frac{1}{4}
\left[\di \theta^2+\sin^2\theta\di \varphi^2+(\di \psi + \cos\theta \di \varphi)^2\right]\,,\nonumber\\
e^{\Phi}&=&(\sin\xi)^{-\frac{5}{6}}\Delta^{\frac{1}{4}}X^{-\frac{5}{4}}\,.\nonumber
\eeqs
The embeddings we are interested in are those in which the D8-brane extends along $x_{1,3}$ and wraps the round $S^3$ inside the deformed $S^4$. Choosing $\xi, \eta$ as the eight and ninth worldvolume coordinates, the D8 embedding is then specified by a function of two variables $\r(\xi, \eta)$. Note that, unlike in the usual case discussed in~\cite{SS} and subsequent papers, here it is not consistent to assume that the embedding is independent of $\xi$ because the metric depends explicitly on this coordinate. The exception is when $\phi=0$, since then the metric becomes $\xi$-independent. We therefore focus on this case and leave a general analysis for future work. 

\begin{figure}[h]
\begin{center}
\begin{picture}(290,150)
\put(11,0){\includegraphics[height=5cm]{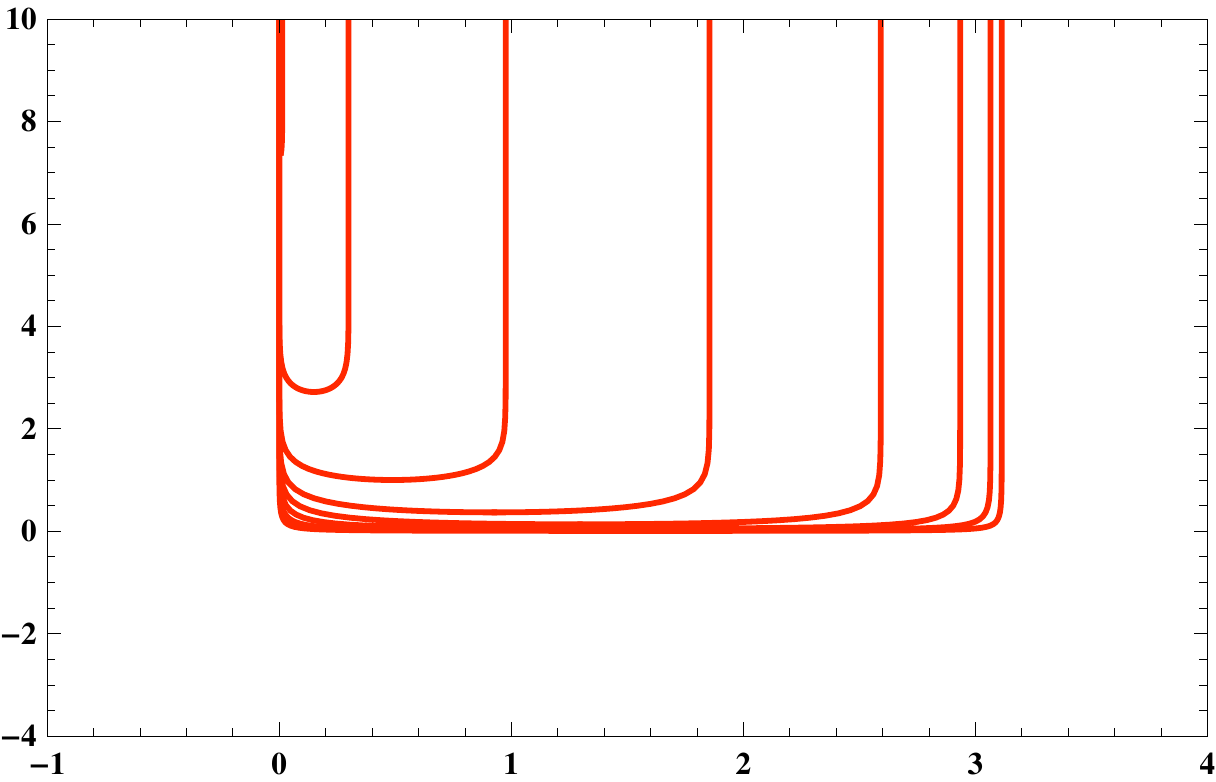}}
\put(0,130){$\r$}
\put(240,6){$\eta$}
\end{picture} 
\caption{ Embedding of the probe D8-brane in the string model, 
in the case of confining solutions with $\phi=0$, for various choices of $\hat{\r}_o>0$, and using $\r_U=35$ as a cutoff.}
\label{Fig:F4brane}
\end{center}
\end{figure}

The $\phi=0$ solution is given by
\beqs
\phi&=&0\,,\nonumber\\
\chi&=&-\frac{1}{6}\log\frac{25}{6}-\frac{1}{5}\log \cosh\left(\frac{5\r}{3}\right)
+\frac{1}{3}\log \sinh\left(\frac{5\r}{3}\right) \,, \nonumber\\
A&=&\frac{1}{6}\log\frac{25}{6}+\frac{4}{15}\log \sinh\left({5\r}{}\right)
+\frac{1}{15}\log \tanh\left(\frac{5\r}{3}\right)\,,\nonumber
\eeqs
Substituting these expressions in the DBI action one finds that the latter is given by 
\beqs
{\cal S}_{\rm D8}&=&
-\frac{36 \pi^2 T_8}{7 g^4}\int\di^4 x\,\int \di \sigma\,
\,e^{4A-4\chi}\sqrt{e^{6\chi}\eta^{\prime\,2}\,+\,\r^{\prime\,2}}\,,
\eeqs
which comparing with \qq{d8}  yields (neglecting the constants):
\beqs
\tilde{F}^2&=&e^{8A-2\chi}\,,\\
\tilde{G}^2&=&e^{8A-8\chi}\,.
\eeqs
The resulting D8 embeddings are shown in Fig.~\ref{Fig:F4brane}. As in \cite{SS}, when $\hat{\r}_o\rightarrow 0$ we recover the antipodal configuration in which  the end-points of the embedding are separated by $\Delta \eta=\pi$, and more generally we find that the separation $\Delta \eta$ is a monotonically decreasing function of 
$\hat{\r}_o$. It would be interesting to use this family of embeddings to  study the meson spectrum following~\cite{SS}.

%
\section{An M-theory model}

In this section we construct the M-theory dual of a multi-scale, confining four-dimensional theory. On the gauge theory side, the four-dimensional theory is obtained by compactifying a 6D CFT on a torus with appropriate boundary conditions.  On the gravity side, the starting point is 11D M-theory. Compactification of this theory on $S^4$ as in \cite{Nastase:1999cb} yields maximal 7D gauged supergravity with $SO(5)$ R-symmetry group~\cite{Pernici:1984xx}. This 7D theory admits two AdS$_7$ backgrounds. One of these solutions is supersymmetric and believed to provide the dual description of the ${\cal N}=(2,0)$ 6D CFT living on a stack of M5-branes. Both solutions, as well as the flow between them, can be described in terms of a seven-dimensional $\sigma$-model, with a single scalar $\phi$, to which the 7D supergravity can be truncated. Further reducing on a circle provides the link with Type IIA supergravity (on the gravity side) and with the SYM theory living on a stack of D4-branes (on the field theory side). Generically the size of the M-theory circle is not constant, and this produces a running dilaton in the ten-dimensional description. 
A final compactification on another circle yields Type IIA string theory on $S^1\times S^4$, or equivalently M-theory on $T^2 \times S^4$. If appropriate boundary conditions are imposed on the $T^2$, the resulting effective four-dimensional field theory provides one of the simplest examples of a confining theory with a gravity dual~\cite{Witten}. On the gravity side, confinement occurs through the smooth shrinking to zero size of the second circle in the IR. This physics can all be described in terms of either the seven-dimensional $\sigma$-model with one scalar mentioned above, or in terms of a five-dimensional $\sigma$-model with three scalars, where the two extra scalars, $\chi$ and $\omega$, arise from the sizes of the two circles in the $T^2$. Fluctuations of these scalars provide a subsector of the $0^{++}$ spectrum of glueballs. The quantum numbers of $\omega$ and $\chi$ were assigned in \cite{gravityspectrum2}. The additional scalar $\phi$ was not considered before, but its quantum numbers must be the same since it mixes non-trivially with the other scalars,  and moreover it is directly related to the ten-dimensional dilaton through the uplift. 

The confining model proposed by Witten~\cite{Witten} corresponds to the simple case in which the scalar $\phi$ is set to zero and hence consistently truncated from the five-dimensional $\sigma$-model. We will show how this construction can be easily generalised to the case in which $\phi$ is retained and acquires a non-trivial bulk profile, thus describing a flow between two fixed points (in the case in which the confinemenent scale vanishes). We will also show how to construct an alternative lift to ten-dimensional Type-IIB, which makes use of non-abelian T-duality based on the symmetries of an $S^3\subset S^4$.
The relationships between the different theories mentioned here are schematically summarised in Fig.~\ref{Fig:Ob}.

\begin{figure}[tpb]
\begin{center}
\includegraphics[height=6.9cm]{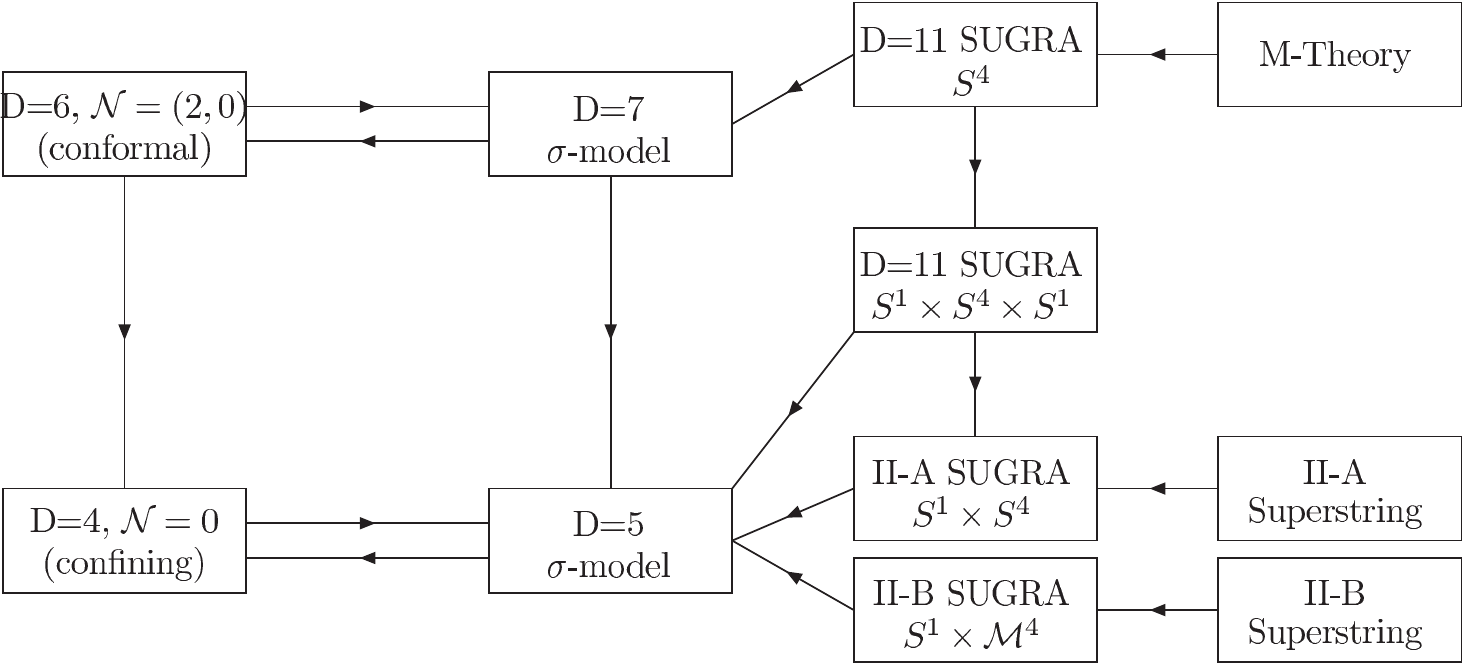}
\end{center}
\caption{Diagramatic representation of the relations between various theories discussed in the paper.}
\label{Fig:Ob}
\end{figure}

\subsection{The model}

\subsubsection{The $\sigma$-model in five dimensions}
\label{lastparag}
We start from the five-dimensional formulation of the model that is used in the computation of the spectrum. In subsequent sections we will give the details of the uplift to M- and string-theory.
The action is given by \qq{form} with $n=3$ scalars, $\Phi^a=(\phi,\omega,\chi)$. The $\sigma$-model metric is 
\beqs
G_{ab}&=&{\rm diag}\,\,\left\{\frac{1}{2},1,\frac{15}{4}\right\}\,,
\eeqs
so that ${\cal G}^a_{\,\,\,\,bc}=0$. The potential is
\beqs
V&=&\frac{1}{2}\,e^{-2\chi}\,{\cal V}(\phi)\,,
\eeqs
with 
\beqs
{\cal V}(\phi)&=&\frac{1}{4}e^{-\frac{8}{\sqrt{5}}\phi} -2 e^{-\frac{3}{\sqrt{5}}\phi}-2e^{\frac{2}{\sqrt{5}}\phi}\,.
\eeqs
Assuming the usual domain-wall type form of the metric 
\beqs
\di s_5^2&=&e^{2A}\di x _{1,3} \,+\,\di r^2\,,\label{dom}
\eeqs
the classical equations that follow from this action are
\beqs
0&=&- e^{-2 \chi (r)} \left(-\frac{2 e^{-\frac{8 \phi
   (r)}{\sqrt{5}}}}{\sqrt{5}}+\frac{6 e^{-\frac{3 \phi
   (r)}{\sqrt{5}}}}{\sqrt{5}}-\frac{4 e^{\frac{2 \phi
   (r)}{\sqrt{5}}}}{\sqrt{5}}\right)+4 A'(r) \phi '(r)+\phi ''(r)\,,  \label{Eq:eq11}\\[2mm]
0&=&   4 A'(r) \omega '(r)+\omega ''(r)\,,  \label{Eq:eq12}\\[2.8mm]
0&=& \frac{4}{15} e^{-2 \chi (r)}
   \left(\frac{1}{4} e^{-\frac{8 \phi (r)}{\sqrt{5}}}-2 e^{-\frac{3 \phi
   (r)}{\sqrt{5}}}-2 e^{\frac{2 \phi (r)}{\sqrt{5}}}\right)+4 A'(r) \chi
   '(r)+\chi ''(r)\,,  \label{Eq:eq13}\\[2mm]
0&=&  6 A'(r)^2+\frac{1}{2} \phi '(r)^2+\omega '(r)^2+\frac{15}{4} \chi '(r)^2+
   e^{-2 \chi (r)} \left(\frac{1}{4} e^{-\frac{8 \phi (r)}{\sqrt{5}}}-2
   e^{-\frac{3 \phi (r)}{\sqrt{5}}}-2 e^{\frac{2 \phi (r)}{\sqrt{5}}}\right)+3
   A''(r)\,, \nonumber\\ \label{Eq:eq2}\\
   0&=&6 A'(r)^2-\frac{1}{2} \phi '(r)^2-\omega '(r)^2-\frac{15}{4} \chi '(r)^2+
   e^{-2 \chi (r)} \left(\frac{1}{4} e^{-\frac{8 \phi (r)}{\sqrt{5}}}-2
   e^{-\frac{3 \phi (r)}{\sqrt{5}}}-2 e^{\frac{2 \phi (r)}{\sqrt{5}}}\right)\,,
   \label{Eq:eq3}
\eeqs
where $^{\prime}$ indicates differentiation with respect to the radial direction $r$. 

\subsubsection{The $\sigma$-model in  seven dimensions}
The 5D $\sigma$-model can be obtained from the 7D $\sigma$-model
\beqs
{\cal S}_7&=&\frac{1}{2}\int\di^7x\sqrt{-g}\left[\frac{}{} R -g^{MN}\partial_{M}\phi\partial_{N}\phi 
-2 {\cal V}(\phi)\frac{}{}\right]\,,
\eeqs
assuming the following ansatz for the seven-dimensional metric:
\be
\di s_7^2 = e^{-2 \chi} ds^2_5 + e^{3\chi+2\omega} \di z^2 + 
e^{3\chi- 2\omega} \di \eta^2 \,.
\label{Eq:ansatz7} 
\ee
We see that $\chi$ and $\omega$ originate from the sizes of the two circles in the $T^2$, with coordinates $z$ and $\eta$. We assume a periodicity of $2\pi$ for both of them. 

The 5D domain-wall ansatz \qq{dom} lifts to the 7D form
\be
\di s_7^2 
= \di \r^2 + e^{2\hat{A}} \di x _{1,3}
 + e^{3\chi+2\omega} \di z^2 + 
e^{3\chi- 2\omega} \di \eta^2 \,,
\ee
where 
\be
d\r = e^{-\chi} dr \sac \hat A = A - \chi \,.
\ee
As we will see, the flow solutions between the two AdS$_7$ spaces enjoy 6D Poincar\'e invariance along the $(x _{1,3}, \,z, \,\eta)$ directions. In the case of confining solutions, the $\eta$ circle will contract to zero size, thus breaking the symmetry to 5D Poincar\'e invariance along the $(x _{1,3},\, z)$ directions. Since the latter  will be a symmetry of all the backgrounds of interest, we will impose the restriction 
\be 
A = \frac{5}{2} \chi+ \omega 
\label{restriction}
\ee
on all our solutions (but \emph{not} on their fluctuations).  Under these conditions the 7D metric takes the form
\be
\di s_7^2 
= \di \r^2 + e^{2\hat{A}}\left(\di x^2_{1,3}+\di z^2 
+e^{-4\omega}\di \eta^2\right)\,. 
\label{takes}
\ee
This suggests that, in 7D,  it is convenient to work with a different linear combination of scalars, $(b,c)$, defined through
\beqs
\omega&\equiv&-\frac{1}{2}b\,,\\
\chi&\equiv&\frac{2}{3}c+\frac{2}{9}b\,,
\eeqs
so that \qq{takes} takes the form
\be
\di s_7^2 
= \di \r^2 + e^{2 \hat A}\left(\di x^2_{1,3}+\di z^2 +e^{2b}\di \eta^2\right)
\sac \hat A = c -\frac{b}{6}  \,.
\label{restricted}
\ee
The motivation for the choice of $b$ is that it directly measures the size of the contractible circle, whereas that of $c$ is fixed by the requirement that the equations of motion simplify in terms of these new variables. This set of equations is given by
\beqs
\partial_{\r}^2 \phi \,+\, 6\partial_{\r}\phi\partial_{\r}c&=&\frac{\partial {\cal V}}{\partial \phi}\,,\label{equi1}\\
\partial_{\r}^2 b\,+\,6\partial_{\r}b\partial_{\r}c &=&0\,,\\
\partial_{\r}^2 c\,+\,6(\partial_{\r} c)^2&=&-\frac{2}{5}{\cal V}\,,
\eeqs
together with the constraint 
\beqs
6(\partial_{\r} c)^2\,-\,\frac{1}{6}(\partial_{\r} b)^2\,-\,\frac{1}{5}(\partial_{\r} \phi)^2&=&-\frac{2}{5} {\cal V}\,.\label{equi4}
\eeqs
Upon use of the restriction \qq{restriction}, eqs.~\qq{equi1}-\qq{equi4} are equivalent to eqs.~\qq{Eq:eq11}-\qq{Eq:eq3}.

Note that the metric \qq{restricted} restricted to the $(\r,\eta)$-plane is, in general, topologically a cylinder. However, if an end-of-space exists in the $\r$ coordinate, at which  the size of the $S^1$ parameterized by $\eta$ vanishes, then one is left with a conical singularity
unless the functions $b$ and $c$ are such that near the end-of-space this becomes the metric of a plane. Since only the derivatives of $b$ and $c$ enter the equations of motion \qq{equi1}-\qq{equi4}, the requirement that no conical singularity be present can always be satisfied, at the only price of fixing an otherwise undetermined integration constant.

\subsubsection{The lift to Type IIA and to M-theory}

The ten-dimensional Type IIA solution can be obtained by lifting the five-dimensional system on $S^{1}\times S^{4}$.
Since maximal supergravity in 7D can be truncated to minimal supergravity by keeping precisely the scalar $\phi$, we will use the considerably simpler embedding formulas of \cite{Minimal7D}. The background has a non-trivial metric, dilaton $\Phi$ and $F_4$  form. The string-frame metric is given by the following equations 
\beqs
\di s_6^2 &=&e^{-\frac{5}{4}\chi+\frac{1}{2}\omega} \di s_5^2 + e^{\frac{15}{4}\chi -\frac{3}{2}\omega} \di \eta^2\,,\\
\di s_{10}^2 &=&\Delta^{\frac{1}{2}}\left[ e^{\frac{3}{4}\chi + \frac{1}{2}\omega} \di s_6^2 + e^{\frac{3}{2}\chi + \omega}\,\frac{1}{g^2}\, \di \tilde{\Omega}_4^2\right]\,, 
\eeqs
where we recall that $0\leq \eta <2\pi$ is the coordinate on the circle. The metric on the internal space reads
\beqs
\di \tilde{\Omega}_4^2&=&X^3 \di \xi^2+\frac{1}{4}X^{-1}\Delta^{-1}\cos^2\xi 
\left[\frac{}{}\di \theta^2+\sin^2\theta \di\varphi^2+(\di\psi +\cos\theta \di \varphi)^2\right]\,,
\eeqs
where we defined the functions
\beqs
X&=&e^{\frac{\phi}{\sqrt{5}}}\,,\\
\Delta &=& X^{4}\sin^2\xi+X\cos^2\xi\,.
\eeqs
The four angles describing the four-dimensional manifold are parameterized by the three angles
describing an $S^3$, i.e.
\be
0\leq\theta\leq\pi \sac 
0\leq\varphi\leq 2\pi \sac
0\leq \psi \leq 4\pi \,,
\ee
and an additional angle $-\frac{\pi}{2}\leq \xi \leq \frac{\pi}{2}$. For $X=\Delta=1$ this manifold is $S^4$, while for generic values of $\phi$ one has a deformed four-sphere with preserved $S^3$. As above, in order to make contact with the lower-dimensional $\sigma$-models one should  set the radius of the internal manifold to unity, $g=1$.

The four-form is given by 
\beqs
F_4&=&\frac{1}{g^3}\,\Delta^{-2}\left(\frac{}{}4X^{-3}-X^{-8}\sin^2\xi+2X^2\cos^2\xi-3X^{-2}\cos^2\xi\right)\epsilon_{(4)}\nonumber\\
&+&\frac{5}{g^3}\,\Delta^{-2}X^{-4}\sin\xi\cos^4\xi\epsilon_{(3)}\wedge \di X\,,
\eeqs
where $\epsilon_{(3)}$ and $\epsilon_{(4)}$ 
are the volume forms of the spheres in three and four dimensions, as in the previous section. 
Finally, the dilaton is given by
\beqs
e^{\frac{4}{3}\Phi}&=&\Delta^{\frac{1}{3}}e^{3\chi+2\omega}\,.
\eeqs
Notice how, for $\phi\neq 0$, the dilaton depends on the radial direction, but also on the internal angle $\xi$.
Remember that the relation between string and Einstein frame is
\beqs
\di s^2_{s}&=&e^{\Phi/2}\di s^2_{E}\,.
\eeqs

The solution can be further lifted to eleven-dimensional supergravity.
In this case, the four-form is unchanged, while the metric is given by a lift which assumes the existence of another $S^1$
parameterized by $0\leq z <2\pi$. The metric is
\beqs
\di s^2_{11}&=&e^{-\frac{2}{3}\Phi}\di s^2_{10}+e^{\frac{4}{3}\Phi}\di z^2\,,\\
&=&\Delta^{1/3}\left(\di s^2_7+\frac{1}{g^2}\,  \di \tilde{\Omega}_4^2\right)\,,
\eeqs
where the external seven-dimensional metric is
given by Eq.~(\ref{Eq:ansatz7}).

\subsubsection{The lift to Type IIB}

The Type IIA background preserves the symmetries of the $S^3$ in the internal space. It is hence natural to 
construct an alternative lift to Type IIB, by non-abelian T-duality along one of the $SU(2)$ factors contained in the $SO(4)$ isometry.

The string-frame metric in IIB takes the same form as in IIA, that is:
\begin{equation}
ds^2_{10}\,=\,\Delta^{\frac12}\,\left[e^{\frac{\hat{A}}{2}}\,ds^2_6+e^{\hat{A}}\,\frac{1}{g^2}\,d\widetilde{\Omega}_4^2\right]\,,
\end{equation}
where again we reinstated the radius of the internal manifold, to be fixed to $g=1$ to recover the lower-dimensional $\sigma$-models. 
The internal manifold is no longer $S^4$ but
\begin{equation}
d\widetilde{\Omega}_4^2\,=\,X^3\,\,d\xi^2+X^{-1}\,\Delta^{-1}\,\cos^2\xi\,\,\frac14\,\left[e^{-4\Theta}\,d\sigma^2+\frac{\sigma^2}{\sigma^2+e^{4\Theta}}\,\left(d\theta^2+\sin^2\theta\,d\varphi^2\right)\right]\,,
\end{equation}
with the warp factor
\begin{equation}
e^\Theta\,=\,\frac{1}{2g}\,e^{\frac{\hat{A}}{2}}\,X^{-\frac12}\,\Delta^{-\frac14}\,\cos\xi\,.
\end{equation} 
Note that the T-dualization has produced a singularity at $\xi\to\pm\frac{\pi}{2}$.
The rest of the NS sector reads
\begin{eqnarray}
e^{-2\Phi}&=&\frac{1}{4g^2}\,e^{-2\hat{A}}\,\Delta^{-1}\,X^{-1}\,\cos^2\xi\,\left(\sigma^2+e^{4\Theta}\right)\,,\\
B&=&-\frac{\sigma^3}{\sigma^2+e^{4\Theta}}\,\,\epsilon_{(2)}\,,
\end{eqnarray}
with $\epsilon_{(2)}=\sin\theta\,d\theta\wedge d\varphi$ the volume
 form of the $S^2$. In addition, the non-vanishing RR sector is
\begin{eqnarray}
F_1&=&-G_1\,,\\
F_3&=&\frac{\sigma^3}{\sigma^2+e^{4\Theta}}\,G_1\wedge\epsilon_{(2)}\,=\,F_1\wedge B\,,
\end{eqnarray}
where we have defined $G_1$ from the IIA (or M-theory) 4-form as
\begin{equation}
F_4\equiv G_1\wedge\epsilon_{(3)}\nonumber\,.
\end{equation}
 
\subsection{Finding solutions}

Once we have defined the model in all relevant dimensions (five, seven,  ten and eleven), we need to solve the corresponding equations and hence fix the backgrounds of interest.

\subsubsection{Fixed points}

First of all, we notice that the potential ${\cal V}$ (as opposed to $V$)
 admits two stationary points $\phi_{UV}$ and $\phi_{IR}$:
\beqs
\phi_{UV}\,=\,0 \,\,\,\,\,\,\,\, 
&\rightarrow& \,\,\,\,\,\,\,\,
{\cal V}(\phi_{UV})\,=\,-\frac{15}{4}\,,\\
\phi_{IR}\,=\,-\frac{\log 2}{\sqrt{5}} \,\,\,\,\,\,\,\,
&\rightarrow& \,\,\,\,\,\,\,\,
{\cal V}(\phi_{IR})\,=\,-\frac{5}{2^{2/5}}\,.
\label{crit}
\eeqs
They define two non-equivalent  AdS$_7$ geometries, which happen to be the two known
critical solutions of the compactification of M-theory on $S^4$.
The labeling comes from the fact that, since ${\cal V}(\phi_{IR})<{\cal V}(\phi_{UV})$, there exist RG flows that start near the latter and end at the former \cite{6DRGFlow}. As above, we will denote by $v\equiv {\cal V}(\phi)$ the value of ${\cal V}$ at either of the fixed points.

With the conventions used here, one finds that the curvature of the AdS$_7$ spaces is given by $R^2=-15/v$, 
which yields 
\be
R_{UV}^2=4 \sac R^2_{IR}=3\cdot 2^{2/5} \,.
\ee
By expanding the potential to quadratic order in $\phi$ one finds that
\be
m^2_{UV}R^2_{UV}=-8 \sac m^2_{IR}R^2_{IR}=12 \,,
\ee
which means that the field $\phi$ is dual  to an operator of dimensions
\be
\Delta_{UV}=4 \sac \Delta_{IR}=3+\sqrt{21} \,.
\ee
Notice that while $\Delta_{UV}$, $6-\Delta_{UV}$ and $\Delta_{IR}$ are positive,
$6-\Delta_{IR}$ is negative. This  implies  that in order for a flow connecting the two to fixed points exist, one has to dial the coefficient of the 
$\Delta_{IR}$ to zero, in such a way as to switch off the VEV of the corresponding dual operator in the IR.

\subsubsection{Flows between the two fixed points}
\label{corres}
We want to start by looking for solutions to the classical equations that realise such a flow
 in the dual six-dimensional theory, as done in \cite{6DRGFlow}.
Hence, we set $b=0$, and $c=\hat{A}$.
The equations of motion for the only non-trivial remaining scalar $\phi$ coupled to gravity are
\beqs
\partial_{\r}^2\phi+6\partial_{\r}\hat{A}\partial_{\r}\phi&=&\frac{\partial {\cal V}}{\partial {\phi}}\,,\\
5\partial^2_{\r} \hat{A}+(\partial_{\r}\phi)^2 &=&0 \,, \\[1.5mm]
30 (\partial_{\r}\hat{A})^2-(\partial_{\r}\phi)^2&=&-2{\cal V}\,.
\eeqs
Note that the last equation is a constraint and that the three equations are not independent: the constraint plus one of the first two imply the other.
We solve this system numerically, by tuning the boundary conditions appropriately.
Asymptotically in the IR, we must impose that the solution behaves as
\beqs
\phi(\r)&\simeq&\phi_{IR}+\tilde{\phi} \, 
e^{-(6-\Delta_{IR})\frac{\r}{L_{IR}}}\,,\\ [1.5mm]
\hat{A}(\r)&\simeq&\frac{\r}{L_{IR}}-\frac{1}{20}\, \tilde{\phi}^2\, 
e^{-2(6-\Delta_{IR})\frac{\r}{L_{IR}}}\,.
\eeqs
Notice that in principle this should depend on three integration constants. However,  we fixed to zero an additive integration constant in $\hat{A}$, and as we explained we have to impose that 
the VEV of the dual field theory operator vanishes, which removes another integration constant in $\phi$.

\begin{figure}[h]
\begin{center}
\begin{picture}(250,150)
\put(20,0){\includegraphics[height=4.6cm]{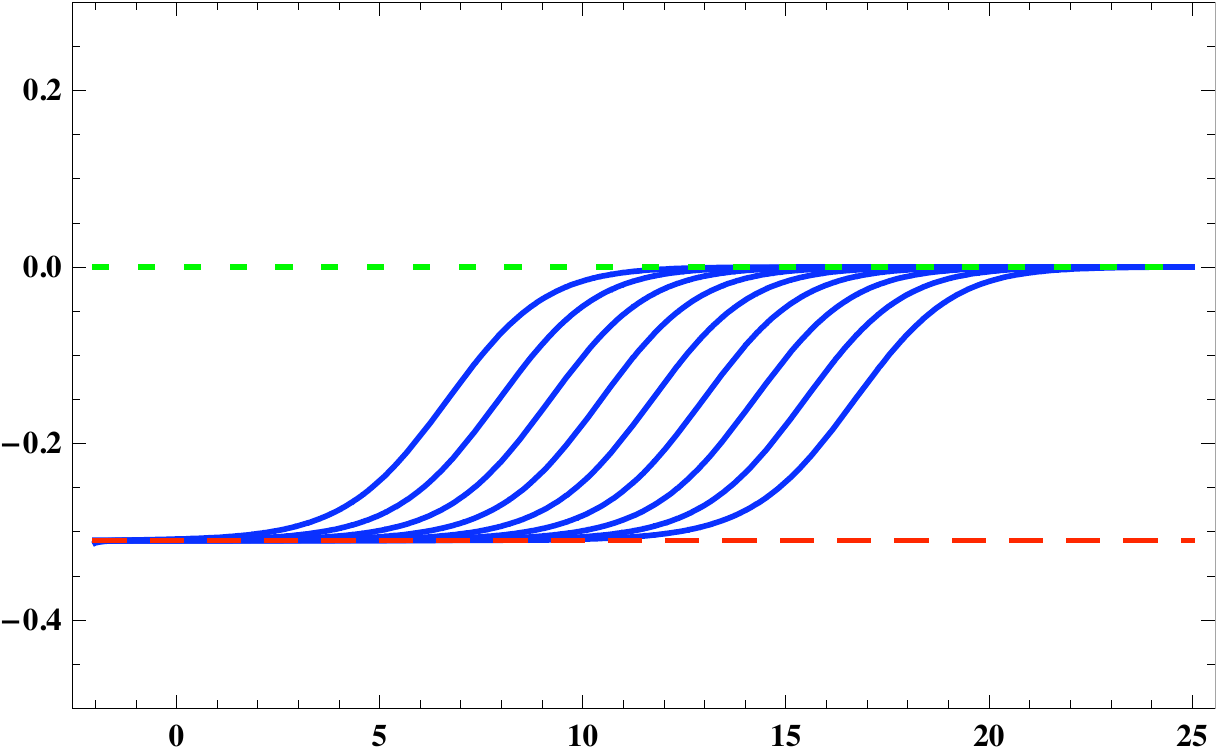}}
\put(11,120){$\phi$}
\put(205,0){$\r$}
\end{picture} 
\caption{The continuous, blue curves are members of the 
one-parameter family of solutions of the $\sigma$-model that 
interpolate between the critical values $\phi_{UV} $ and $\phi_{IR}$. The long-dashed, red line corresponds to the IR fixed point.
The short-dashed green line corresponds to the UV fixed point.}
\label{Fig:scalar}
\end{center}
\end{figure}

The result is the one-parameter family depicted in Fig.~\ref{Fig:scalar}.
By varying $\tilde{\phi}>0$ one finds a whole family of possible solutions that interpolate between the values of $\phi$ at the two critical points \qq{crit}, and represent the RG flows between the fixed points of the dual six-dimensional field theory. The choice of $\tilde\phi$ is equivalent to the choice of the scale  at which the transition takes place. Since this is the only scale in the theory, all the solutions in this family differ merely by a choice of units and are therefore  physically equivalent.

\subsubsection{Simple confining solutions}
\label{simple}
Let us focus now on solutions that have constant $\phi=\phi_0$ corresponding to the fixed points of the seven-dimensional potential. In each of these solutions we can place a horizon at the bottom of the AdS$_7$ geometry and perform a  double analytic continuation to obtain a confining solution \`a la \cite{Witten}, the so called AdS-soliton.
In these solutions $b$ varies in such a way that the $\eta$-circle closes off smoothly at some $\r=\r_0$ that we choose to set to $\r_0=0$ without loss of generality.  
In this case, the general solution is given by 
\beqs
\phi&=&\phi_0\,,\\[1mm]
c&=&\frac{1}{6}\log\left[\frac{1}{x}\sinh (x \r)\right]\,,\\[1mm]
b&=&\log \left[\frac{2}{x}\tanh\left(\frac{x}{2}\r\right)\right]\,,
\eeqs
where 
\be
x=\sqrt{-\frac{12 v}{5}} \sac v={\cal V}(\phi_0) \,.
\ee
This solution is simply the seven-dimensional AdS-soliton written in an unusual radial coordinate, and in it a number of integration constants have been fixed. In particular, an integration constant has been chosen to avoid a conical singularity in the IR (see the last paragraph of Sec.~\ref{lastparag}), and another integration constant has been fixed arbitrarily without loss of generality, since it can be absorbed in a rescaling of the gauge theory coordinates. In  five-dimensional language, this solution takes the form
\beqs
\phi&=&\phi, \qquad\qquad \omega = -\frac{1}{2}b, \qquad\qquad \chi = \frac{2}{3}c +\frac{2}{9}b\,, \nonumber \\
A&=&\frac{1}{18}b+\frac{5}{3}c\,,\qquad\qquad \di \r = e^{-\frac{2}{3}c-\frac{2}{9}b} \di r\,.
\eeqs

\subsubsection{Multi-scale confining solutions}
In the previous section, we obtained confining solutions by placing a black hole at the bottom  of each of the two AdS$_7$ solutions of our model. Consider now the one-parameter family of solutions which interpolate between these two fixed points, as reviewed in Sec.~\ref{corres}. In the CFT these are RG flows between two CFTs. A given flow is parametrized by the scale at which the `transition' between the UV CFT and the IR CFT takes place. In each flow we can again place a black hole at the bottom of the geometry  and perform a double analytic continuation to obtain a confining theory. This is now characterized by two scales, the confinement scale and the scale at which the flow takes place, but the physics will only depend on the ratio of these two scales. 

In order to find the solutions, we consider the IR expansion around the point where the space smoothly closes off. As above, without loss of generality we choose the boundary conditions so that this happens at $\r=0$. As explained above, we also fine-tune the choice of integration constants in such a way as to forbid the IR-relevant deformation. This means that in looking for non-trivial solutions to the equations we will allow for  only one additional integration constant in the system, the scale $s_{\ast}$. 
Because we will keep the end-of-space fixed at $\r_0=0$, the value of $s_{\ast}$ encodes the ratio between the confinement and the scale at which the flow takes place, so physical quantities will only depend on $s_{\ast}$. Under these circumstances the solution near $\r=0$ takes the form 
 \beqs
\phi(\r)&=&-\frac{\log (2)}{\sqrt{5}}+\tilde{\phi}-\frac{e^{-\frac{8
   \tilde{\phi}}{\sqrt{5}}} \left(2-3 e^{\sqrt{5} \tilde{\phi}}+e^{2 \sqrt{5}
   \tilde{\phi}}\right) \r^2}{2^{2/5} \sqrt{5}}\nonumber
   \\ &&\nonumber +
\frac{e^{-\frac{16 f_0}{\sqrt{5}}} \left(-2+e^{\sqrt{5} \tilde{\phi}}\right)
   \left(-1+e^{\sqrt{5} \tilde{\phi}}\right) \left(-18+17 e^{\sqrt{5} \tilde{\phi}}+6
   e^{2 \sqrt{5} \tilde{\phi}}\right) \r^4}{20 \,\,2^{4/5} \sqrt{5}}
   +{\cal O}(\r^6)\,, \nonumber \\
   b(\r)&=&
 \log (\r) -\frac{e^{-\frac{8 \tilde{\phi}}{\sqrt{5}}}
   \left(-1+4 e^{\sqrt{5} \tilde{\phi}}+2 e^{2 \sqrt{5} \tilde{\phi}}\right) \r^2}{5\,\,
   2^{2/5}} \nonumber   \\ \nonumber
   &&\nonumber 
    +\frac{e^{-\frac{16 \tilde{\phi}}{\sqrt{5}}} \left(31-128 e^{\sqrt{5} \tilde{\phi}}+162
   e^{2 \sqrt{5} \tilde{\phi}}+76 e^{3 \sqrt{5} \tilde{\phi}}+34 e^{4 \sqrt{5}
   \tilde{\phi}}\right) \r^4}{250\,\, 2^{4/5}}+{\cal O}(\r^6)\,,\nonumber\\
   c(\r)&=&
  +\frac{\log (\r)}{6}
 +\frac{e^{-\frac{8
   \tilde{\phi}}{\sqrt{5}}} \left(-1+4 e^{\sqrt{5} \tilde{\phi}}+2 e^{2 \sqrt{5}
   \tilde{\phi}}\right) \r^2}{15\,\, 2^{2/5}}  \nonumber\\ && 
     -\frac{1}{375} \sqrt[5]{2} e^{-\frac{16 \tilde{\phi}}{\sqrt{5}}} \left(13-44
   e^{\sqrt{5} \tilde{\phi}}+51 e^{2 \sqrt{5} \tilde{\phi}}-2 e^{3 \sqrt{5}
   \tilde{\phi}}+7 e^{4 \sqrt{5} \tilde{\phi}}\right) \r^4
   +{\cal O}(\r^6)\,,
   \label{wesee}
\eeqs
where $\tilde{\phi}$ is the free constant that parametrizes the family of solutions. Notice that in order for the solutions to approach in the UV the $\phi=0$ solution (the UV fixed point), one has to 
 choose $0\leq \tilde{\phi} \leq  \log 2/\sqrt{5}$. It is therefore convenient to parametrize $\tilde \phi$ as 
\SP{
	\tilde{\phi}=\frac{\log 2}{2\sqrt{5}}\left(1-\tanh\left(\frac{s_{\ast}}{2}\right)\right)\,,
}
The limit $s_{\ast}\rightarrow +\infty$ corresponds to $\tilde \phi \to 0$. In this limit we see from \qq{wesee} that the IR value of $\phi$ is $\phi(\rho=0)\to - \log 2/\sqrt{5}$. In other words, the scalar reaches the value of the IR fixed point before the theory confines. In the gauge theory language this means that the confinement scale is much smaller than the scale at which the transition between fixed points takes place. As a consequence there is some energy range in the IR in which the dynamics is that dictated by the IR CFT, and below this range the theory eventually confines. In the gravitational language this limit corresponds to the case in which the size of the black hole at the bottom of the flow is much smaller than the radial position of the kink in the geometry. In the opposite limit, $s_{\ast}\rightarrow -\infty$, we see that 
$\tilde \phi \to \log 2/\sqrt{5}$ and therefore $\phi(\rho=0) \to 0$. In this limit the confinement scale is much larger than the scale at which the flow would have taken place, the scalar $\phi$ remains approximately constant and equal to its UV value, and confinement takes place before the theory can probe any physics associated to the IR fixed point.

\begin{figure}[h]
\begin{center}
\begin{picture}(460,300)
\put(10,150){\includegraphics[height=4.6cm]{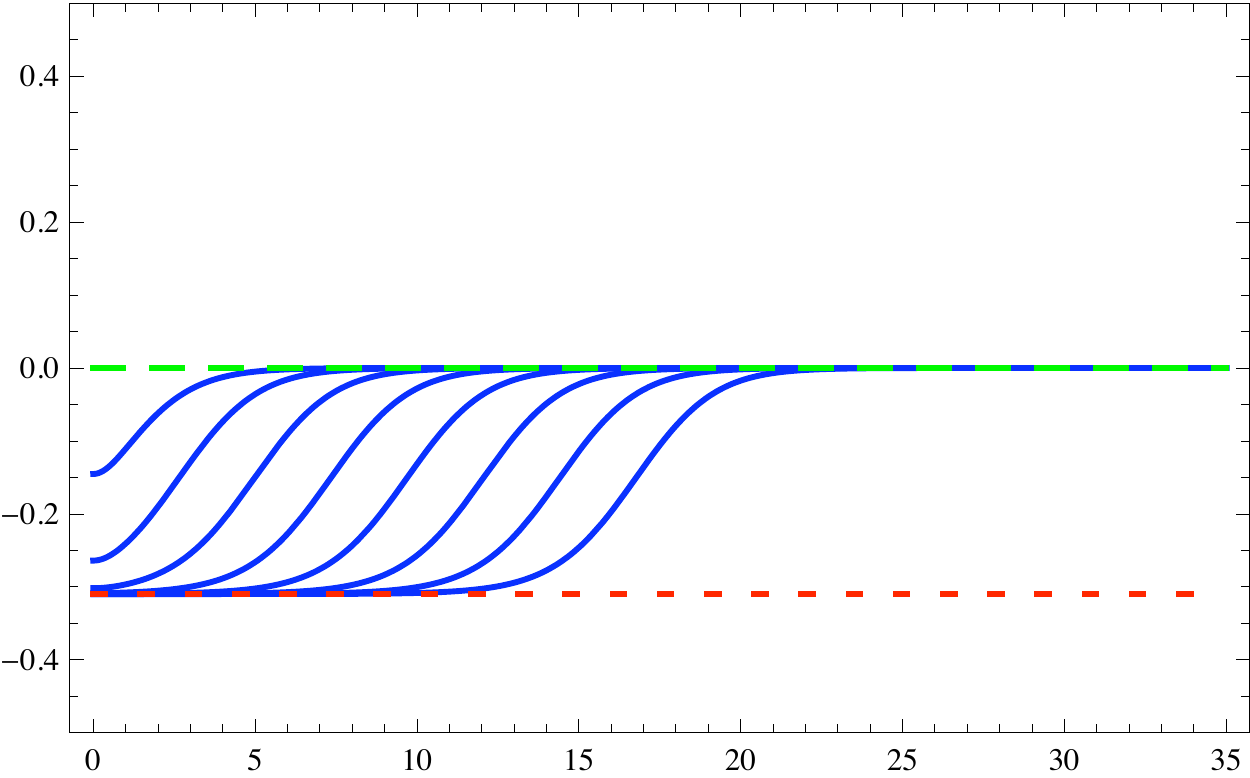}}
\put(240,150){\includegraphics[height=4.6cm]{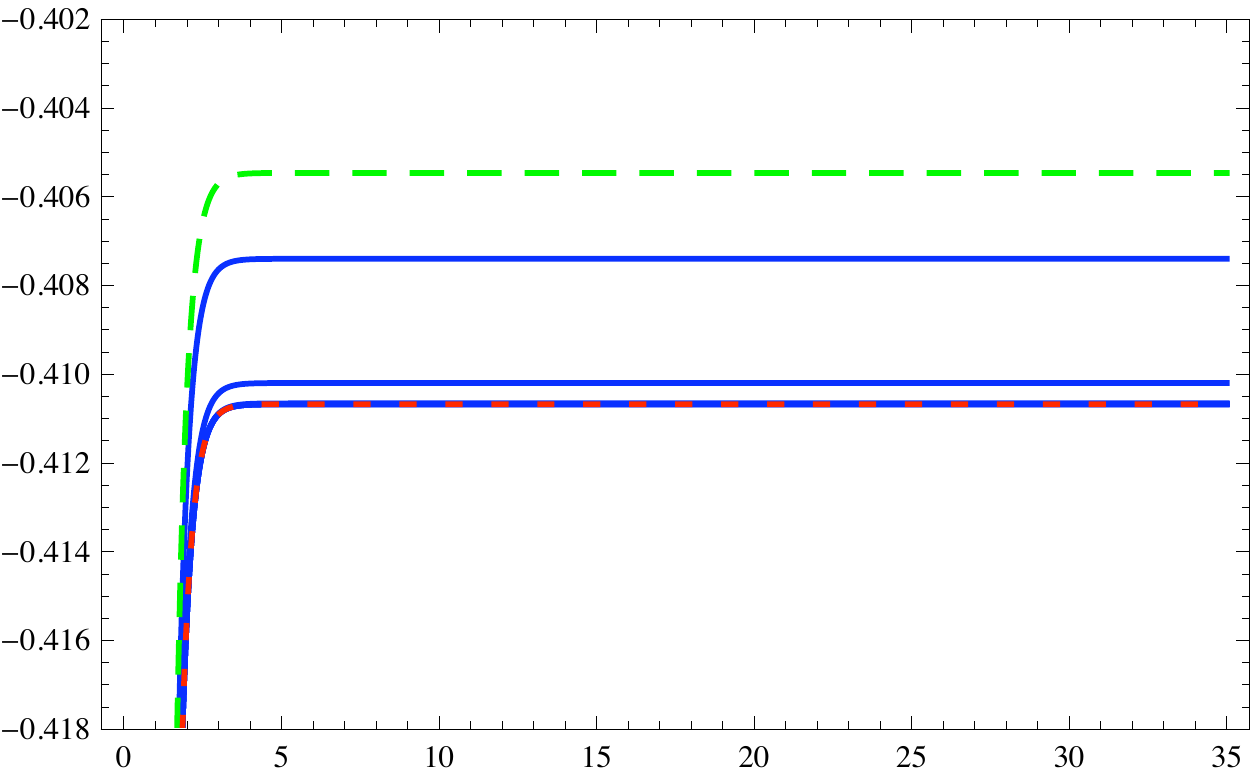}}
\put(10,4){\includegraphics[height=4.6cm]{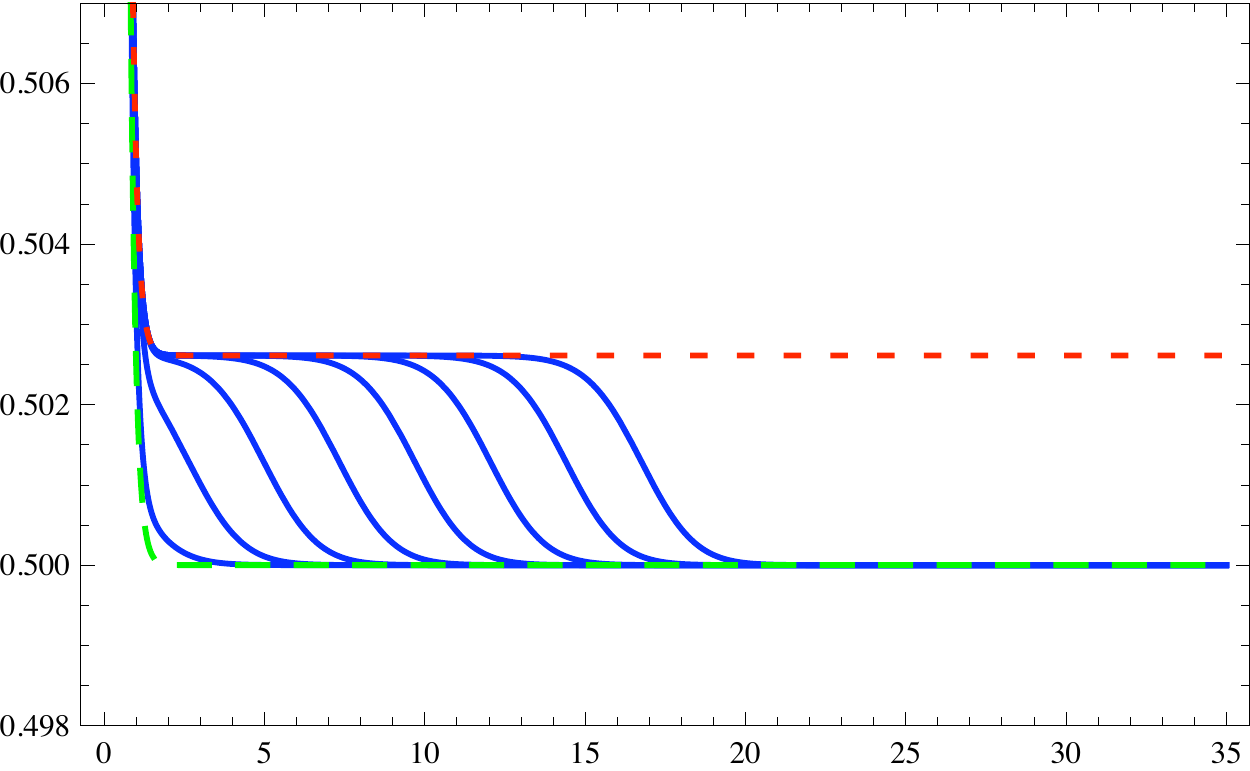}}
\put(244,4){\includegraphics[height=4.5cm]{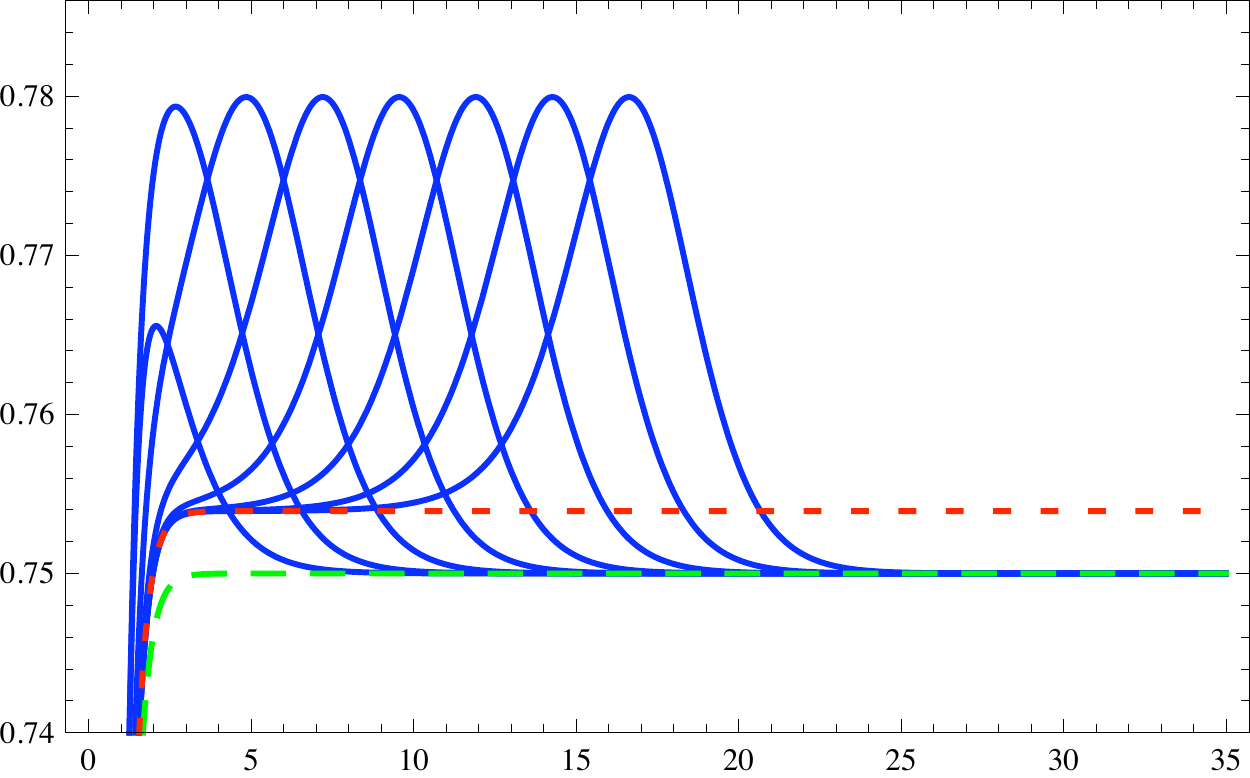}}
\put(5,280){$\phi$}
\put(5,130){$\partial_{\r}c$}
\put(235,280){$b$}
\put(235,130){$\partial_{\r}\Phi$}
\put(200,0){$\r$}
\put(200,147){$\r$}
\put(430,0){$\r$}
\put(430,147){$\r$}
\end{picture} 
\caption{Multi-scale confining solutions for several values of $\tilde\phi$. We show
the functions $\phi$, $\partial_{\r}c$, $b$ and $\partial_{\r}\Phi$ as a function of $\r$. For the latter we have assumed $\xi=\frac{\pi}{2}$, in which case $\Phi=\frac{\phi}{\sqrt{5}}+\frac{3}{2}c-\frac{1}{4}b$.}
\label{Fig:plotsolutions}
\end{center}
\end{figure}
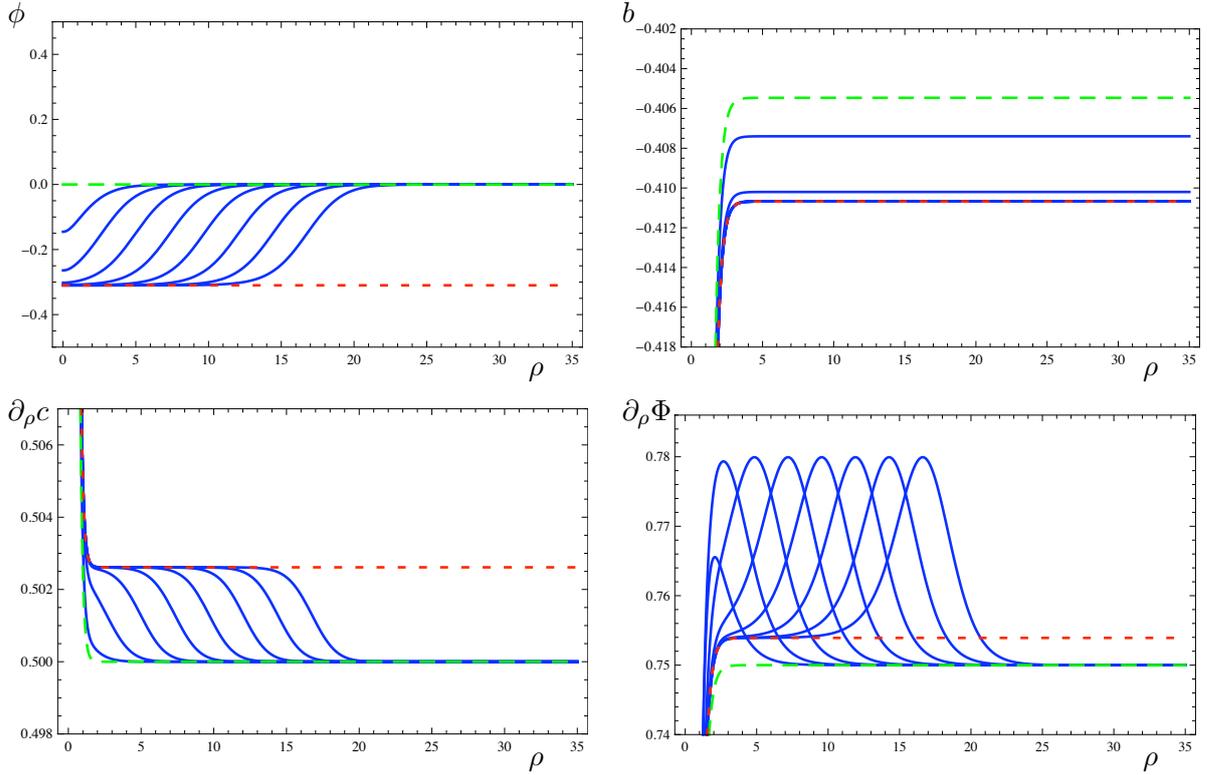

We use the expansion \qq{wesee}  to set up the IR boundary conditions for the numerical study of the backgrounds. The result of the procedure is shown in Fig.~\ref{Fig:plotsolutions}. One can clearly see that all the solutions interpolate between the two simple confining solutions described in Sec.~\ref{simple}, with
the parameter $\tilde{\phi}$ controlling the scale at which the transition  takes place. Note that  these two solutions are very close to each other, because the  two critical points have values of ${\cal V}$ that are themselves very close to each other, and that all the solutions have a linear-dilaton behaviour in the far UV.

\subsection{Spectrum of four-dimensional scalar bound states}
\label{sect2}

Since the $\sigma$-model metric is trivial $G_{ab}={\rm diag}\,\,\left\{\frac{1}{2},1,\frac{15}{4}\right\}$ also for this model, the same simplifications take place as for the model of the previous section, namely that the covariant derivative 
with respect to the $\sigma$-model is just the partial derivative, and also that ${\cal R}^a_{\,\,\,\,bcd}=0$. After changing the radial variable according to 
$\frac{\partial}{\partial_r}=e^{-\chi}\frac{\partial}{\partial_{\r}}$, one thus finds the same formal expressions for the equations of motion and the boundary conditions for the fluctuations as those given in Eqs.~\eqref{eq:eomsflucs1}, \eqref{eq:eomsflucs2}, and \eqref{Eq:bcinf}. 

\begin{figure}[t]
\begin{center}
\begin{picture}(320,210)
\put(10,6){\includegraphics[height=7cm]{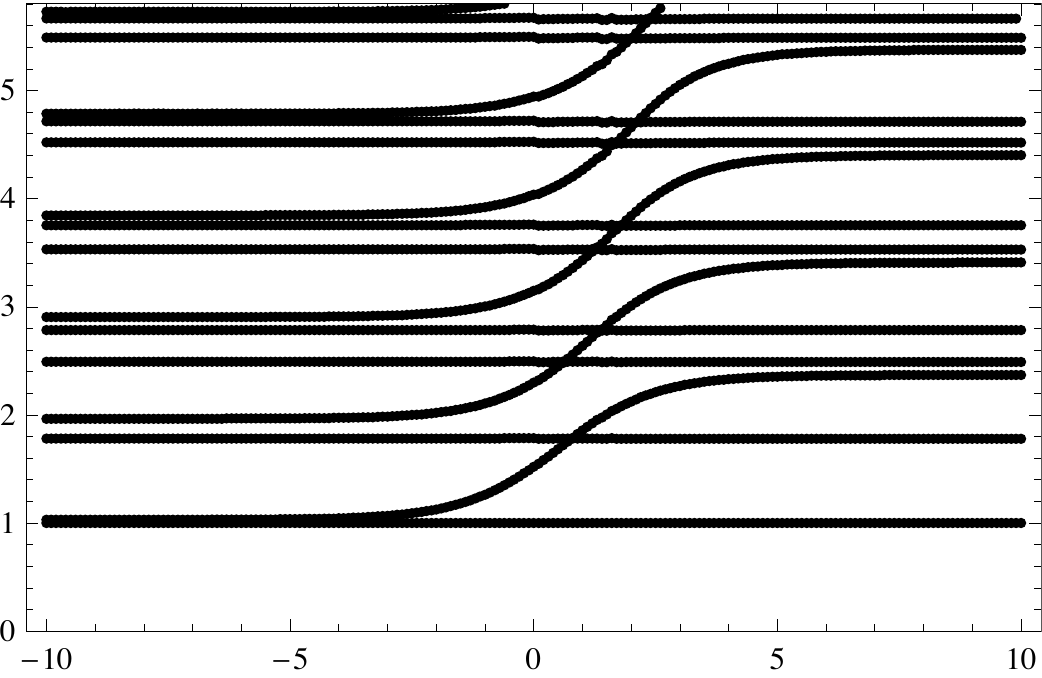}}
\put(-15,110){$M$}
\put(160,-5){$s_*$}
\put(280,210){$\longrightarrow$ IR}
\put(5,210){UV $\longleftarrow$}
\end{picture} 
\caption{Spectrum for the confining solutions in the M-theory model 
as $s_*$ is varied, with $\r_I=0.001$, $\bar{\r}=4$ and $\r_U=15$. The spectrum at the two fixed points is recovered in the limits $s_\ast \to -\infty$ (UV fixed point) and $s_\ast \to + \infty$ (IR fixed point), as indicated in the figure. The overall normalization is chosen so that the first state among those that are insensitive to $s_*$ has unit mass.}
\label{Fig:plotscalars3}
\end{center}
\end{figure}
We  perform a numerical study of the complete spectrum for different values of $s_{\ast}$. The result is shown in Fig.~\ref{Fig:plotscalars3}. 
We see that in the limits $s_{\ast}\rightarrow \pm \infty$ the spectrum converges smoothly to the cases in which $\phi$ is constant. The numerical values of the glueball masses at the fixed points are shown in Table~\ref{masses2}. We also note that, despite the fact that for non-constant $\phi$  there is mixing among all the scalars, this mixing 
is small, and as a result only one of the three towers of states shows an explicit dependence on $s_{\ast}$. This is the tower of states corresponding to fluctuations of $\phi$, whose bulk equations 
depend on which critical point the background is close to.

\begin{table}[tbp]
   \centering
\begin{tabular}{ | c | c| }
  \hline
  IR ($s_\ast \to + \infty$) & UV ($s_\ast \to -\infty$) \\ \hline
  1 & 1 \\
   & 1.02 \\
    1.77 & 1.77 \\
     &  1.96 \\
  2.36 &  \\
  2.50 &  2.51 \\
  2.78 & 2.78 \\
   & 2.90 \\
  3.40 &  \\
  3.55 & 3.55\\
  3.75 & 3.75 \\
   & 3.84 \\
     4.40 &  \\
  4.54 & 4.55 \\
   4.71& 4.72 \\
      & 4.78 \\
   5.38 & \\
   5.52 & 5.52\\
   5.67 & 5.67\\
  \hline
\end{tabular}
 \caption{Numerical values of the masses of scalar bound states at each of the two fixed points in the M-theory model.}
   \label{masses2}
\end{table}

\subsection{Other physical quantities}

In this section we look at various other quantities that may be used to characterize the long-distance behavior of the dual field theory,
all of which are studied by using extended objects as probes of the dynamics.
\subsubsection{Wilson loops}
To compute the expectation value of the Wilson loop, we assume that we have a 
solution for $c$, $\phi$ and $b$ of the class in Fig.~\ref{Fig:plotsolutions}.
We then write explicitly the string-frame metric
\beqs
\di s^2_{10} &=& e^{\frac{2}{3}\Phi}\Delta^{1/3}\left[\frac{}{}e^{2\left(c-\frac{b}{6}\right)}\di x_{1,3}^2+\di \r^2+e^{2\left(c+\frac{5}{6}b\right)}\di \eta^2+\di \tilde{\Omega}_4
\frac{}{}\right]\nonumber\\
&=&\Delta^{\frac{1}{2}}e^{\left(c-\frac{b}{6}\right)}\left[\frac{}{}e^{2\left(c-\frac{b}{6}\right)}\di x_{1,3}^2+\di \r^2+e^{2\left(c+\frac{5}{6}b\right)}\di \eta^2+\di \tilde{\Omega}_4
\frac{}{}\right]\nonumber\\
&=&\left(e^{\frac{4\phi}{\sqrt{5}}}\sin^2\xi+e^{\frac{\phi}{\sqrt{5}}}\cos^2\xi\right)^{\frac{1}{2}}e^{\left(c-\frac{b}{6}\right)}\left[\frac{}{}e^{2\left(c-\frac{b}{6}\right)}\di x_{1,3}^2+\di \r^2+\cdots
\frac{}{}\right]\,.
\label{pre}
\eeqs
For simplicity, we consider a string whose endpoints lie at $\xi=\frac{\pi}{2}$. 
Unlike in the previous section, in this case the entire string lies at this value of $\xi$. This follows from the fact that, for the allowed values of $\phi \leq 0$, the minimum of the $\xi$-dependent prefactor in the metric \qq{pre} always lies at $\cos^2 \xi = 0$. We therefore set $\xi=\frac{\pi}{2}$ in the following. 

\begin{figure}[h]
\begin{center}
\begin{picture}(290,300)
\put(29,150){\includegraphics[height=5cm]{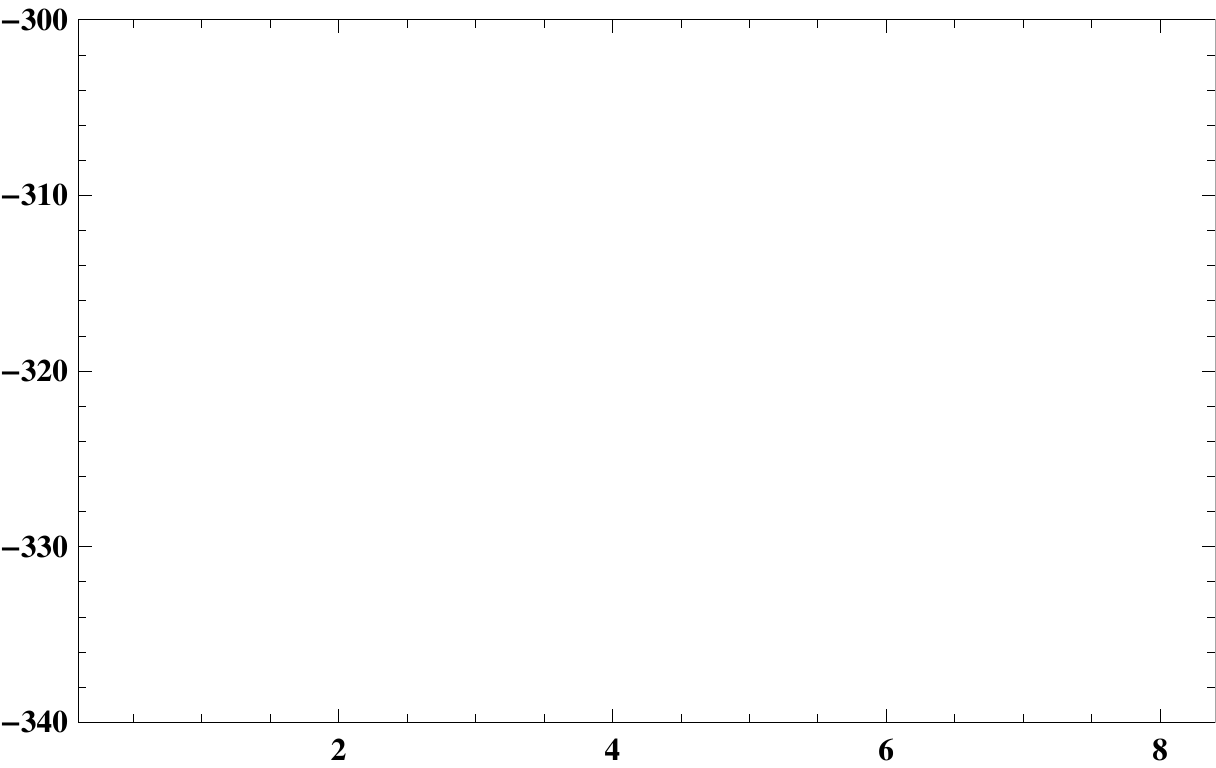}}
\put(21,0){\includegraphics[height=4.9cm]{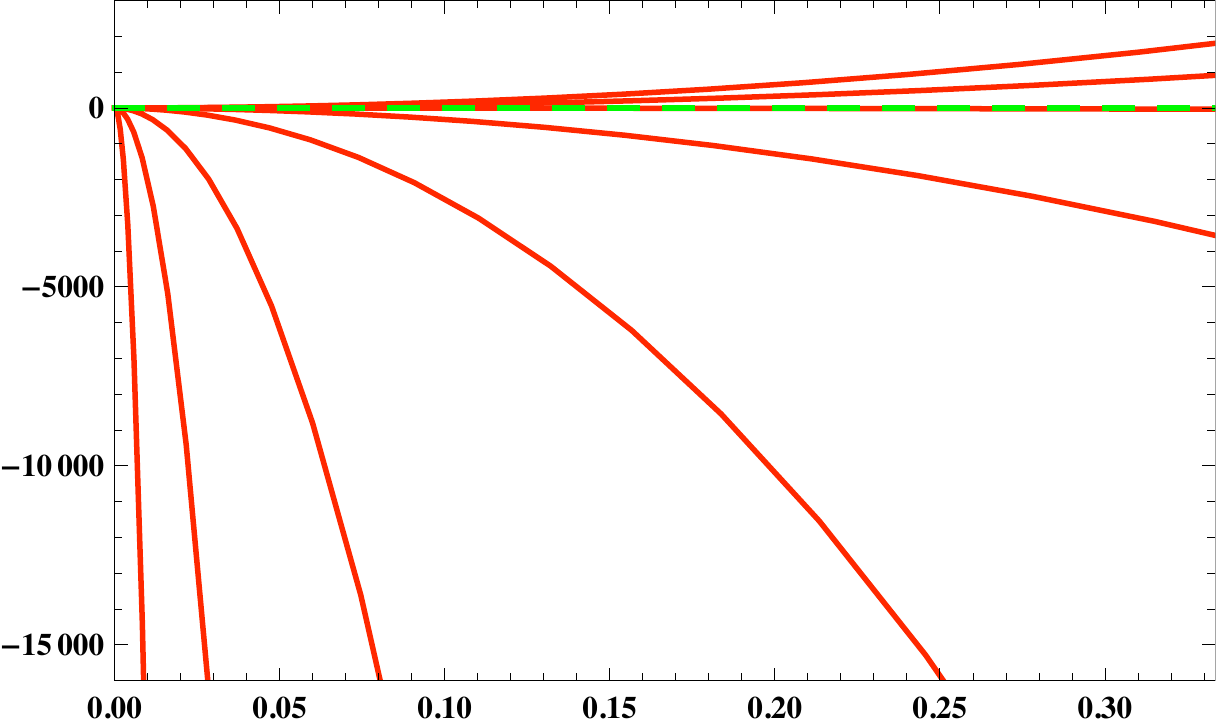}}
\put(0,280){$\bar{E}_{QQ}$}
\put(0,130){$\bar{E}_{QQ}L_{QQ}^2$}
\put(260,156){$L_{QQ}$}
\put(260,6){$L_{QQ}$}
\end{picture} 
\caption{ The result of the study of the Wilson loop in the M-theory model
for confining solutions which differ by the choice of $s_{\ast}$.}
\label{Fig:Wilson7D}
\end{center}
\end{figure}

Under these circumstances, the functions controlling the behavior of the string embedding are
\beqs
F^2&=&g_{tt}g_{xx}\,=\,e^{4\frac{\phi}{\sqrt{5}}+6c-b}\,,\\
G^2&=&g_{tt}g_{\r\r}\,=\,e^{4\frac{\phi}{\sqrt{5}}+4c-\frac{2}{3}b}\,.
\eeqs
The result of the numerical study of the Wilson loop is shown in Fig.~\ref{Fig:Wilson7D}.
The main result is that at large $L_{QQ}$ we find the expected linear behavior. The long-distance behavior is governed by the 
the string tension $\sigma =F(0)$, which depends on the IR behavior of the background. In particular, for the extreme case $\phi=0$ one has $\sigma=1$, while $\phi=-\frac{\log 2}{\sqrt{5}}$  yields $\sigma=2^{-\frac{2}{5}}$. The short-distance behavior is  $\bar{E}_{QQ}\propto 1/L_{QQ}^2$. 
The origin of this is that, despite the fact that the ten-dimensional geometry is not asymptotically AdS in the UV, the eleven-dimensional lift to M-theory is.

\subsubsection{Gauge coupling}

\begin{figure}[h]
\begin{center}
\begin{picture}(290,300)
\put(11,150){\includegraphics[height=4.9cm]{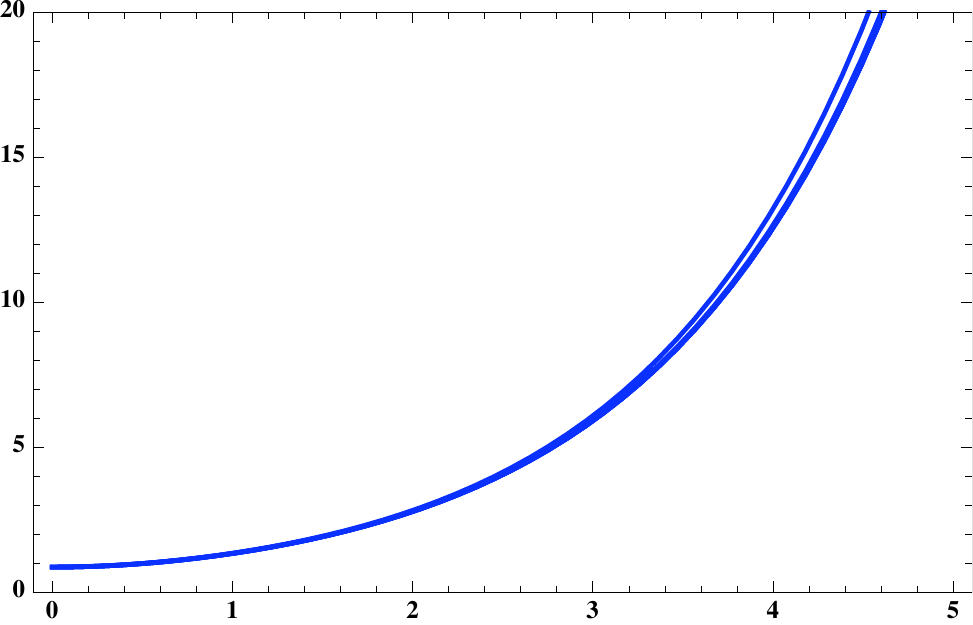}}
\put(14,0){\includegraphics[height=4.9cm]{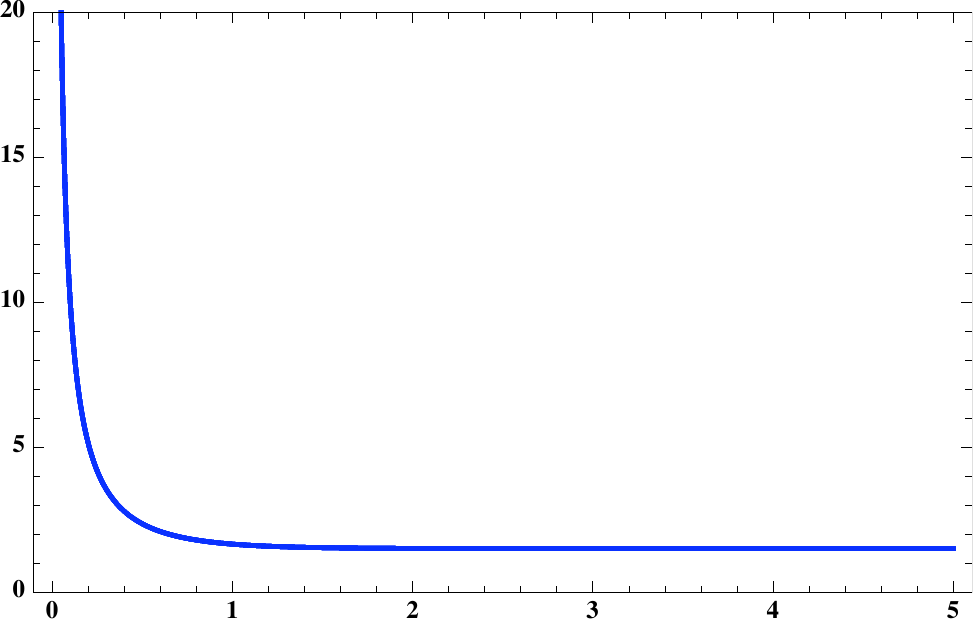}}
\put(-15,270){$\frac{g_5^2}{16\pi^3\ell_s}$}
\put(-15,125){$\frac{g_4^2\,R_5}{8\pi^2\ell_s}$}
\put(237,156){$\r$}
\put(237,6){$\r$}
\end{picture} 
\caption{ The five-dimensional gauge coupling $g_5^2$ and the four-dimensional gauge coupling $g_4^2$ in the M-theory model 
as a function of $\r$, for a sample of confining solutions that differ by the choice of scale $s_{\ast}$.}
\label{Fig:gauge7D}
\end{center}
\end{figure}

We compute the gauge coupling of the dual field theory by embedding 
D4-branes extended in the four-dimensional Minkowski directions
as well as along the $S^1$ spanned by $\eta$.  We assume that the probes have a configuration
for which $\xi=\frac{\pi}{2}$ and setting $\alpha^{\prime}=1$ for convenience we find that:
\beqs
\frac{1}{g_5^2}&=&\frac{e^{-\Phi}}{16\pi^3 \ell_s}\,=\,\frac{e^{-\frac{\phi}{\sqrt{5}}-\frac{3}{2}c+\frac{1}{4}b}}{16\pi^3 \ell_s}\,,\\ [1.5mm]
\frac{1}{g_4^2}&=&\frac{2\pi R_5}{g_5^2}\sqrt{\tilde{g}_{\eta\eta}}
\,=\,\frac{e^{b} R_5}{8\pi^2 \ell_s}\,.
\eeqs
We show in Fig.~\ref{Fig:gauge7D} the result of using these equations with a set of solutions which differ only by the choice of $s_{\ast}$.
Several features are worth noticing.
First of all, the difference between the various backgrounds is barely visible.
Second, the fact that the dilaton diverges linearly in the UV means that the 
five-dimensional coupling does so as well.
Finally, contrary to what happened in the string model, the asymptotic value of the $g_4$ coupling is a large constant. Again, this is the result of the linear-dilaton behaviour of the background.

\subsubsection{D$8$ embedding}

We focus on the Type IIA lift and embed a probe-D8 in the background.
We recall the expressions of the metric (in string-frame) and of the dilaton:
\beqs
\di s_{10}^2 &=&\left(e^{\frac{4\phi}{\sqrt{5}}}\sin^2\xi
+e^{\frac{\phi}{\sqrt{5}}}\cos^2\xi\right)^{\frac{1}{2}}e^{\left(c-\frac{b}{6}\right)}
\left[\frac{}{}e^{2\left(c-\frac{b}{6}\right)}\di x_{1,3}^2+\di \r^2+e^{2(c+\frac{5}{6}b)}\di \eta^2+\di\tilde{\Omega}_4^2
\frac{}{}\right]\,, \nonumber\\
\di \tilde{\Omega}_4^2&=&e^{\frac{3\phi}{\sqrt{5}}} \di \xi^2+\frac{1}{4}e^{-\frac{\phi}{\sqrt{5}}}
\left(e^{\frac{4\phi}{\sqrt{5}}}\sin^2\xi
+e^{\frac{\phi}{\sqrt{5}}}\cos^2\xi\right)^{-1}\cos^2\xi 
\left[\frac{}{}\di \theta^2+\sin^2\theta \di\varphi^2+(\di\psi +\cos\theta \di \varphi)^2\right]\,,\nonumber\\
e^{\frac{4}{3}\Phi}&=&\left(e^{\frac{4\phi}{\sqrt{5}}}\sin^2\xi
+e^{\frac{\phi}{\sqrt{5}}}\cos^2\xi\right)^{\frac{1}{3}}e^{2c-\frac{1}{3}b}\,.\nonumber
\eeqs
We use an ansatz in which the brane extends in the four Minkowski dimensions and wraps the
internal $S^4$.  The embedding is then specified by the functions $\r(\sigma)$ and $\eta(\sigma)$.
As clear from the expression of the metric, the $\xi$-dependence is somewhat problematic,
in the sense that it makes the integration over this angle complicated.
For our purposes, we hence focus only on the case
\beqs
\phi&=&0\,,\\
c&=&\frac{1}{6}\log\left[\frac{1}{3}\sinh(3\r)\right]\,,\\
b&=&\log\left[\frac{2}{3}\tanh\left(\frac{3 \r}{2}\right)\right]\,,
\eeqs
that is, the original Witten model~\cite{Witten}.  This means that we are reproducing the result of Sakai and Sugimoto~\cite{SS}, which we include here for completeness. 
The DBI action becomes
\beqs
{\cal S}_{\rm D8}&=&-\frac{8 \pi^2 T_8}{3}\int \di^4 x 
\int \di\sigma\sqrt{\tilde{F}^2\eta^{\prime\,2}+\tilde{G}^2\r^{\prime\,2}}\,,
\eeqs
where
\beqs
\tilde{F}^2&=&e^{16 c-\frac{2}{3}b}\,,\\
\tilde{G}^2&=&e^{14 c -\frac{7}{3}b}\,.
\eeqs

\begin{figure}[h]
\begin{center}
\begin{picture}(290,150)
\put(11,0){\includegraphics[height=5cm]{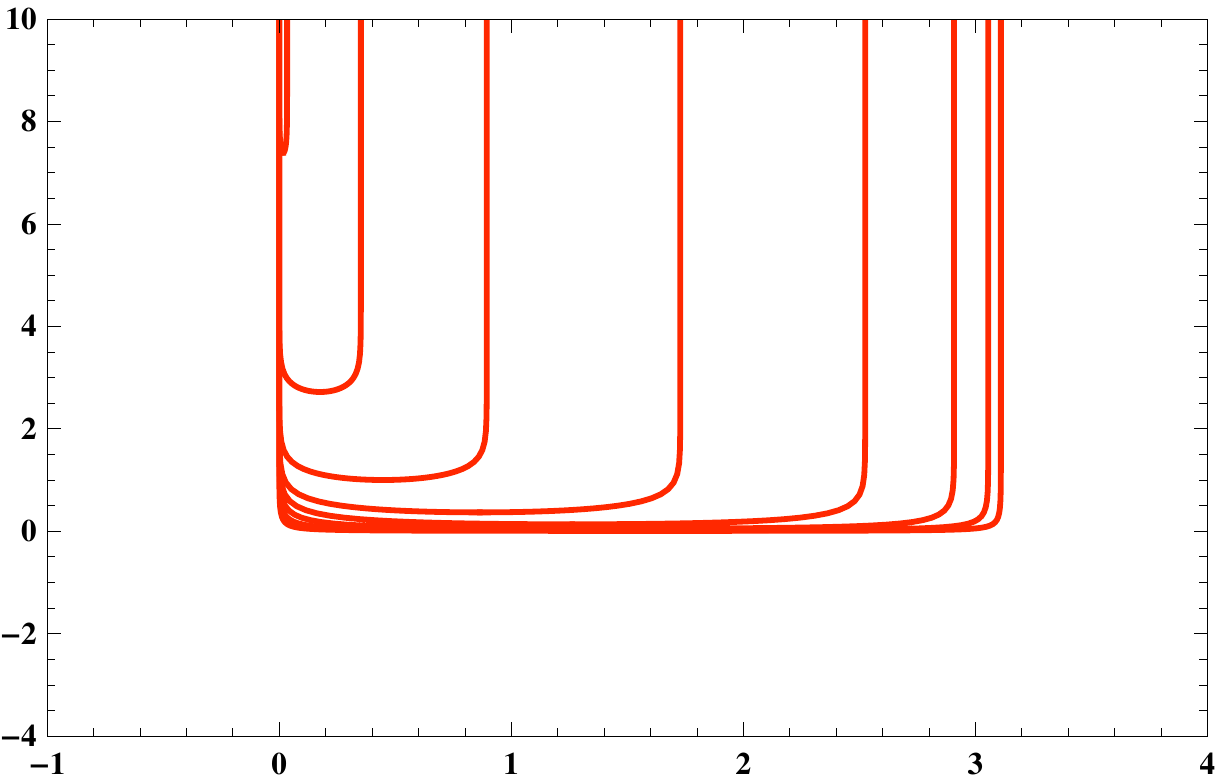}}
\put(0,130){$\r$}
\put(240,6){$\eta$}
\end{picture} 
\caption{ Embedding of the probe D$8$ in the Witten model, 
in the case of confining solutions with $\phi=0$, for various choices of $\hat{\r}_o>0$,
and using $\r_U=35$ as a cutoff.}
\label{Fig:7Dbrane}
\end{center}
\end{figure}

The result of the numerical study is shown in Fig.~\ref{Fig:7Dbrane}. The equivalent embedding can be found for the confining solution based on the non-susy fixed point.
Notice the clear similarity with what we found in the string model.

\section{Discussion}

This section is devoted to the discussion of the physical implications of 
the work presented in the main body of the paper.
In particular, we critically compare results obtained with different models, and 
address the two fundamental questions we started with, 
related to the concept of universality of gauge/gravity results,
and to the physics of the four-dimensional dilaton.

\subsection{Comparing spectra of glueballs}

\begin{figure}[h]
\begin{center}
\begin{picture}(290,220)
\put(-60,10){\includegraphics[height=9cm]{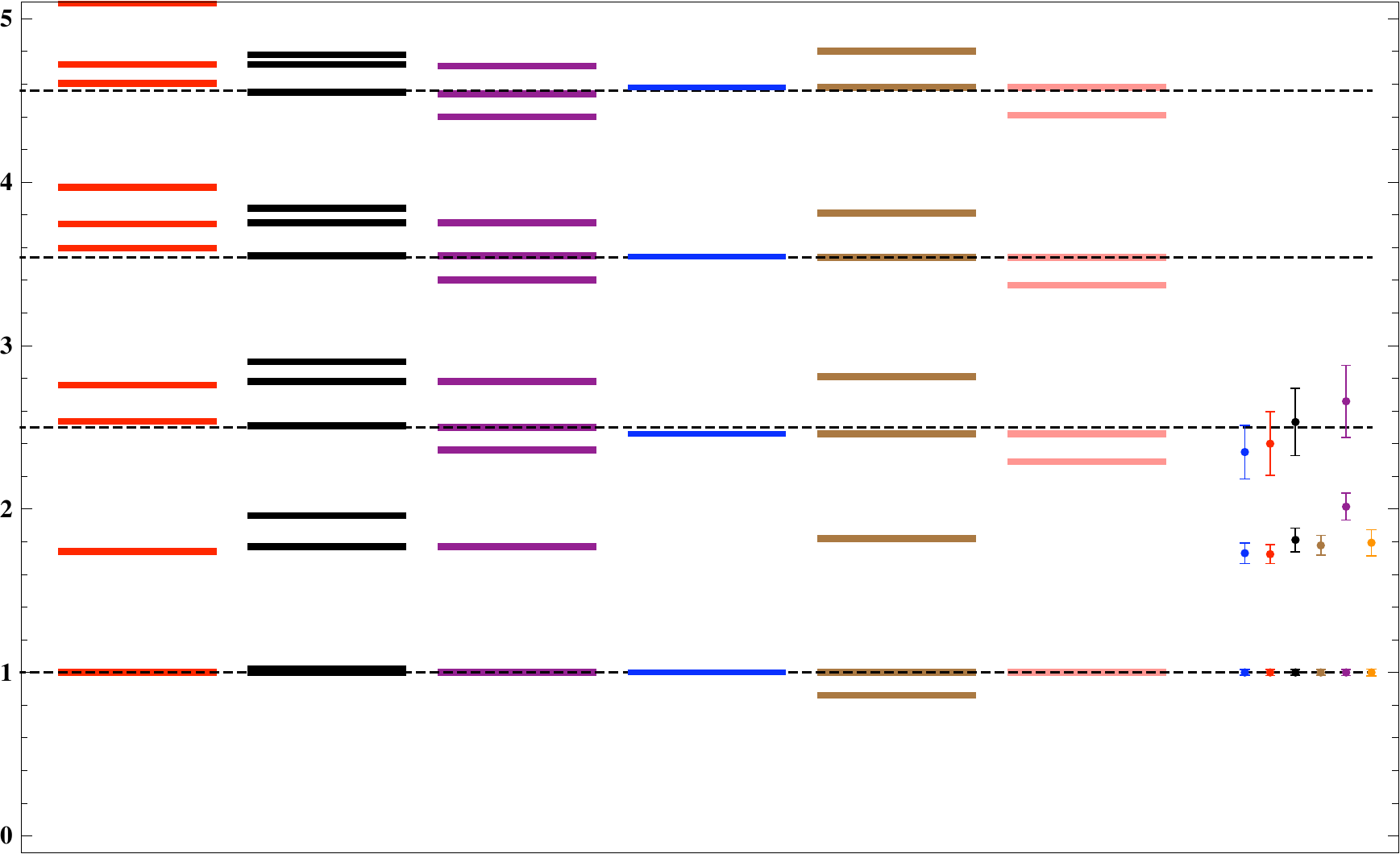}}
\put(-80,150){$M$}
\end{picture} 
\caption{ Spectrum of $0^{++}$ glueballs according to several different calculations. 
Left to right:  the spectrum of the Witten model (the M-theory model at the UV fixed point) according to~\cite{gravityspectrum2} (red) and according to our calculations in Section
 \ref{sect2} (black), the spectrum obtained from the M-theory model at the IR, non-supersymmetric fixed-point, also discussed in Section \ref{sect2} (purple), the spectrum obtained by considering the string model at either fixed point and truncating $\phi$ from the spectrum \cite{Wen} (blue),  the spectrum obtained from the string theory model
at the UV, supersymmetric fixed point discussed in Section \ref{sect1} (brown) and at the IR, non-supersymmetric fixed point, also discussed in Section \ref{sect2} (pink). 
The normalisation  of the spectra is discussed in the main text.
The dotted lines, which lie at $M= \{1, 2.5, 3.54, 4.56\}$, highlight the subset of states that is common to all the models (within numerical precision). 
Finally, in the last column we report lattice results for $SU(N_c)$ Yang-Mills, with $N_c=3\,,\ldots\,,8$ (left to right),
as discussed in the text.}
\label{Fig:spectracompared}
\end{center}
\end{figure}

In Fig.~\ref{Fig:spectracompared} we show the comparison between six different calculations of the glueball spectrum of a QCD-like (or, rather, Yang-Mills-like) theory based on gauge/gravity dualities
in the large-$N_c$ limit,\footnote{We do not compare with the results from  \cite{gravityspectrum1}  since this reference worked in the probe approximation. Despite existing similarities in the spectra, we also do not pursue a detailed comparison with semi-phenomenological models such as \cite{Gursoy:2009jd}.} as well as some lattice results. The normalization for the second and third (fifth and sixth) columns is that explained in Section \ref{sect1} (Section \ref{sect2}), namely the mass of the lightest state that is common to the UV and IR fixed points is set to unity. The first column is the spectrum of the Witten model, i.e.~the UV fixed point of the M-theory model of Section \ref{sect2}, according to \cite{gravityspectrum2}, so for consistency we have simply set the mass of the lightest state to unity. The fourth column is the spectrum obtained by considering the string model at either fixed point and truncating $\phi$ from the spectrum \cite{Wen}.

Fig.~\ref{Fig:spectracompared} exhibits several interesting features. First, there is a set of states that \cite{gravityspectrum2} included in their calculation that we have truncated away, and viceversa. For the states that both \cite{gravityspectrum2} and we kept, the agreement is fairly good; it is possible that the small discrepancies are simply  due to the numerical nature of the calculations. 
Similarly, our calculation for the string-theory model agrees well with~\cite{Wen} for the common part of the spectrum, but in our calculation we kept one more tower of states.

Second, the six towers shown in Fig.~\ref{Fig:spectracompared} agree to a remarkable degree on
 the subset of the spectrum highlighted with dotted lines, suggesting a certain universality of this part of the spectrum. 
Establishing whether this is an artifact of the specific mechanism implementing confinement on the gravity side or a truly generic feature of large-$\nc$ Yang-Mills theories would require further study.

Third, the lightest state is a bit lighter than what one might have expected based on the splitting within the rest of states in the common subset. This splitting is $\delta M \simeq 1.03$, as can be seen from the dotted lines in Fig.~\ref{Fig:spectracompared}, which is somewhat smaller than the $\Delta M \simeq 1.5$ between the first and the second state. As explained in the main body of the paper, the origin of this fact is clear on the gravity side: the physical states are admixtures of fluctuations of the five-dimensional scalars 
and of the four-dimensional dilaton, and the lightest of such states is also the one which contains the largest dilaton component, making it somewhat special. Thus, although none of the models discussed in this paper admits a naturally light dilaton,  the physics of the lightest state is appreciably affected by the dilaton dynamics.

Fourth, all the towers of states have masses that asymptotically
behave as $M_n\simeq M_0+ n \, \delta M$, i.e.~the masses (rather their squares) are linear in the excitation number, failing to reproduce the Regge behaviour. This is of course expected in this type of supergravity backgrounds.

It  would be very interesting to know whether such pattern also emerges  in the large-$N_c$ extrapolation of lattice calculations (see~\cite{Lucini:2012gg}). Since this extrapolation is difficult (see below),  in the last column of Fig.~\ref{Fig:spectracompared} we compare  with finite-$\nc$ lattice data taken from~\cite{Lucini:2010nv}.
We show the results for the states that the authors of \cite{Lucini:2010nv} refer to as $A_1^{++}$, obtained from lattice calculations with finite lattice spacing ($N_L=12$) for  $SU(N_c)$ Yang-Mills theories with $N_c=3\,,\ldots\,,8$ (left to right  in the plot). Where available, we included the first three excited states.
 Notice that in the process we selected only states that have negligible mixing with  lattice artifacts such as multi-glueballs and bi-torelon states (see \cite{Lucini:2010nv} for a discussion), and we did not include the extrapolation to large-$\nc$.
 Again, we normalise the lightest state to unity, and show also the statistical uncertainty of the lattice results.
 
 The outcome of this comparison, while encouraging, is at  present inconclusive. On the one hand, it looks as if the first excited universal state that we identified on the supergravity side, with $M\simeq 2.5$, 
also appears in the lattice data. On the other hand, this data also seems to show that  an intermediate state with $M\sim 1.8-2$ is present, which would agree with the calculations in the M-theory model, but not in the string model. However, several considerations must be kept in mind when interpreting this. For example, it is possible that this mode is present in the full string model, as opposed to the subset of modes that we truncated the model to. Also, the lattice data show that the extrapolation to large-$\nc$  is difficult.
 Notice for example how the excited states for odd and even $N_c$ exhibit a different pattern. This might well be just a statistical fluctuation, or instead an indication of the fact that  the lattice data is still too far from the large-$\nc$ limit, and that we are therefore misidentifying the states. Furthermore, the error quoted is only statistical,  the continuum limit has not been implemented, and it might be that the intermediate state is actually a two-glueball state. (Although the analysis 
 in~\cite{Lucini:2010nv} would disfavour this possibility, it does show that some contamination is present.)
 It would be very useful if lattice studies such as the one in~\cite{Lucini:2010nv} 
 could be improved systematically.


\subsection{Dilaton, fine-tuning and naturalness}

All the models of strongly-coupled confining theories that we have  studied have a spectrum of glueballs that resembles what is expected from a generic Yang-Mills theory, in which no parametrically light pseudo-dilaton appears. However, it is interesting to investigate the degree of fine-tuning that would be needed in order to make the lightest state parametrically light.
 
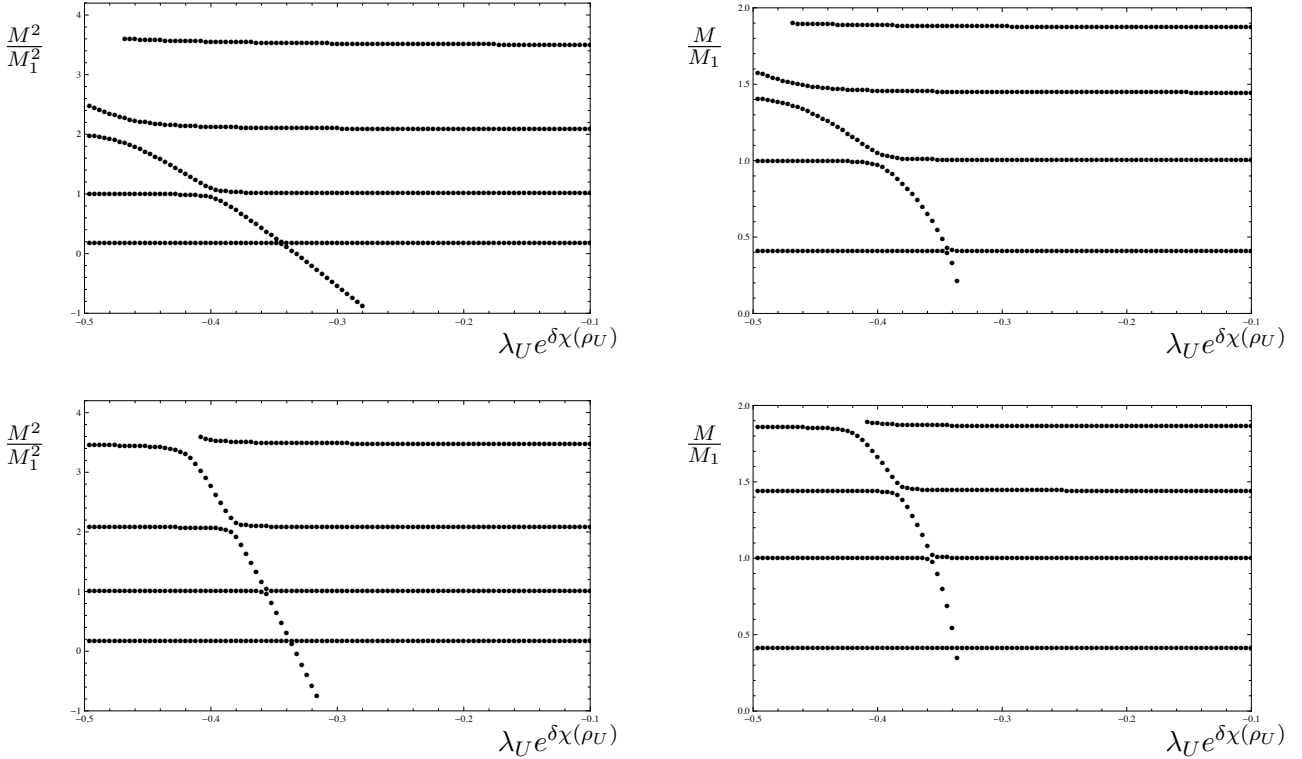
\begin{figure}[h,t]
\begin{center}
\begin{picture}(500,310)
\put(0,157){\includegraphics[height=4.3cm]{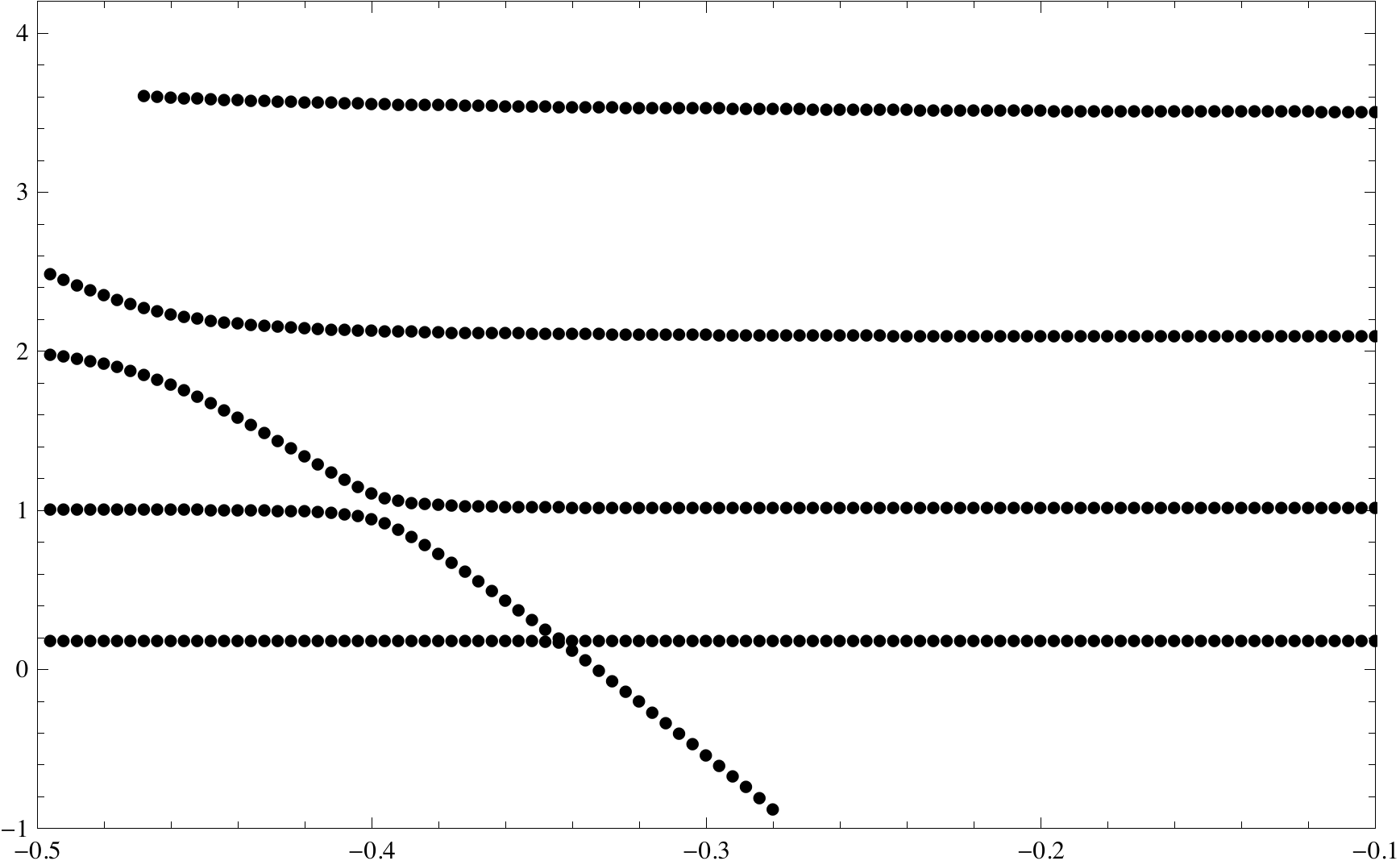}}
\put(250,157){\includegraphics[height=4.3cm]{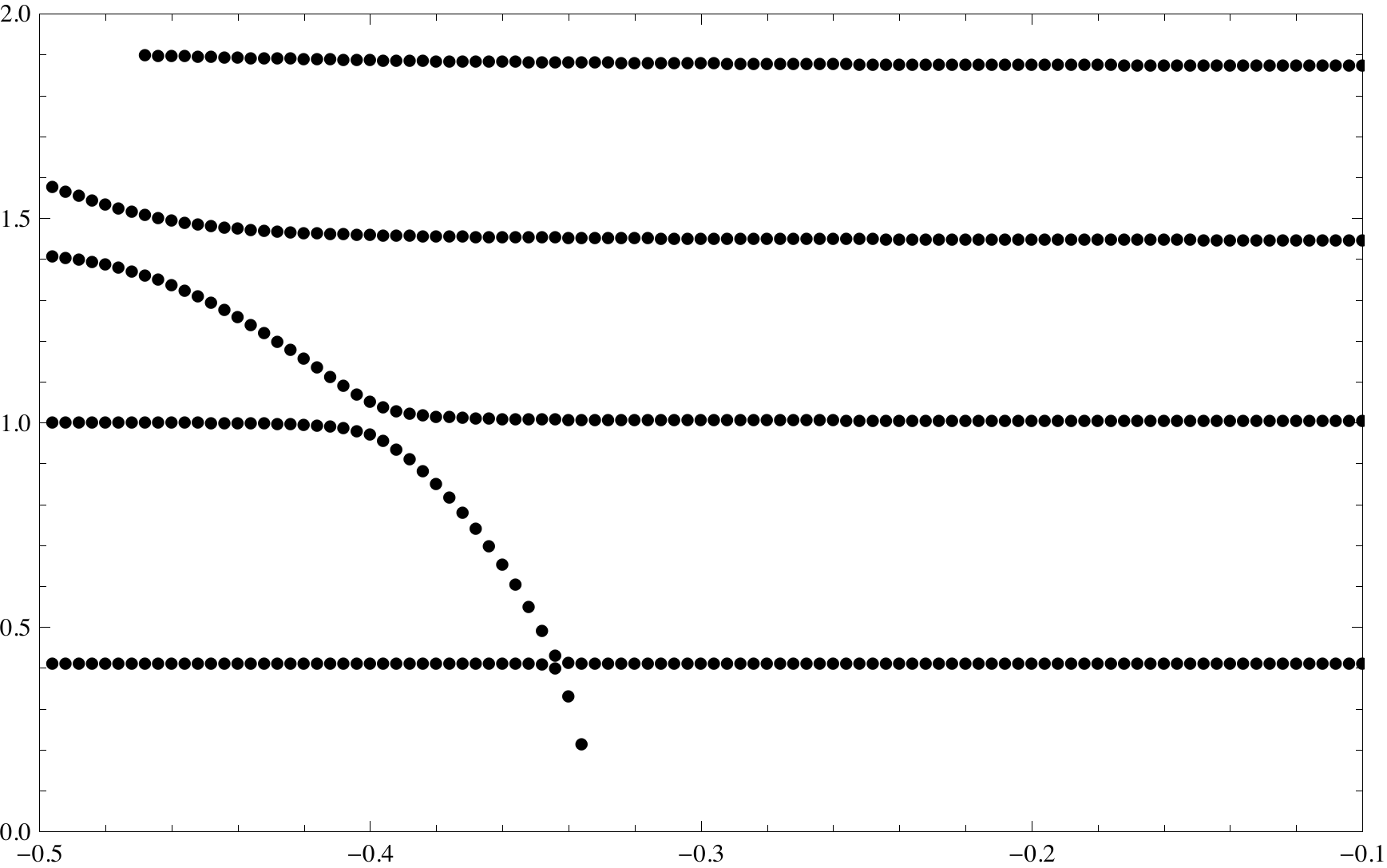}}
\put(0,7){\includegraphics[height=4.3cm]{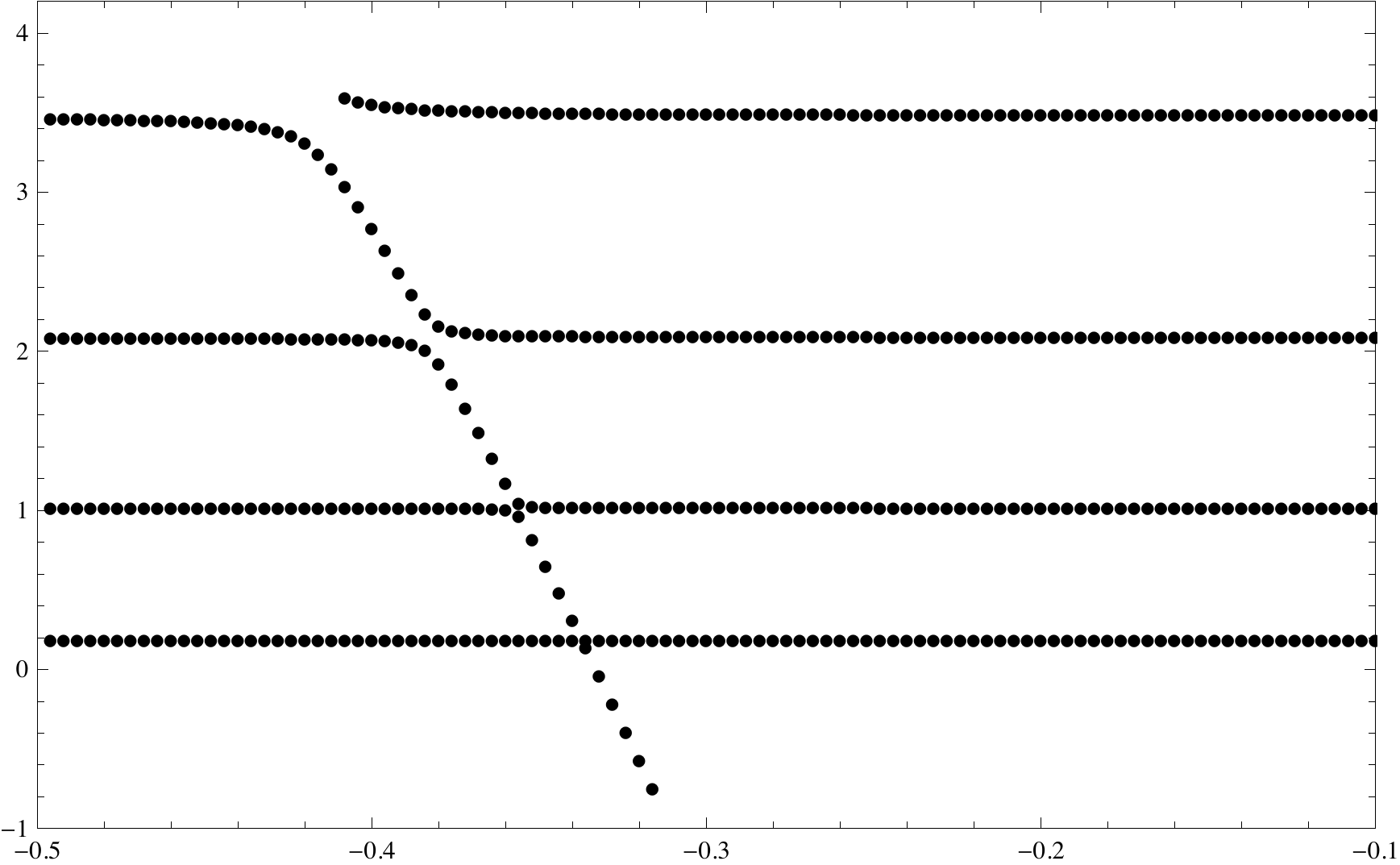}}
\put(250,7){\includegraphics[height=4.3cm]{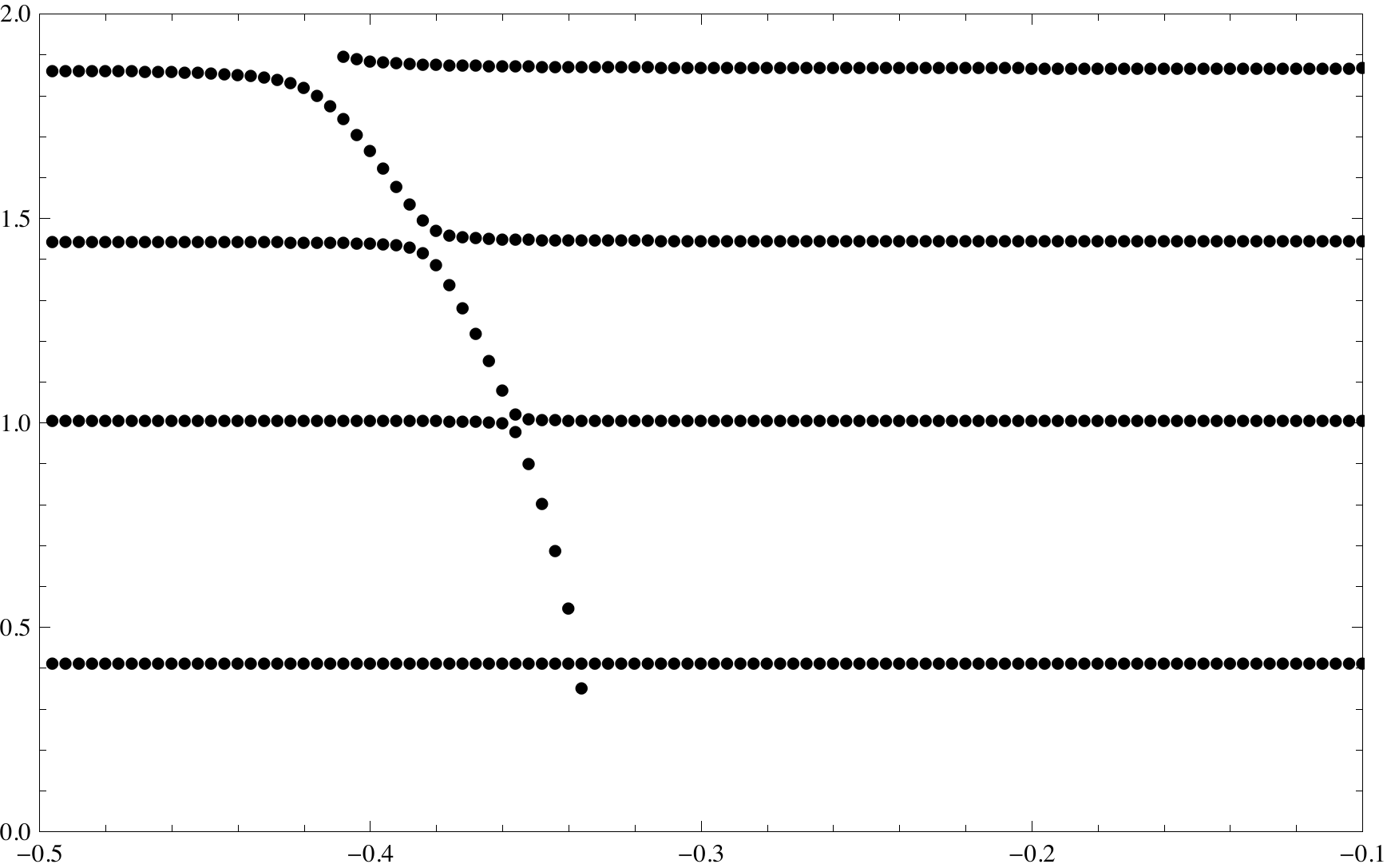}}
\put(230,260){$\frac{M}{M_1}$}
\put(230,110){$\frac{M}{M_1}$}
\put(-25,260){$\frac{M^2}{M^2_1}$}
\put(-25,110){$\frac{M^2}{M^2_1}$}
\put(160,-3){$\lambda_Ue^{\delta\chi(\r_U)}$}
\put(160,147){$\lambda_Ue^{\delta\chi(\r_U)}$}
\put(410,-3){$\lambda_Ue^{\delta\chi(\r_U)}$}
\put(410,147){$\lambda_Ue^{\delta\chi(\r_U)}$}
\end{picture}
\caption{Mass of the scalar glueballs in the string theory model with $\phi=0$, computed keeping a finite value of the $\lambda_U$ term in the UV boundary conditions, and for finite UV cutoff. The two top panels show the result with $\r_U=3$, the two lower panels with $\r_U=3.5$. In all cases the background is the confining one,
and we used $\r_I=10^{-3}$ in the numerical study. $\delta$ is a model-dependent parameter that in the present case is $\delta \simeq 0.4$.}
\label{Fig:finetuning}
\end{center}
\end{figure}

 With this aim, we perform an exercise that to some extent is a repetition of a similar one in~\cite{EP}, in the context of a model inspired by the Goldberger--Wise mechanism. However, we are dealing with a more realistic set-up. Since all the theories we studied share a common subset of glueball states, we concentrate for simplicity on the string model with $\phi=0$ and consider the truncation to $\chi$. We redo the calculation of the spectrum, but now we change the boundary conditions in the UV,
 allowing for a finite value of the constant $\lambda_U$ (rather than taking the limit $\lambda_U\rightarrow -\infty$,
 as in the body of the paper). This is implemented by holding in place a finite UV-cutoff $\r_U$, 
 chosen to be  small.
 We show some of  the results of this study in Fig.~\ref{Fig:finetuning}.

Before we explain its physical meaning, let us briefly summarise the results
of the exercise. We show in the figure how the spectrum gets modified by the choice of 
the UV-localized boundary potential.  We explained earlier in the paper
 that this boundary-localised term is the gravity formulation
of the details about the regulator procedure one needs to implement in the UV.
It is important to notice that for generic values of $\lambda_U$ one finds very good agreement
with the results we showed in the main body of the paper, for all the theories we discussed,
in spite of the fact that we are here using  low values of the UV cutoff.
This corroborates that the physical results do not depend significantly 
on the regulator procedure adopted in the UV. In particular it confirms that, had we carried out the
procedure with other (allowed) choices of the boundary potential, we would have gotten the same results.

There is a dramatic exception: there exists a very special value of $\lambda_U\simeq -\frac{1}{3}e^{-\delta\chi(\r_U)}$
(where $\delta\simeq 0.4$) for which the lightest mass can be made arbitrarily light. Actually, 
one can make the lightest state into a tachyon, signifying that not all possible choices of $\lambda_U$ are admissible.
For a generic $\lambda_U$, the result we quoted in the paper about the mass of the lightest state
is in fact an upper bound.
Yet, reducing the mass requires to finely adjust the choice of $\lambda_U$ to lie in a very narrow range
below this special value.
The range depends strongly on the UV cutoff one chooses, as can be seen
in Fig.~\ref{Fig:finetuning} from the comparison between $\r_U=3$ and $\r_U=3.5$. Already for $\r_U\simeq 4$ it 
becomes difficult to find this range numerically. Remember that $\r_U$ is 
related to the logarithm of the cutoff scale in the dual theory,
and that we used always $\r_U > 8 $ in the main body of the paper.

What is the origin of this phenomenon, at the technical level?
With some algebra, one can convince oneself that, irrespectively of the
bulk dynamics, the bulk equations for the gauge-invariant fluctuations $\mathfrak{a}^a$
always admit 
a solution with $M^2=0$, given by $\mathfrak{a}\propto \frac{\Phi^{\prime\,a}}{A^{\prime}}$.
This being a system of  second-order  linear equations, there exists also another
independent massless solution, which is not known in closed form,
and any superposition of the two is still a massless solution.
Effectively, this solution results from the undoing of the 
mixing between the fluctuations of the five-dimensional scalar fields with the dilaton $h$.
But in general this mode is unphysical, since it does not satisfy the boundary conditions.
The actual physical states all result form the mixing of $h$ with the $\sigma$-model scalars.

By fine-tuning the boundary potential in the UV, one can effectively achieve the cancellation of this
mixing. As a result, there always exists, for any  $\sigma$-model and as long as one works 
with a finite UV cutoff, a special range of values  of the
$\lambda^a_{\,\,\,\,|c}$ matrix in the UV boundary conditions such that the 
resulting spectrum contains a very light scalar, the composition of which is dominantly $h$.
But this is accomplished at a cost: this range of possible numerical values
 shrinks to zero very fast once we take the UV cutoff to infinity, 
and ultimately reduces to a pathological singular value.

Let us now explain the physical meaning of this exercise.
As suggested in the title of this subsection, we want to explain what role the concept of fine-tuning has in all of 
this paper.
All the complete models we considered are based on backgrounds that asymptotically in the far-UV 
are AdS. The dual field theories are hence UV-complete. Irrespectively of the fact that 
one might consider them not to be fully realistic models of nature 
(for example, the UV fixed points live in higher dimensions), these are not effective field theories but UV-complete. As such, there is no sense in which a fine-tuning problem can arise.

This statement is however tainted by the observation that we have been considering the models in complete isolation.
If we were to suggest that these models be used to construct 
the electroweak symmetry breaking sector of the Standard Model (in the spirit of technicolor), 
we would need to add to the system new elementary degrees of freedom
(quarks, leptons and gauge bosons) and new weakly-coupled interactions
(the $SU(3)\times SU(2)\times U(1)$ of the Standard Model, 
and some couplings replacing the Yukawa interactions).
Setting aside complications related with realistic model-building and with holographic renormalisation, 
the simplest possible way to achieve this
would be to allow the additional
 degrees of freedom to propagate on the UV-boundary,
and to add specific localised couplings.
In field theory terms, this would render the theory as a whole UV-incomplete, because most of the 
Standard-Model interactions do not admit a UV fixed point.
Hence, the whole procedure we carried out would have to be considered
 in the framework of an effective theory, and the UV-cutoff $\r_U$ would be given a physical meaning.

Let us pretend that we are in this scenario, and ask ourselves how it affects the UV boundary conditions.
Perturbative corrections, coming from loops of the weakly-coupled sector, would affect 
the boundary potential of those scalars that couple to the standard-model fermions and gauge bosons.
Hence, one in general expects that the physical result of the calculation of any observable quantity
(in particular, the spectrum of scalar states), will be modified with respect to what results from 
the analogous calculations performed for the strongly-coupled sector in isolation.
In particular, it was emphasised in~\cite{Foadi:2012bb} that the mass of the lightest glueball
might be significantly affected,
because this scalar contains a significant contamination from the dilaton, and the latter will necessarily couple 
to the fermions and gauge bosons of the standard model, in analogy with what happens for the Higgs particle
of the Standard Model.
In particular, it might turn out that the resulting physical 
mass is actually lighter than one would have expected on the basis of the strong-coupling calculation.

The exercise in this subsection shows a concrete realisation of this possibility: by dialing the
value of what is effectively a localized mass term we could make the mass of the lightest state arbitrarily light.
However, the exercise we performed shows explicitly  two remarkable facts that we want  to stress.
First of all, the relation between the physical mass of the lightest state 
 in the spectrum, and the boundary-localised mass term
(which one might want to infer from the divergent loop diagrams of the weakly-coupled sector)
 is highly non-linear (because the strongly-coupled interactions are),
and does not allow for simple naive dimensional analysis estimates (because there is no
way to treat the scalar glueballs as weakly-coupled objects in a low-energy effective theory).
Most importantly, the degree of fine-tuning one has to accept in order to carry out 
such cancellation procedure is catastrophically large: it takes a very accurate choice of $\lambda_U$
in order to reduce the mass of the lightest state by a factor of two, and this choice is rendered 
practically impossible by its strong dependence on the UV cutoff. Unless  the UV cutoff is very low,
with all the other problems that would come with it.

The spectrum we quoted in the paper is the result one gets for any typical choice
of the boundary term $\lambda_U$, and hence it is the {\it natural} result to be expected also once
the strongly-coupled sector is coupled to the Standard-Model (or to any other external physical sector).
It is still possible that by accurately choosing $\lambda_U$ one can make the (pseudo-)dilaton light, 
irrespectively  of the strong dynamics, hence invoking some conspiracy between the
strongly-coupled and weakly-coupled sectors of the full theory.
 But in the models we looked at in this paper, 
this can be done only at the price of a large fine-tuning.

If one is to accept that some fine-tuning is to be expected, 
this is actually technically possible.
Indeed, in this case  the lightest state is predominantly a dilaton,
and as such its leading order couplings to the standard-model fields are qualitatively the same
as those of the Higgs particle recently discovered by the LHC.

We are not going to pursue this line of arguments any further, since
one main reason for looking at strongly-coupled models of electroweak symmetry breaking 
is to solve the naturalness problem.
It would be satisfactory to find a model where the lightness 
of the scalar state has a natural dynamical explanation, rather than
being the result of fine-tuning.
We conclude by reminding the reader that in the special case studied in~\cite{ENP},
a light state is present in regions of the parameter space of the strongly-coupled theory, without 
having implemented any fine tuning with the boundary conditions ($\lambda_U\rightarrow -\infty$ in 
those calculations). 

\subsection{Future directions}

We close by summarising a few directions for future investigation suggested by our results:

\begin{itemize}

\item We proposed two one-parameter families of models, within Type IIA supergravity, which generalise the 
Witten model, and we showed that the U-shaped 
D8 embedding by Sakai and Sugimoto exists also for all these models. In would be  interesting to repeat the exercise of computing the spectrum of mesons, baryons, 
glueballs with spin, and in general all
the excited states in these new contexts. 

\item We proposed alternative lifts to Type IIB supergravity for all the models considered,
constructed by making use of non-abelian T-duality. It would be interesting to study these lifts, 
and understand whether there is any concrete sense in which the resulting 
models are phenomenologically interesting.

\item We showed that, at the price of fine-tuning, it is possible to make the lightest glueball 
in these models parametrically light compared to the rest of the spectrum.
In this limit, we expect the physics of this state to look similar to that of the Higgs 
boson recently discovered at the LHC.
It would be interesting to study  in more detail the properties of such a light state in this regime,
and possibly build a phenomenological model of electroweak symmetry breaking based on this scenario.
In particular, it would be interesting to compute the decay constant of the light scalar, 
and to compare to the LHC data.

\item We showed that a special sequence of $0^{++}$ glueballs appears in the spectrum of all the models we constructed. It would be interesting to understand to what extent this is a robust feature of large-$\nc$ Yang-Mills theories.

\item
While the models of flow between fixed points that we considered yield a 
spectrum of glueballs with no parametrically light scalar,
and yielding no distinctive features in the string and D8  probes, these results are most likely model-dependent.
It would be interesting to find other models of holographic 
flows of this type and see whether their physics  shows 
more dramatic effects due to the intrinsic multi-scale nature of the dynamics. For example, we expect this to be the case in models in which the scales of explicit and spontaneous breaking of scale invariance can be parametrically separated. 

\end{itemize}

\vspace{1.0cm}
\begin{acknowledgments}
MP would like to thank Biagio Lucini and Luigi Del Debbio for discussions about the current status of lattice calculations of glueball spectra. DE is supported by the DOE through the grant \protect{DE-SC0007884}.
 DM and AF are supported by grants 2009-SGR-168, MEC FPA2010-20807-C02-01, MEC FPA2010-20807-C02-02, CPAN CSD2007-00042 Consolider-Ingenio 2010, and  ERC Starting Grant ``HoloLHC-306605''. 
The work of CH is supported in part by the Israeli Science Foundation Center
of Excellence, and by the I-CORE program of Planning and Budgeting Committee and the Israel Science Foundation (grant number 1937/12). The work of MP is supported in part by WIMCS and by the STFC grant ST/J000043/1.

\end{acknowledgments}
\appendix
\section{About the spectra of the string model}
In this appendix we collect some technical aspects regarding the computation of the spectra
of scalar bound states in the theory defined by the confining backgrounds
of the string model.

It is of some interest to write explicitly the bulk equations, specifying for  the case in which we fix the background scalar $\phi=0$:
\beqs
0&=&\left[\partial_{\r}^2\,+\,\frac{10}{3}\coth\left(\frac{10 \r}{3}\right)\partial_{\r}+\frac{8}{3}
+{2^{-\frac{14}{15}}\cosh^{\frac{4}{5}}\left(\frac{5\r}{3}\right)}M^2\right]
\mathfrak{a}^1\,,\\
0&=&\left[\partial_{\r}^2\,+\,\frac{10}{3}\coth\left(\frac{10 \r}{3}\right)\partial_{\r}+\frac{1000}{3}\left(1+4\cosh\left(\frac{5\r}{3}\right)\right)^{-2}
+{2^{-\frac{14}{15}}\cosh^{\frac{4}{5}}\left(\frac{5\r}{3}\right)}M^2\right]
\mathfrak{a}^2\,.\nonumber
\\ 
\eeqs
Since $\phi$ is constant in the background, there is no mixing between $\mathfrak{a}^1$ and $\mathfrak{a}^2$. 
 Moreover, its fluctuations coincide with $\mathfrak{a}^1$ and obey Dirichlet boundary conditions:
\beqs
\left.\mathfrak{a}^1
\right|_{\r_i}&=&0\,,
\eeqs
while the boundary conditions for $\mathfrak{a}^2$ are unilluminating and we do not reproduce them here.

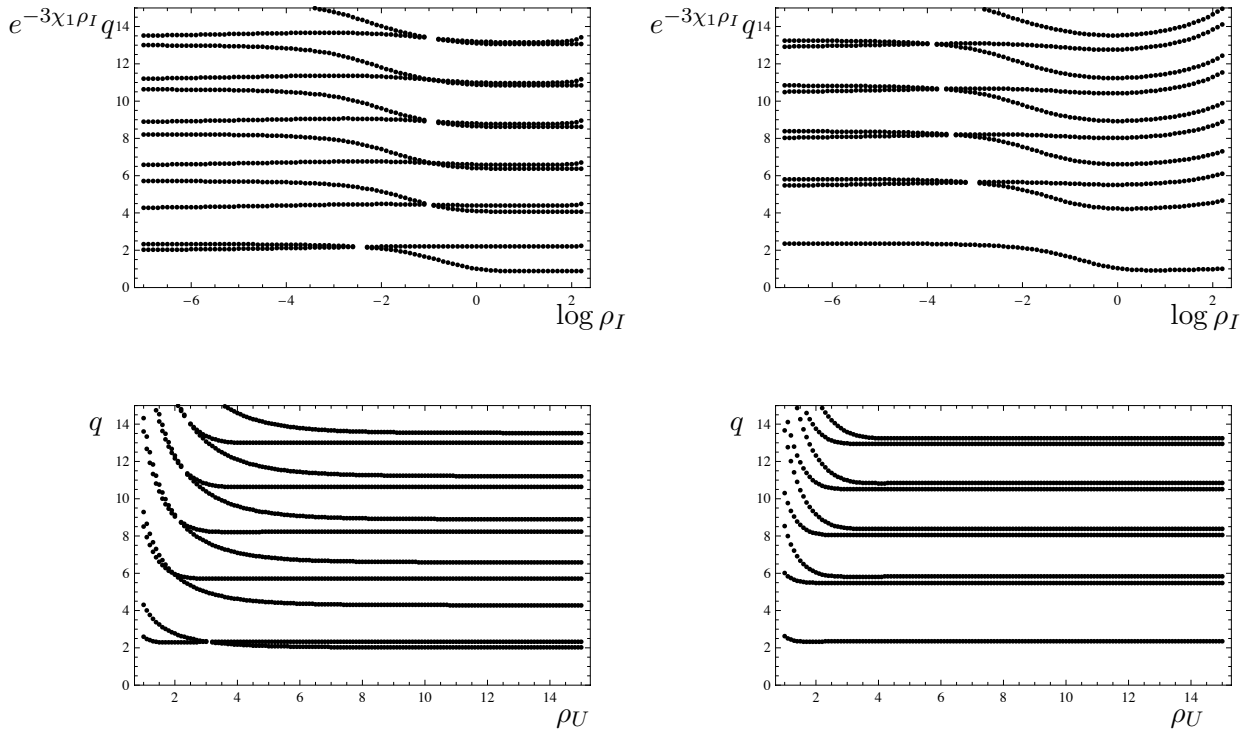
\begin{figure}[t]
\begin{center}
\begin{picture}(680,300)
\put(15,156){\includegraphics[height=4cm]{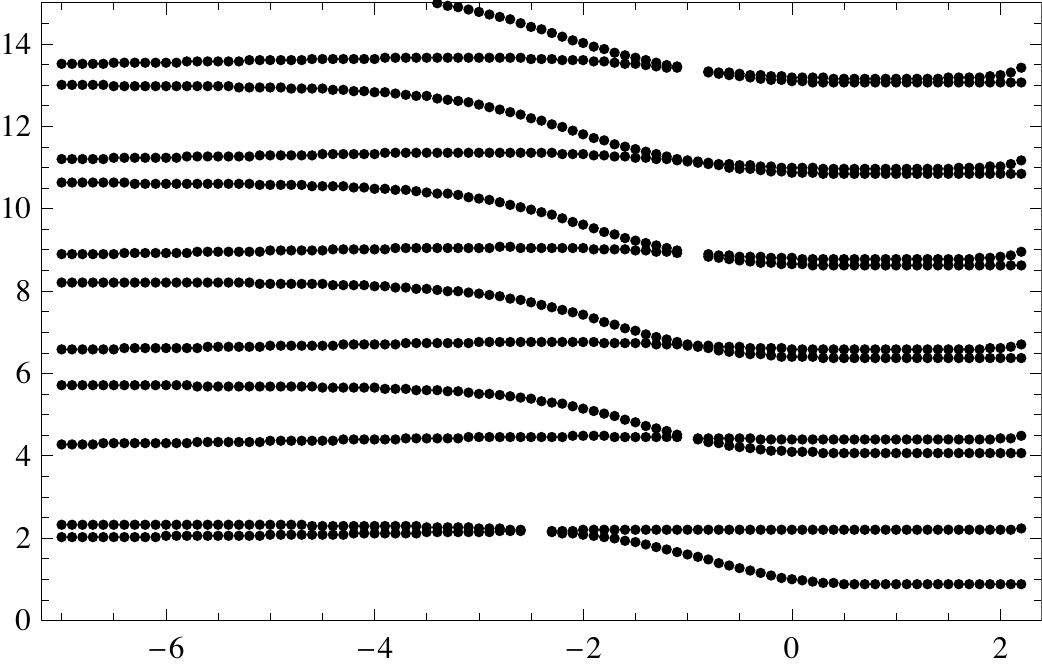}}
\put(255,156){\includegraphics[height=4cm]{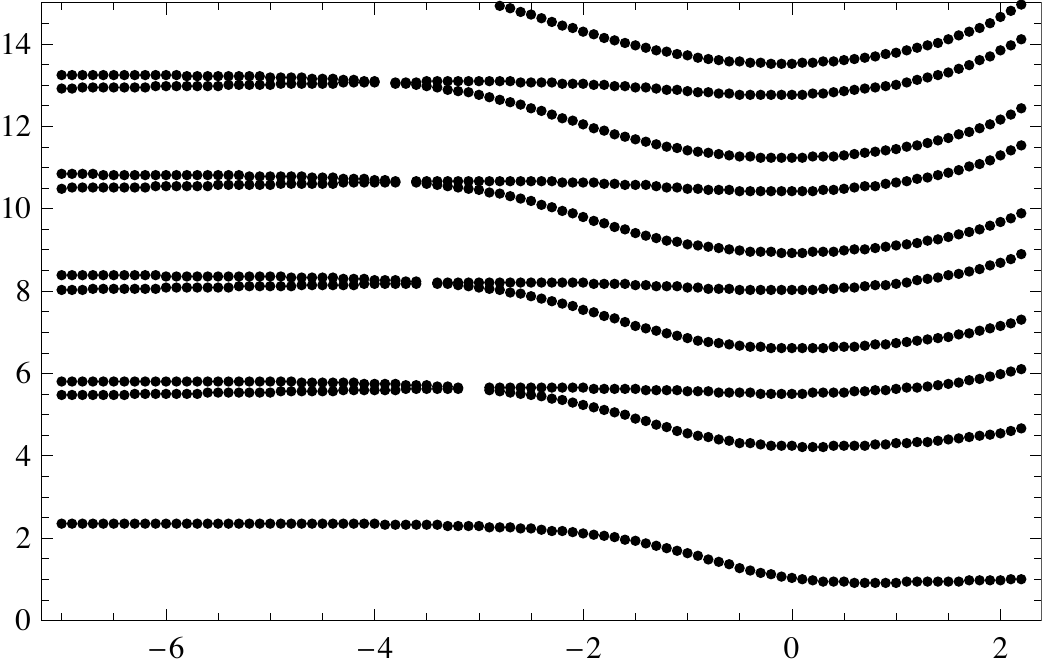}}
\put(15,6){\includegraphics[height=4cm]{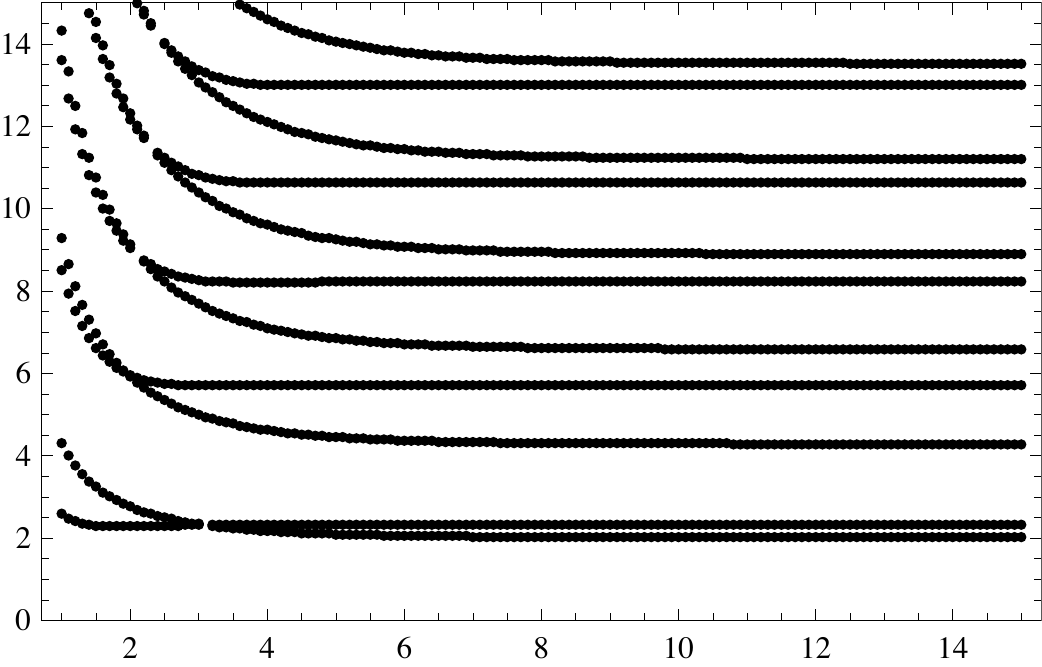}}
\put(255,6){\includegraphics[height=4cm]{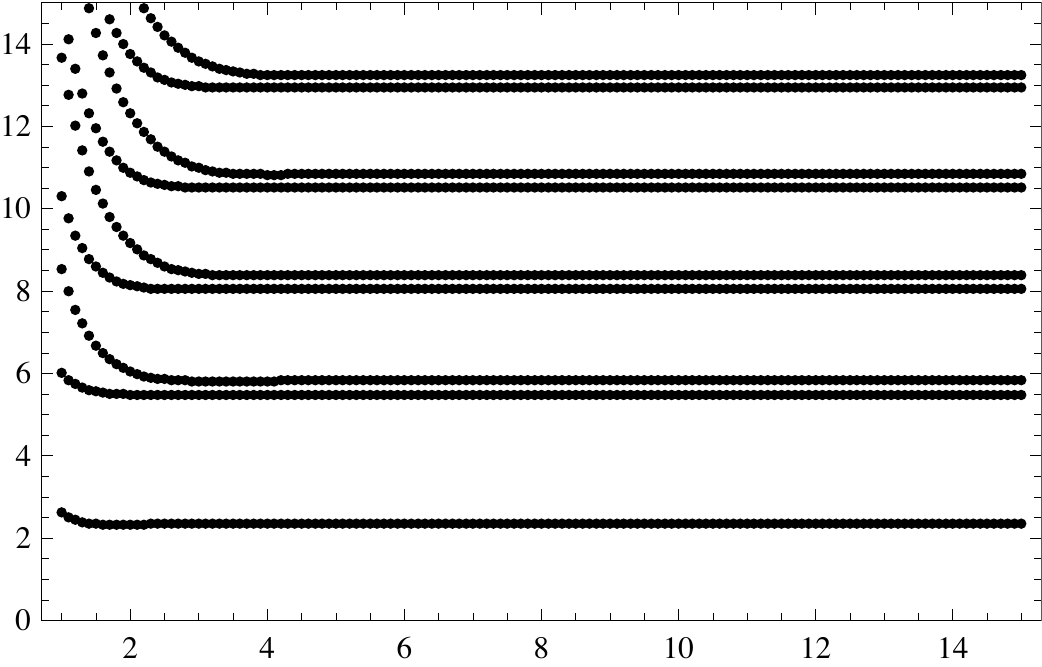}}
\put(-25,260){$e^{-3\chi_1 \r_I} q$}
\put(215,260){$e^{-3\chi_1 \r_I} q$}
\put(5,110){$ q$}
\put(245,110){$ q$}
\put(180,150){$\log\r_{I}$}
\put(410,150){$\log\r_{I}$}
\put(180,0){$\r_{U}$}
\put(410,0){$\r_{U}$}
\end{picture} 
\caption{Spectrum of the confining solution in the string model.
In the top panels we show the dependence of the spectrum on $\r_{I}$,
keeping $\r_U=15$ fixed.
In the bottom two plots we show the spectrum as a function of the UV cutoff $\r_U$,
while keeping $\r_I=10^{-3}$ fixed.
The left panels are for $\phi=0$, while the right panels have $\phi=-\frac{\log 3}{4}$. }
\label{Fig:F4SpectrumFPcomment}
\end{center}
\end{figure}

We report on the results obtained by computing the spectrum of bound states for the critical solutions
in which $\phi$ is constant (i.e.~in the limits in which the scale $s_{\ast}=\pm\infty$).
These are displayed in Fig.~\ref{Fig:F4SpectrumFPcomment}.
 We want to show how the results depend on the cutoffs $\r_I$ and $\r_U$.
First of all, we notice from the top panels in the figure that by keeping $\r_U$ large and varying $\r_I$,
for $\r_I\gg 0$ the lightest state does show some suppression.
Also, we observe that when $\r_I$ is close to the end-of-space, 
the whole spectrum converges smoothly to the physical results.

By looking at the bottom panels in the figure, 
in which we kept $\r_I =10^{-3}$ very close to the end-of-space,
but varied $\r_U$, we see that for very small values of $\r_U$ 
the spectrum looks distorted, with various unphysical features emerging.
 Also in this case the convergence towards the physical result is smooth, 
 and in particular the values of $\r_I$ and $\r_U$ that we use 
in the body of the paper are safely in the region where the discrepancy between the numerical 
results and the extrapolation to the physical limit is negligibly small.

One can think of this part of the study in analogy to what is done numerically on the lattice. Given that 
we must work with (unphysical) cutoffs, we perform our calculations for finite values of $\r_I$ and $\r_U$,
and then extrapolate the results from finite $\r_U$ to the physical $\r_U\rightarrow +\infty$, which 
broadly speaking corresponds to the continuum-limit extrapolation (in the language of lattice calculations).
Also, we must extrapolate to the limit $\r_I\rightarrow 0$. This is 
somewhat reminiscent of the procedure that allows to evaluate 
and possibly remove finite-size effects on the lattice, 
in the sense that for finite $\r_I$ the model contains two IR-scales: the physical confinement 
scale and the scale set by the finite volume of the system. 
The latter being unphysical, we want to perform our calculations 
in the regime where finite-volume effects are negligible, and hence we need to check what choices of $\r_I$
are close enough to the end of space to yield physical results.
We performed this kind of study for all the models discussed in the paper, but we report 
only on the case of the $F(4)$ theory, since the others are similar, and 
would add nothing to the present paper.


\end{document}